\documentclass[preprint2]{aastex}

\shorttitle{Observations and Modelling of EC2}
\shortauthors{Ruffle et al.}

\begin{document}

\title{Galactic Edge Clouds I: \\ 
       Molecular Line Observations and \\ 
       Chemical Modelling of Edge Cloud 2}

\author{P. M. E. Ruffle\altaffilmark{1}, T. J. Millar\altaffilmark{2}, H. Roberts\altaffilmark{2},}
\affil{School of Physics and Astronomy, The University of Manchester, UK}
\email{pruffle@nrao.edu}

\author{D. A. Lubowich,}
\affil{Department of Physics and Astronomy, Hofstra University, Hempstead, New York}

\author{C. Henkel,}
\affil{Max-Planck-Institut f\"ur Radioastronomie, Bonn, Germany}

\author{J. M. Pasachoff and G. Brammer\altaffilmark{3}}
\affil{Department of Astronomy, Williams College, Williamstown, MA}

\altaffiltext{1}{now at NRAO, Green Bank, WV.}
\altaffiltext{2}{now at School of Mathematics and Physics, Queen's University Belfast.}
\altaffiltext{3}{now at Department of Astronomy, Yale University, New Haven, CT.}

\begin{abstract}

Edge Cloud 2 (EC2) is a molecular cloud, about 35 pc in size, with one of the largest galactocentric distances known to exist in the Milky Way. We present observations of a peak CO emission region in the cloud and use these to determine its physical characteristics. We calculate a gas temperature of 20\,K and a density of $n(\mathrm{H}_{2}) \sim 10^{4}$\,cm$^{-3}$. Based on our CO maps, we estimate the mass of EC2 at around $10^{4}\,M_{\sun}$ and continuum observations suggest a dust-to-gas mass ratio as low as 0.001. Chemical models have been developed to reproduce the abundances in EC2 and they indicate that: heavy element abundances may be reduced by a factor of five relative to the solar neighbourhood (similar to dwarf irregular galaxies and damped Lyman alpha systems); very low extinction ($A_\mathrm{V} < 4$\,mag) due to a very low dust-to-gas ratio; an enhanced cosmic ray ionisation rate; and a higher UV field compared to local interstellar values. The reduced abundances may be attributed to the low level of star formation in this region and are probably also related to the continuing infall of primordial (or low metallicity) halo gas since the Milky Way formed. Finally, we note that shocks from the old supernova remnant GSH 138--01--94 may have determined the morphology and dynamics of EC2.

\end{abstract}


\section{Introduction}

Observations of CO emission at large galactocentric distances have detected a number of molecular clouds, among which is Edge Cloud 2 (hereafter EC2), with an estimated kinematic galactocentric distance of $R \sim 28$\,kpc \citep{digel94}. Although this kinematic distance for the far-outer Galaxy should be treated with caution, EC2 appeared to be up to 6\,kpc further away than Edge Cloud 1 (EC1), the next most distant molecular cloud detected by \citeauthor{digel94}, and much further than the extent of the optical disk of the Milky Way, $\sim$~15--19\,kpc \citep{fich8907,robin9211}. This is almost as far as the most distant H\,{\sc i}, detected at $\sim$~30\,kpc \citep{kulkarni8208}. More recently \citet{levine0606} have traced the spiral structure in the southern half of the Galaxy out to at least 25\,kpc, implying a minimum radius for the gas disk. 
EC2 has an effective radius of $\sim$20 pc and is situated $\sim$350 pc below the
distant warped Galactic plane \citep{digel96iau1}. The CO luminosity of
EC2 is at least a factor of 2 larger than those of the other 10 clouds
detected at large distances, and is comparable to that of the Taurus Giant
Molecular Cloud \citep{digel94}. EC2 is also the only edge cloud detected in
the high-density tracer CS \citep{digel96iau2}.  The CO maps of EC2 show
that it has sub-structure. It is in a region of extremely low gas pressure and very small spiral arm perturbation.

EC2 was also found to have an associated H\,{\sc ii} region excited by an early B star (MR1) \citep{degeus93} that appears to have triggered star formation. Studies of metallicity as a function of galactocentric distance have shown that there is a galactic metallicity gradient. Spectra of MR1 \citep{smartt96} indicate significant metal depletion -- with elemental abundances reduced on average by some 0.5 dex -- and that its galactocentric distance is more likely to be in the range 15-19\,kpc (placing it at the limit of the optical disk). Subsequently, \citet{rolleston0011} have calculated metal depletions of about five for C,~N, and O. \citet{kobayashi00} used NIR observations to argue that MR1 has triggered the formation of young stellar objects in EC2. \citet{snell02} argue that it is the most distant cloud in the Milky Way, with evidence for massive star formation, although this distinction is claimed for WB89-789 by \citet{brand0702}. More recently, two embedded young star clusters have been associated with EC2 \citep{kobayashi05}. \citet{yasui0610} identified 52 members in one of the clusters, suggesting that the cluster is a T Tauri association. \citet{stil0112} associate EC2 with the approaching side of the H\,{\sc i} shell from supernova remnant (SNR) GSH 138--01--94, thereby setting the galactocentric distance $R \sim 23.6$\,kpc for EC2. Because of the uncertainty of $R$ for EC2, we have used two heliocentric distances, 14 and 20\,kpc ($R = 22$ and 28\,kpc) for any distance sensitive calculations in this work.

Current models of galacto-chemical evolution include time and spatial variations in the halo gas infall and star formation rate whereby the Galactic halo, bulge, and thick disk formed first, separately from the thin disk, in two infall episodes. Chemical abundances of the interstellar medium and their radial variation across galactic disks provide a fundamental set of constraints for theories of disk formation and evolution. The most accepted mechanism to explain the existence of abundance gradients in disk galaxies is the so-called `biased-infall' \citep{chiappini99}, where infall of gas occurs at a faster rate in the innermost regions than in the outermost ones (`inside-out' disk formation). Growing observational evidence for slow formation of disks in spiral galaxies associated with continuing infall of primordial (or low metallicity) gas over the lifetime of the disk \citep[e.g.][]{braun9901} seems to give support to the above scenario. \citet{lubowich0006} demonstrated that continuous infall of low-metallicity gas is occurring in the Galactic Center. 

Of particular relevance to test these models are the abundances in the very outer galactic disks. Chemical evolution models for abundance gradients and the formation of the Milky Way \citep{chiappini0106} show that the steepness of the outer gradients are particularly sensitive to thresholds in star formation, to the halo-thick disk enrichment history, and to the radial variation of the disk formation timescales. They concluded that some abundance ratios increase substantially toward the outermost disk regions but that `more observations at large galactocentric distances are needed to test these predictions'. The fact that N is almost constant with the galactocentric distance up to 18\,kpc \citep{chiappini0302} reflects the high N production in AGB stars. The primary N contribution from AGB stars is however very uncertain. At such large galactocentric distances a N production threshold may be operating due to the paucity of late-type stars. 

Galactic chemical evolution also predicts that the abundances of C, N, O, $^{13}$C, and $^{15}$N will be lower in clouds at the edge of the Galaxy than in any other cloud of the Galactic disk \citep{maciel9905}. The composition of these edge clouds should be similar to that of the early Galactic disk modified by infall from the halo. Thus the metallicity in clouds such as EC2 is expected to be similar to dwarf irregular galaxies, giving us an unique opportunity to study gas from the early stages of the formation of the Galactic disk \citep{kobayashi00}. 


We have, therefore, searched for emission from a number of molecules towards EC2, concentrating on the northern CO peak \citep[which is coincident with the T Tauri cluster identified by][]{yasui0610}, in order to constrain its physical conditions and chemical composition. Such molecular evolution in low metallicity Galactic clouds also provides a local laboratory in understanding molecular gas in extragalactic sources such as high-$z$ quasars. In particular, since deuterated molecules are sensitive tracers of physical conditions and chemical pathways, we have also searched for a number of deuterium (D) bearing molecules, noting that \citet{rogers0509} have detected D{\sc i} towards the Galactic anticenter. 
We also examine the role of SNR GSH 138--01--94 in providing the energy input for both the chemistry and star formation observed in EC2. 

\section{Observations}\label{ec2obs} 

We have completed over 210 hours of observations toward this object using the 12\,m Arizona Radio Observatory (ARO) telescope (formerly part of the National Radio Astronomy Observatory), the 100\,m Max-Planck-Institut f\"{u}r Radioastronomie (MPIfR) Effelsberg telescope, the 15\,m James Clerk Maxwell Telescope (JCMT), and the 30\,m Institut de Radio Astronomie Millim\'{e}trique (IRAM) telescope. 
The line observations range in frequency from 4.83 GHz (H$_{2}$CO 1$_{1,0}$--1$_{1,1}$) to 492 GHz (C{\sc i} $^3$P$_{1}$--$^3$P$_{0}$). 
EC2 is extended in CO and we have observed a number of different positions within $\sim$40\arcsec\ of the three peaks seen in it by \citet{digel94}. Fig. \ref{ec2_map_positions} shows observed positions A through K on our contour map of CO 2--1 emission. Positions A, J and K correspond to positions {\it copk1, copk2} and {\it copk3} respectively from \citet{digel94}.
In this work we present the results of our line search at position A 
 ($\alpha_{\rm 2000} = 02^\mathrm{h}\,48^\mathrm{m}\,38.5^\mathrm{s}$, $\delta_{\rm 2000} = 58\arcdeg\,28\arcmin\,28.1\arcsec$), 
which at the time showed the strongest line spectrum based on a twelve position HCO$^+$ map. 
Our subsequent CO maps centered on position E, 
($\alpha_{\rm 2000} = 02^\mathrm{h}\,48^\mathrm{m}\,38.5^\mathrm{s}$, $\delta_{\rm 2000} = 58\arcdeg\,28\arcmin\,58.3\arcsec$), 
covering the northern peak in emission, are also presented. These maps revealed that the northern peak in CO emission was $\sim$30\arcsec\ north-east of position A, and $\sim$25\% stronger. 
See \citet{lubowich04carn} for observations of positions B and C. 
Full details of observed line spectra for positions other than A, and CO maps covering the southern peak (centered at 
$\alpha_{\rm 2000} = 02^\mathrm{h}\,48^\mathrm{m}\,23.6^\mathrm{s}$, $\delta_{\rm 2000} = 58\arcdeg\,23\arcmin\,59.0\arcsec$) 
are available in \citet{ruffle06phd}. 

Observations at the ARO 12\,m were made between June 2002 and January 2006 at 2 and 3\,mm. Data were obtained in beam switching mode. The beam size ($\theta_\mathrm{b}$) of the 12\,m telescope varies between 93\arcsec\ at 72 GHz and 43\arcsec\ at 150 GHz. Observing conditions were generally good, with system temperatures ranging from 140 to 350\,K for the 2002 June observations. For subsequent observations, system temperatures ranged from 170 to 820~K. Dual channel SIS receivers were used in conjunction with two spectrometer systems: Filter Banks 1 and 2 (FB1 and FB2), and a digital Millimeter Auto Correlator (MAC1). The 256 channel 100 and 250~kHz filter bank spectrometers were used simultaneously, with 128 channels allocated to each polarisation, giving bandwidths of 12.8 and 32 MHz respectively. The two polarisations were subsequently averaged during data reduction. The MAC1 was used to confirm filter bank data and set to a resolution of 50, 100, 200 or 400~kHz and bandwidths of 300 or 600~MHz. The ARO 12\,m produces spectral line data calibrated to the $T^*_\mathrm{R}$ (K) corrected antenna temperature scale. 

We used the Effelsberg 100\,m telescope in June 2002, December 2002, January 2003, May 2004 and November 2005 at 1.0, 1.3, 2, and 6\,cm to observe transitions of SO, NH$_3$, H$_2$O, and H$_2$CO with beamwidths ($\theta_\mathrm{b}$) of about 30$''$, 40$''$, 55$'''$ and 160$''$ and typical system temperatures of 400, 150, 80, and 40\,K on a main beam brightness temperature scale ($T_{\rm mb}$). $T_{\rm mb}$ to flux density conversion ratios were 1.6, 1.4, 1.8, and 2.4\,K/Jy, respectively. For the 1.0\,cm SO 1$_0$--0$_1$ line and the 2\,cm H$_2$CO 2$_{1,1}$--2$_{1,2}$ transition, single channel primary focus HEMT (High Electron Mobility Transistor) receivers were employed. At 1.3\,cm, for H$_2$O and NH$_3$, a primary focus dual channel HEMT receiver was used, allowing us to add signals with orthogonal linear polarization, thus reducing the effective system temperature given above by $\sim$30\%. At 6\,cm, H$_2$CO 1$_{1,0}$--1$_{1,1}$ observations were taken with a four channel secondary focus receiver. Position switching with appropriate offsets, alternating between east and west in azimuth, was used for all measurements. An `AK90' autocorrelator included four spectrometers with 2048 channels and bandwidths of 20\,MHz each. Pointing was accurate to about 10\arcsec\ or better. 

In June 2004--July 2005, we used the JCMT to map the 2--1 and 3--2 transitions of several isotopes of CO and to observe the 492 GHz transition of C{\sc i} with beam sizes ($\theta_\mathrm{b}$) which range from 23\arcsec\ at 220 GHz to 10\arcsec\ at 492 GHz. We also searched unsuccessfully for the HCO$^+$ 3--2 transition at 267 GHz. Data were obtained in a position-switching mode in a tracking coordinate frame with a chop throw of 600\arcsec\ at an angle of 270\degr. For the CO 2--1 and HCO$^+$ observations, the single channel A3 receiver was used yielding system temperatures from 260--330\,K. For the CO 3--2 transitions the dual channel B3 receiver was used with two low-noise SIS mixers in single-sideband mode, yielding system temperatures from 420--800\,K. Receiver W was used to observe the C{\sc i} transition in single sideband with a system temperature of 2400--4100\,K. The JCMT produces spectral line data calibrated to the $T^*_\mathrm{A}$ (K) antenna temperature scale, and this was converted to $T^*_\mathrm{R}$ using the JCMT's $\eta_\mathrm{fss}$ (forward scattering and spillover) efficiencies. 

Finally, in October-November 2005 we used the MAMBO II bolometer at the IRAM 30\,m telescope to make a 1.2\,mm dust emission map of EC2. The bolometer comprises 117 pixels with a resolution of 11\arcsec. Zenith opacity ranged from 0.07--0.27 during the observations and the sky noise difference (between on and off source) was generally low, varying between 34 and 90\,mJy/beam/2\,Hz. We made five maps of EC2 centered on $\alpha_{2000}$=02:48:41.0, $\delta_{2000}$=58:29:27.9, with an rms which ranged from 8.8--11.5\,mJy per beam. The five maps were combined using MOPSIC.

\section{Analysis}\label{analysis} 

\begin{figure} 
\includegraphics[width=75mm]{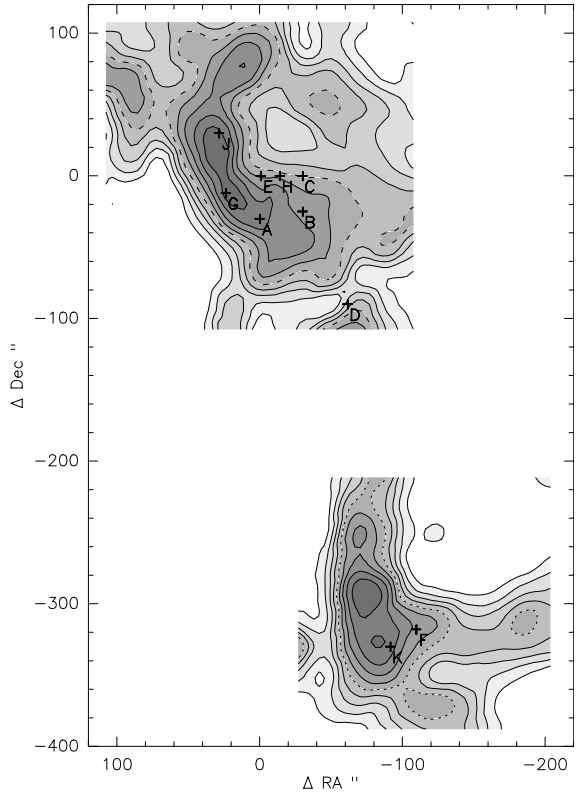} 
\caption{JCMT 15m May-July 2005 combined contour map of observed CO 2--1 intensities, 
	 showing observed positions A through K. 
	 Positions A, J and K correspond to positions {\it copk1, copk2} and {\it copk3} respectively from \citet{digel94}.
	 Axis offsets in arcsec from position E 
	($\alpha_{2000}=$ 02:48:38.5, $\delta=$ 58:28:58.3, $v_\mathrm{rad} = -103.70 $ km\,s$^{-1}$). 
	{\it EC2MAP1} (top) CO 2--1: peak $T^*_\mathrm{R}$ = 6.906\,K, rms = 0.321\,K. 
	{\it EC2MAP2} (bottom) CO 2--1: peak $T^*_\mathrm{R}$ = 7.124\,K, rms = 0.22\,K.}
\label{ec2_map_positions}
\end{figure}

\begin{figure} 
\hbox{
\includegraphics[width=72mm]{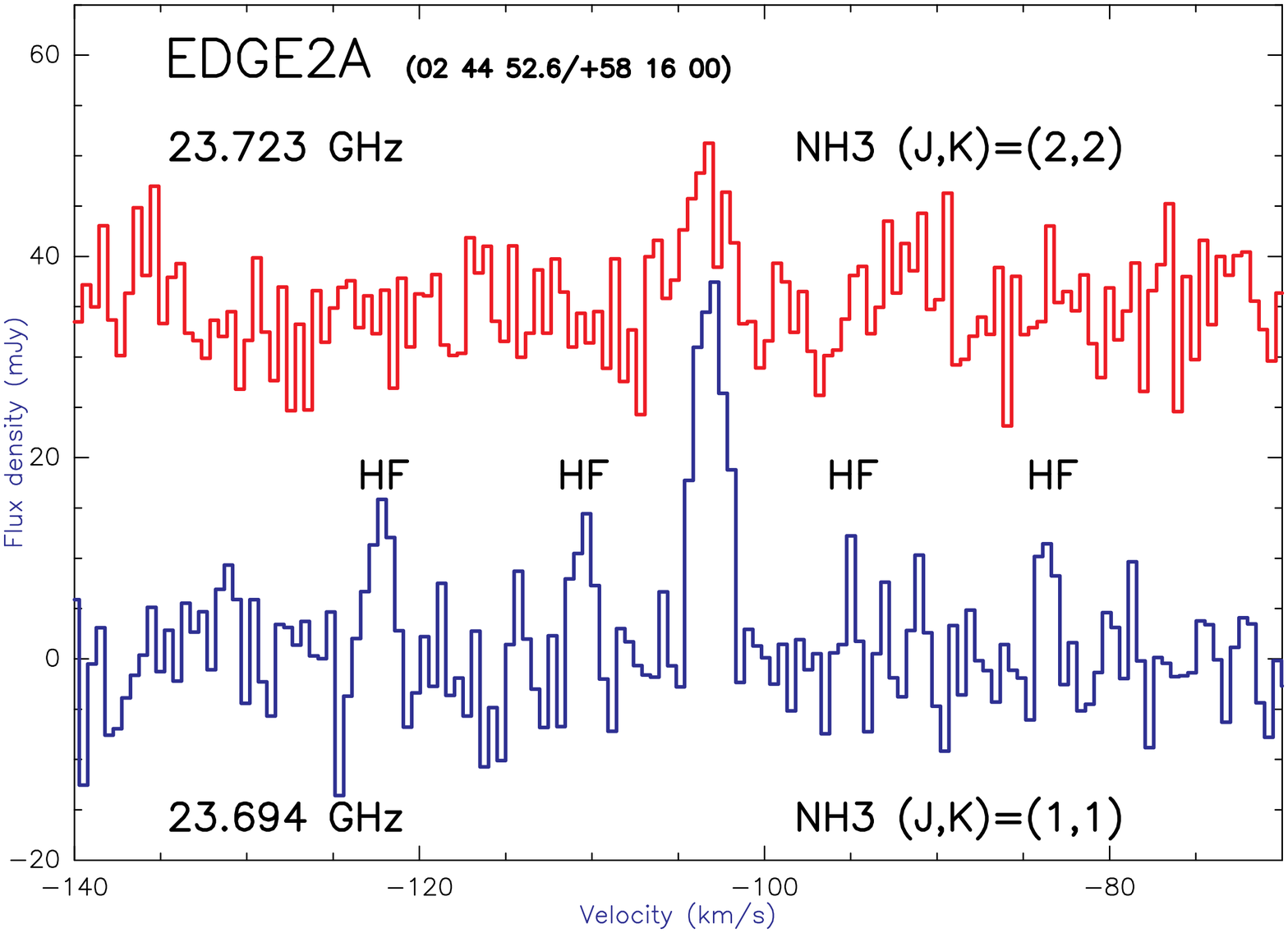}}
\vspace{6mm}
\hbox{
\includegraphics[width=72mm]{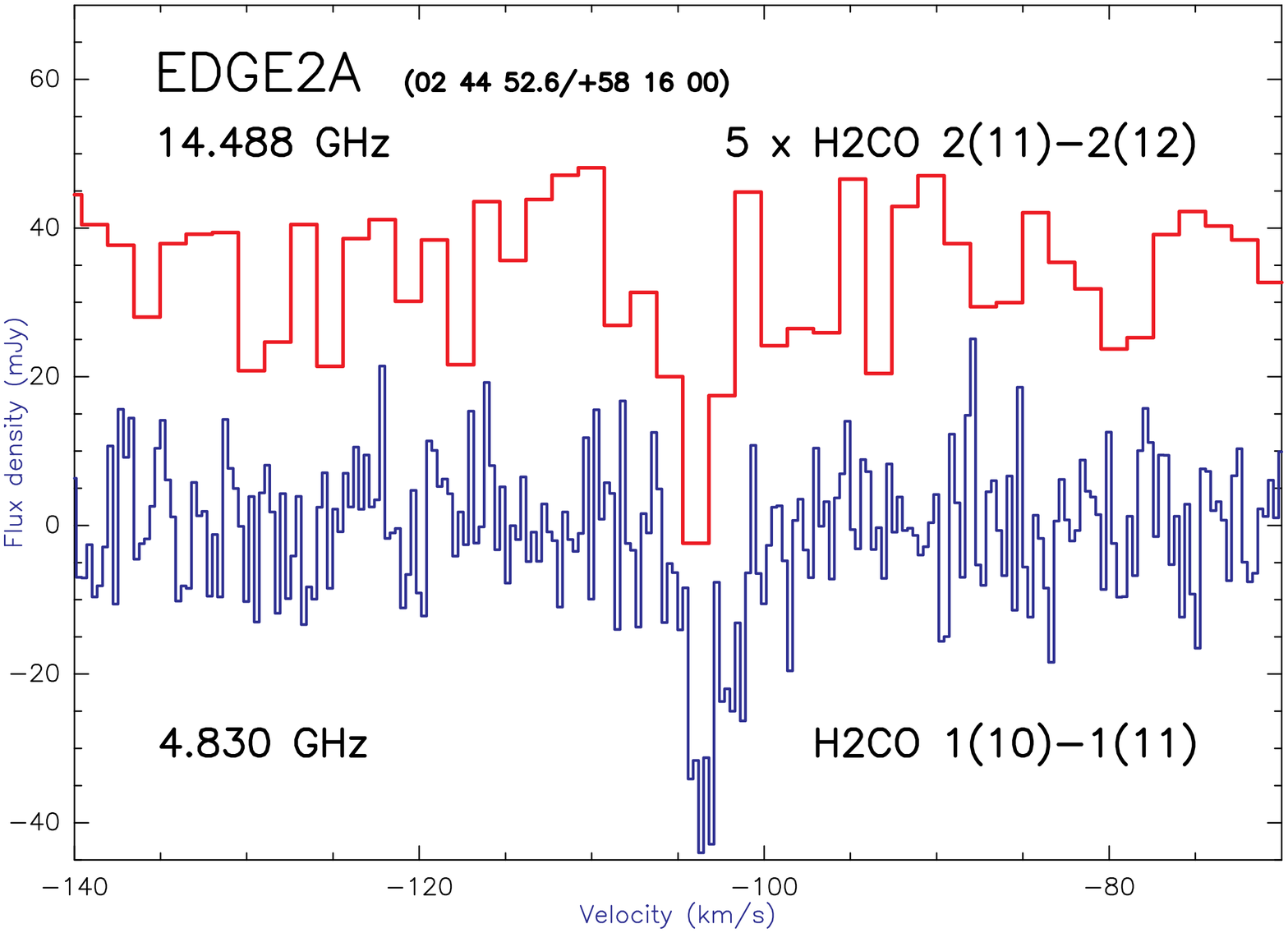}}
\vspace{1mm}
\hbox{
\includegraphics[width=43mm,angle=-90]{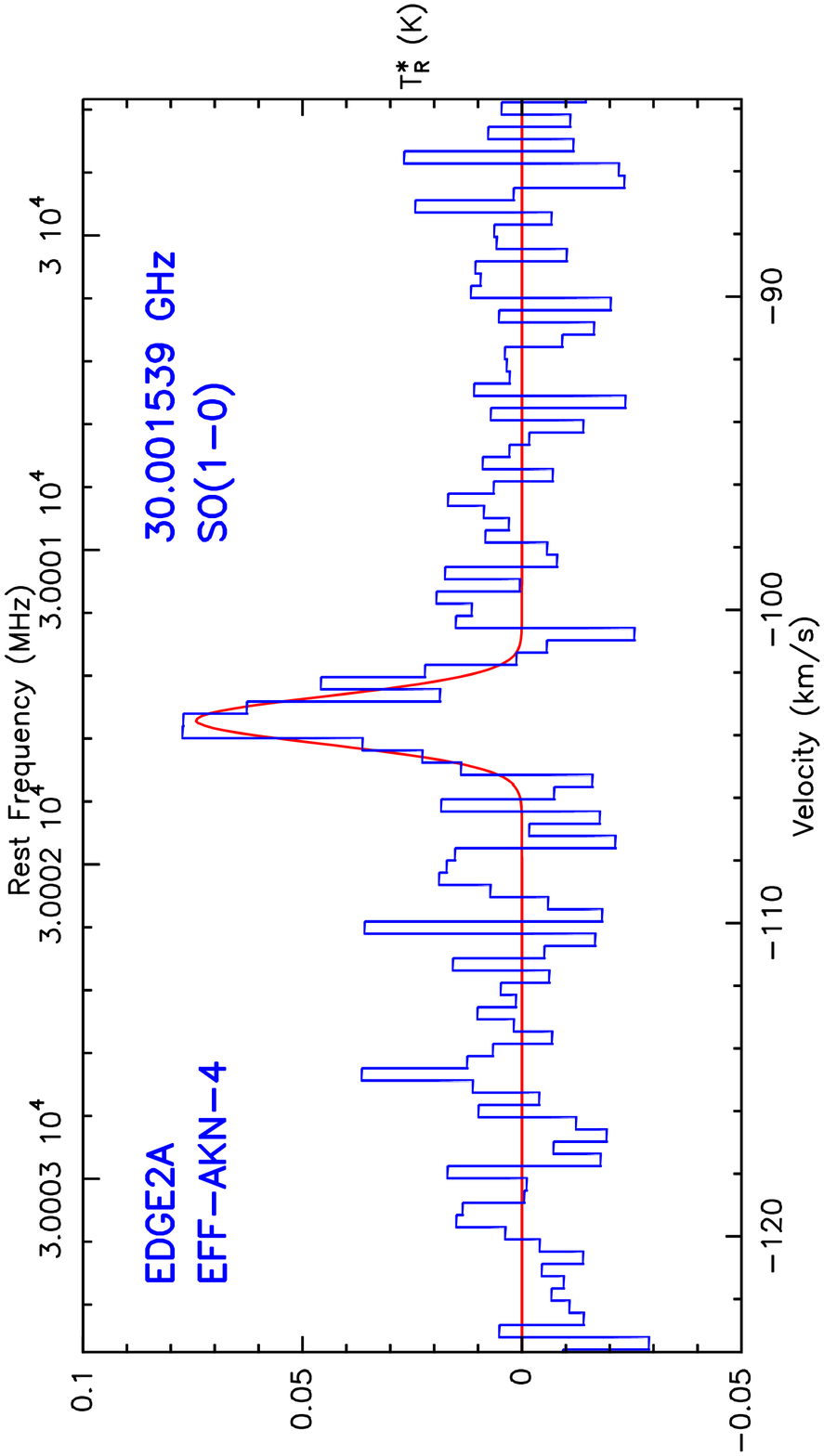}}
\caption{EC2 spectra observed at the MPIfR Effelsberg 100\,m position A: 
         (top) NH$_{3}$ (J,K) = (1,1) and (2,2) (HF = group of satellite hyperfine components). 
         (middle) H$_{2}$CO 2$_{1,1}$--2$_{1,2}$ and 1$_{1,0}$--1$_{1,1}$. 
         Note that the scale for the 2$_{1,1}$--2$_{1,2}$ transition has been multiplied by 5. 
         (bottom) SO 1$_{0}$--0$_{1}$.
\label{bonn}}
\end{figure}

\begin{figure*} 
\hbox{ 
\includegraphics[scale=0.20,angle=-90]{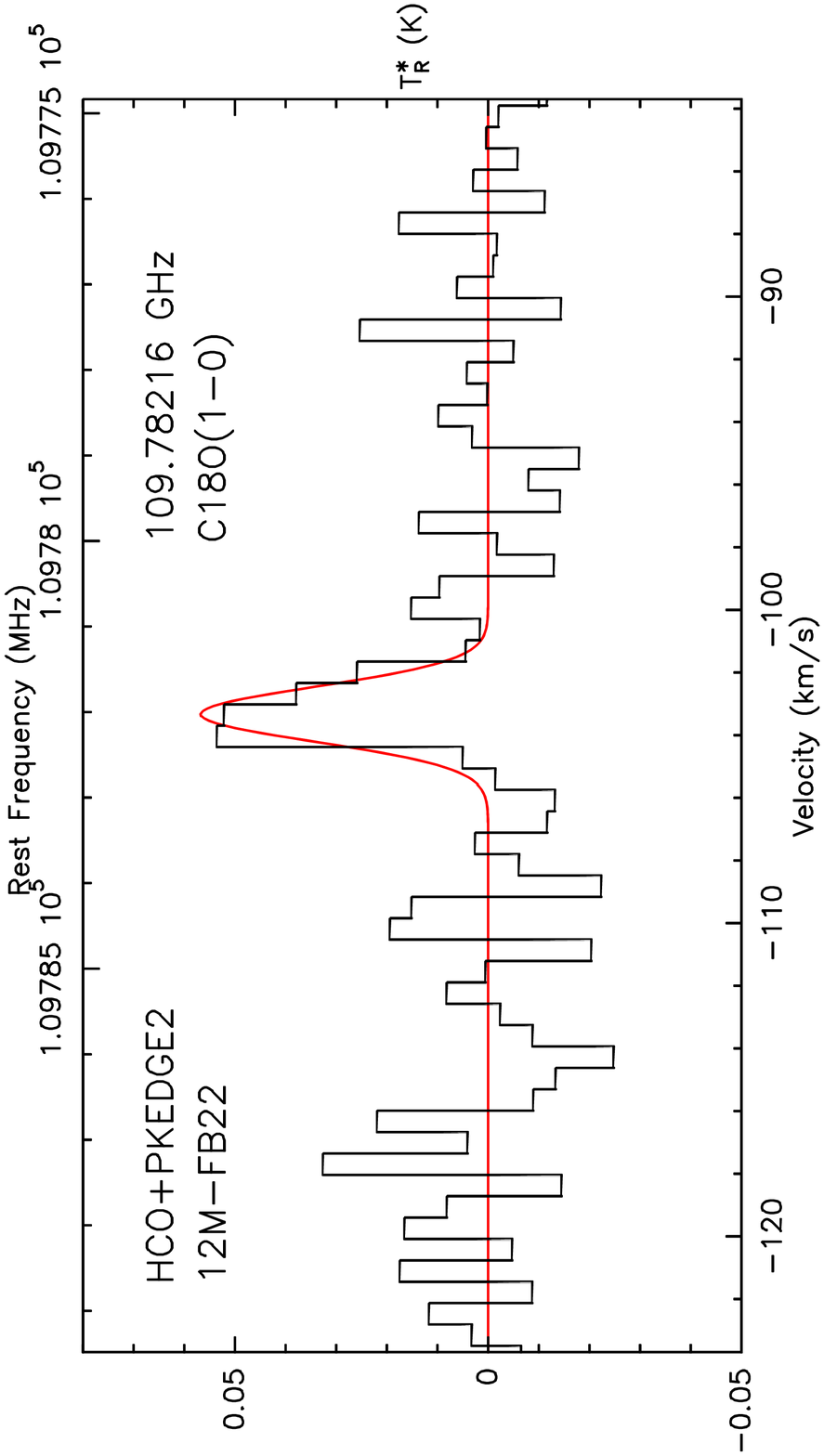}%
\hspace{3mm}%
\includegraphics[scale=0.20,angle=-90]{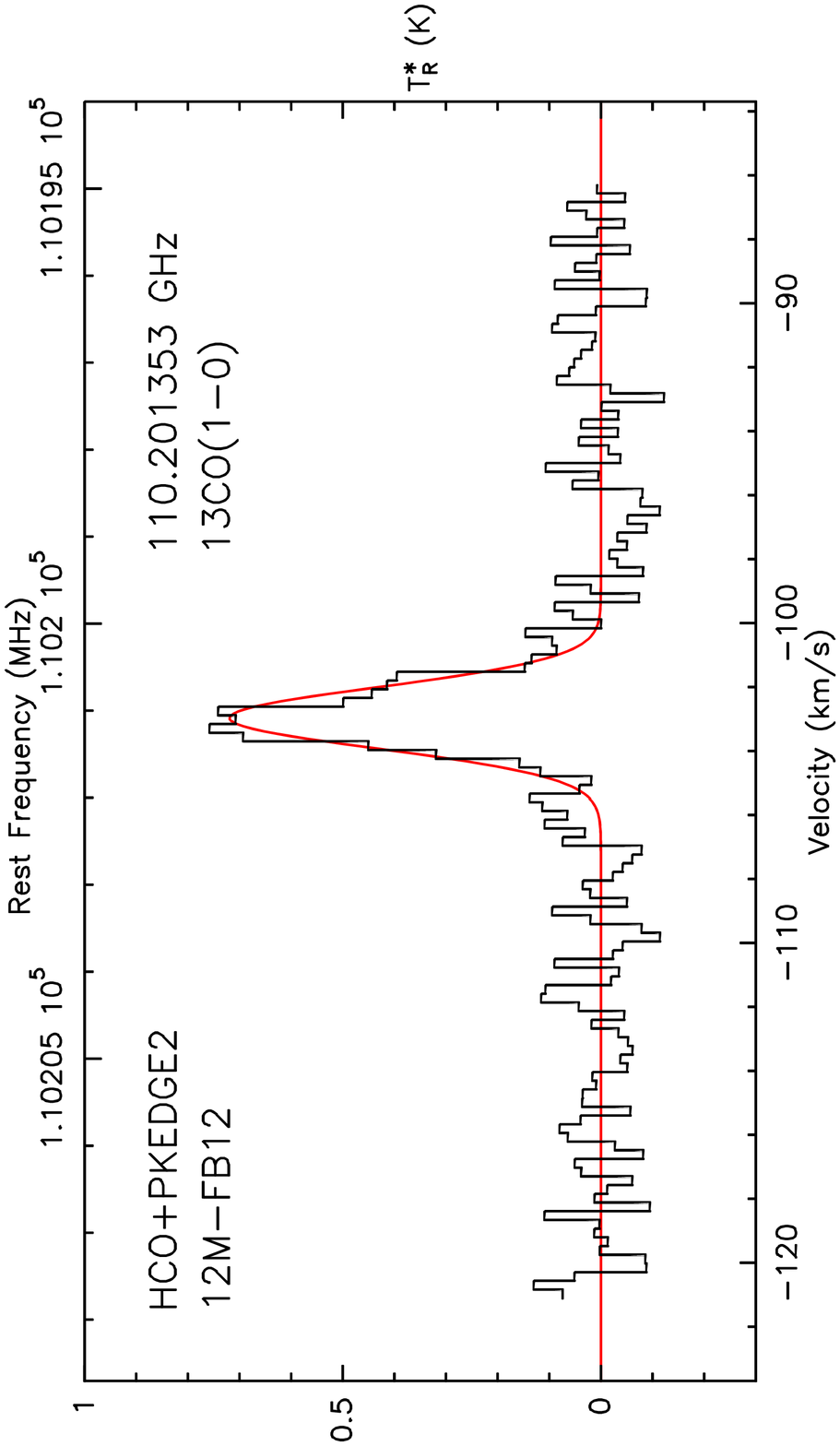}%
\hspace{3mm}%
\includegraphics[scale=0.20,angle=-90]{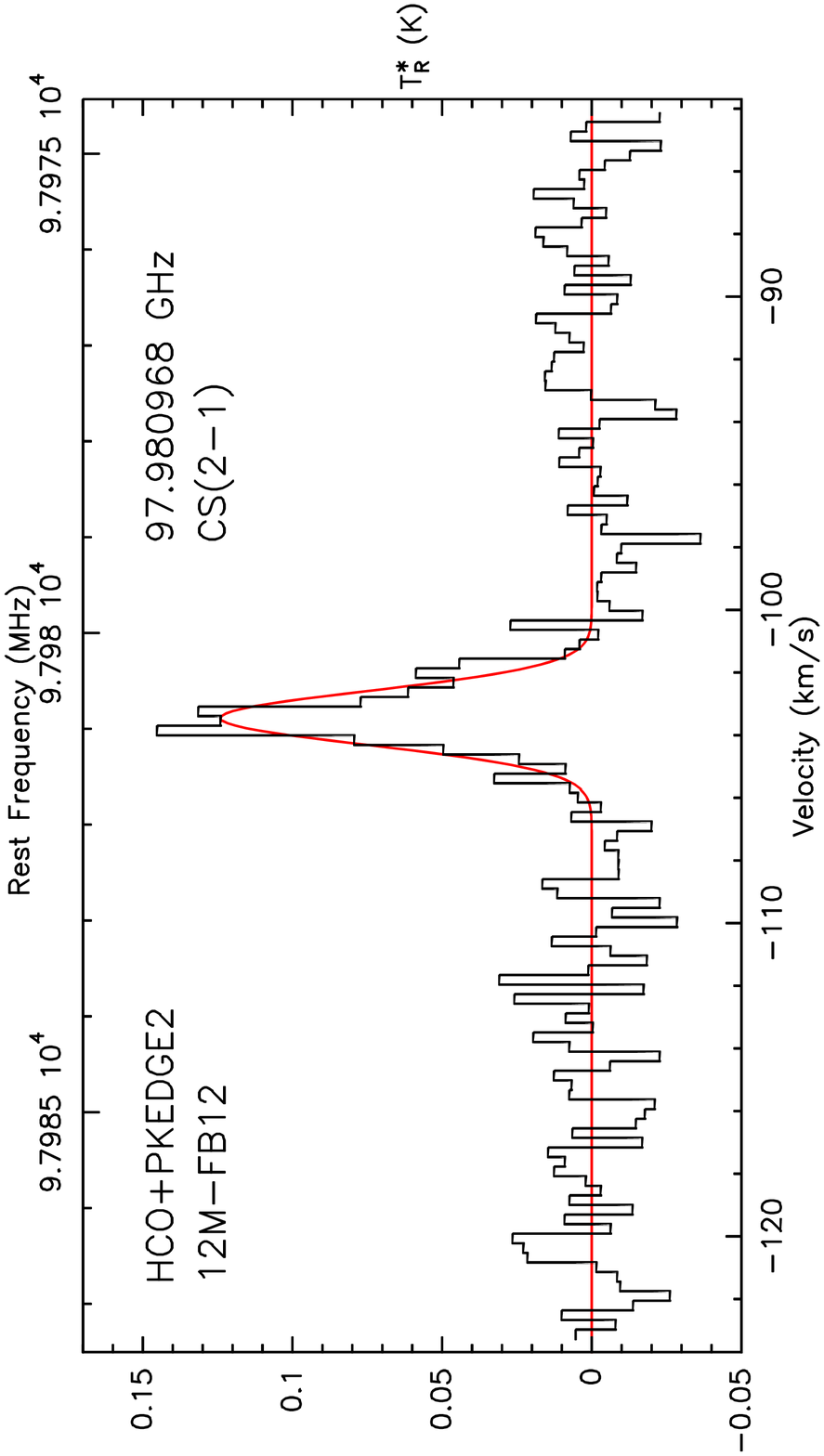}%
} 
\vspace{5mm} 
\hbox{ 
\includegraphics[scale=0.20,angle=-90]{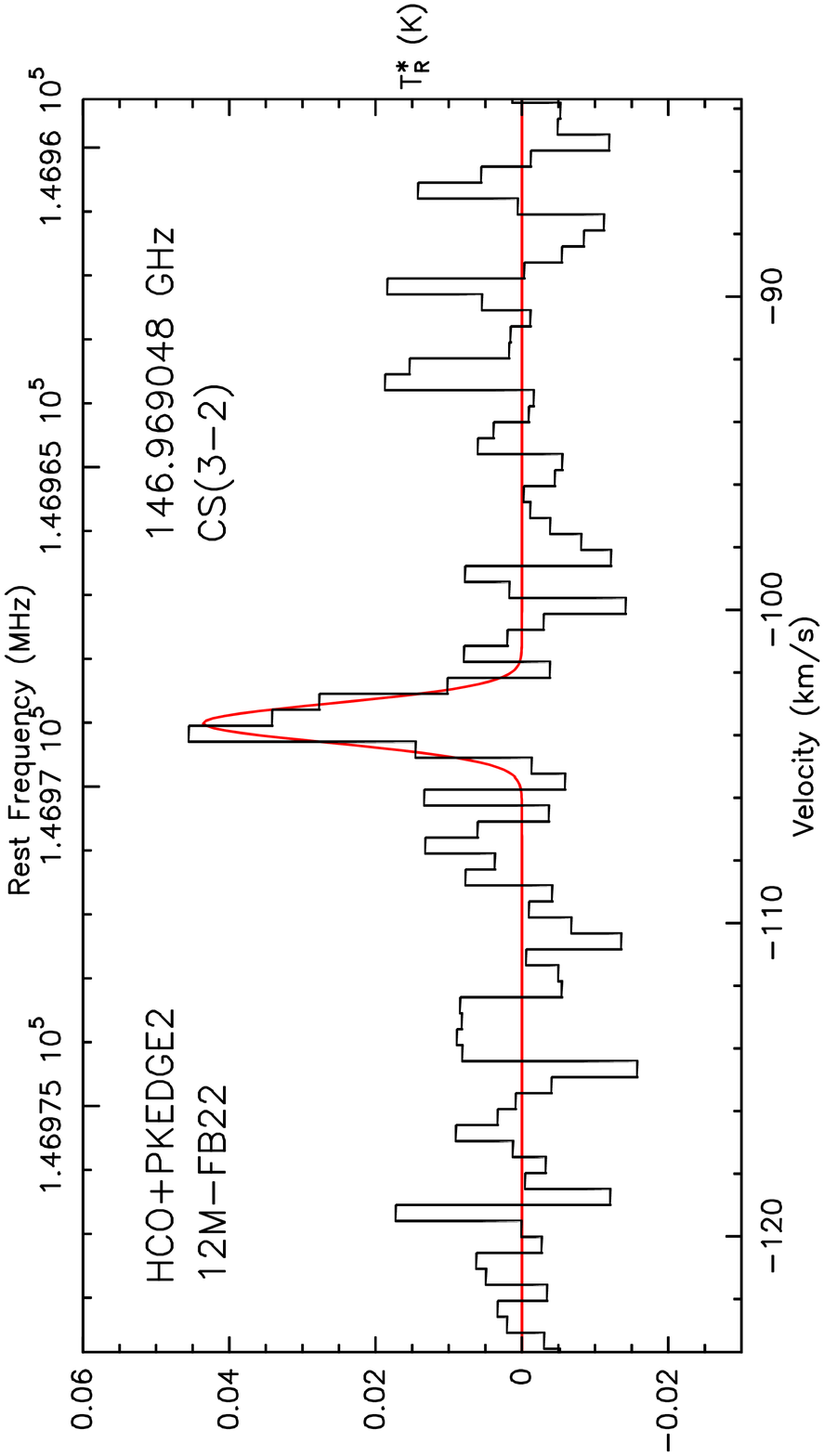}%
\hspace{3mm}%
\includegraphics[scale=0.20,angle=-90]{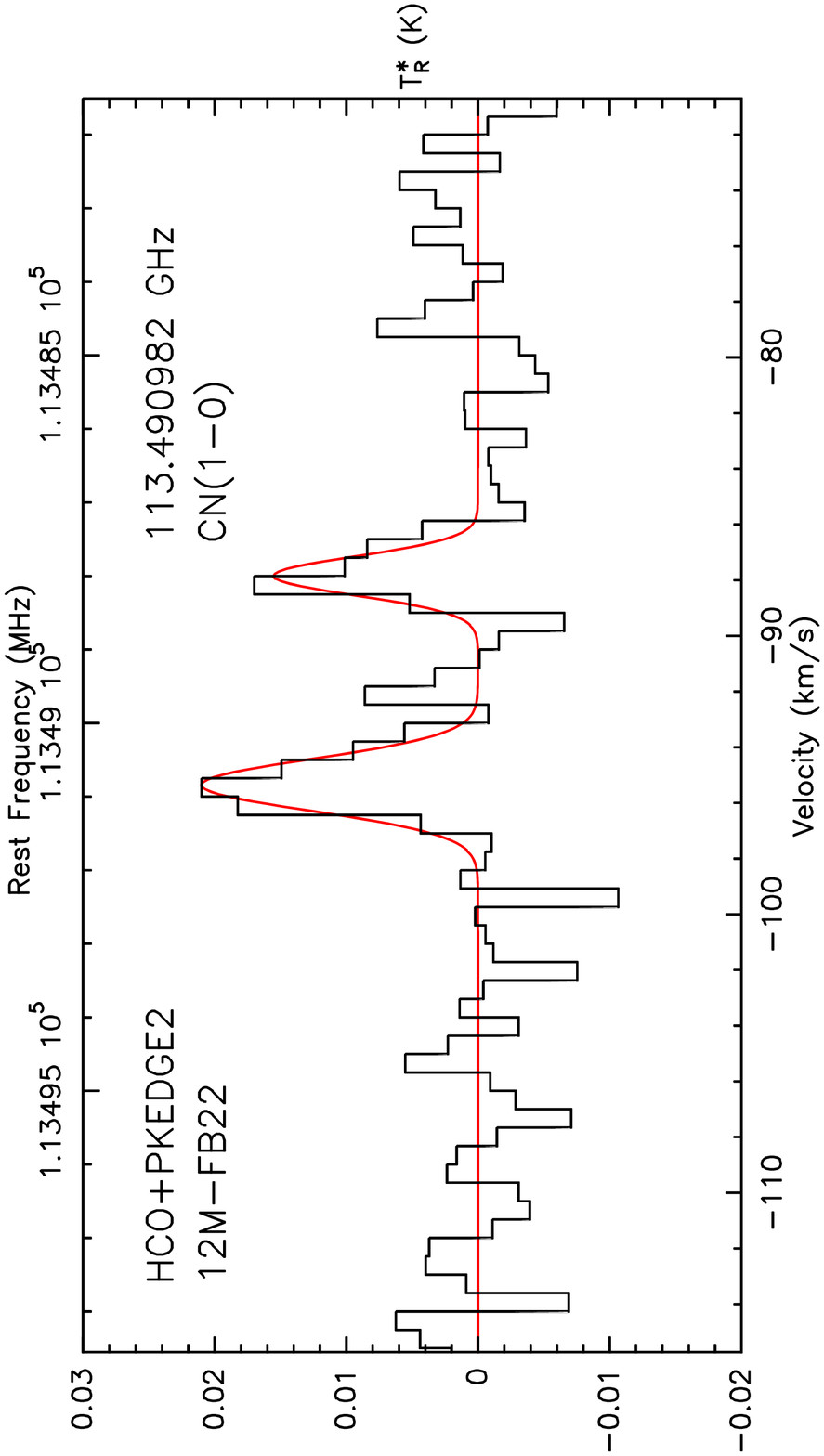}%
\hspace{3mm}%
\includegraphics[scale=0.20,angle=-90]{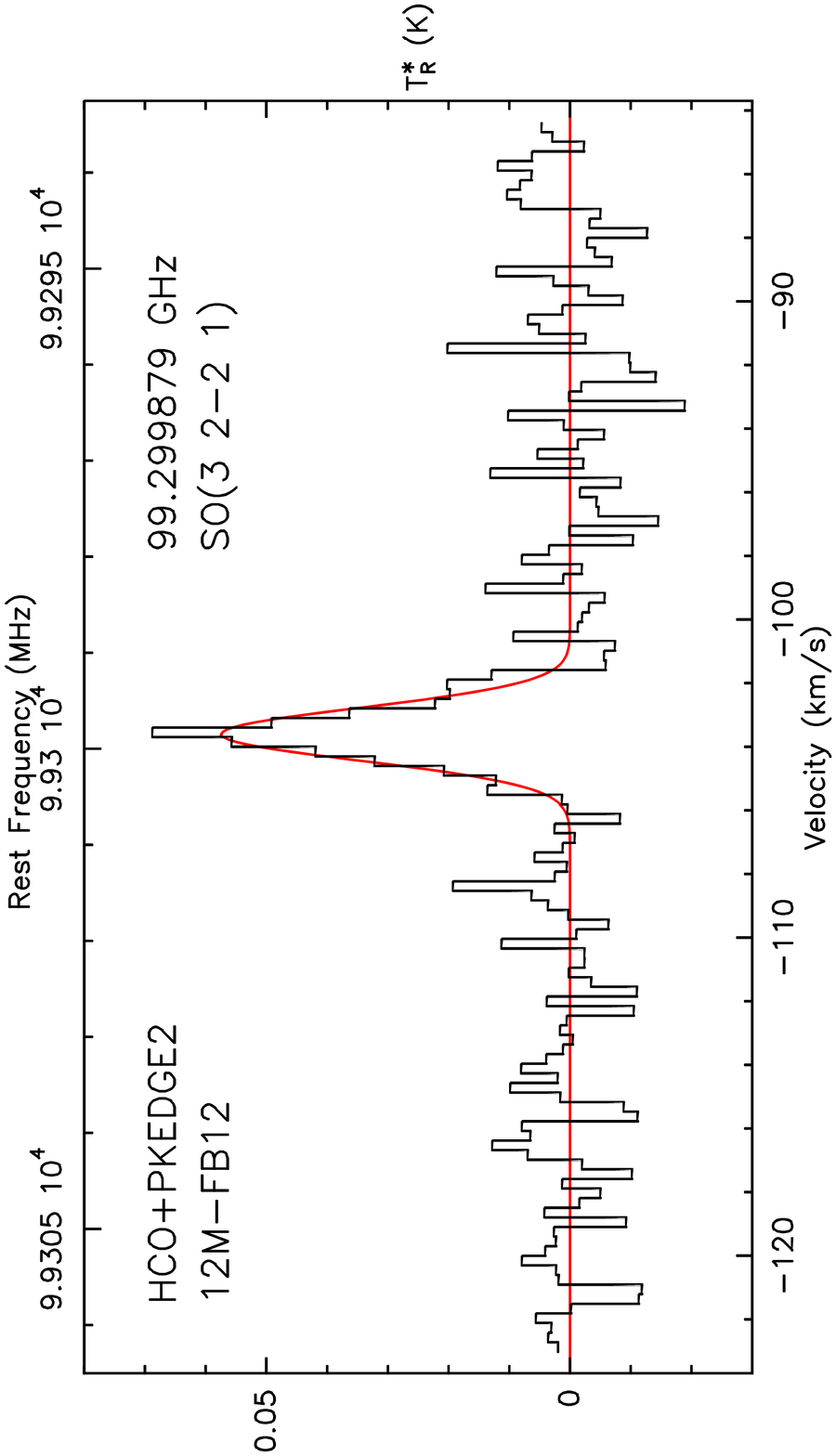}%
} 
\vspace{5mm} 
\hbox{ 
\includegraphics[scale=0.20,angle=-90]{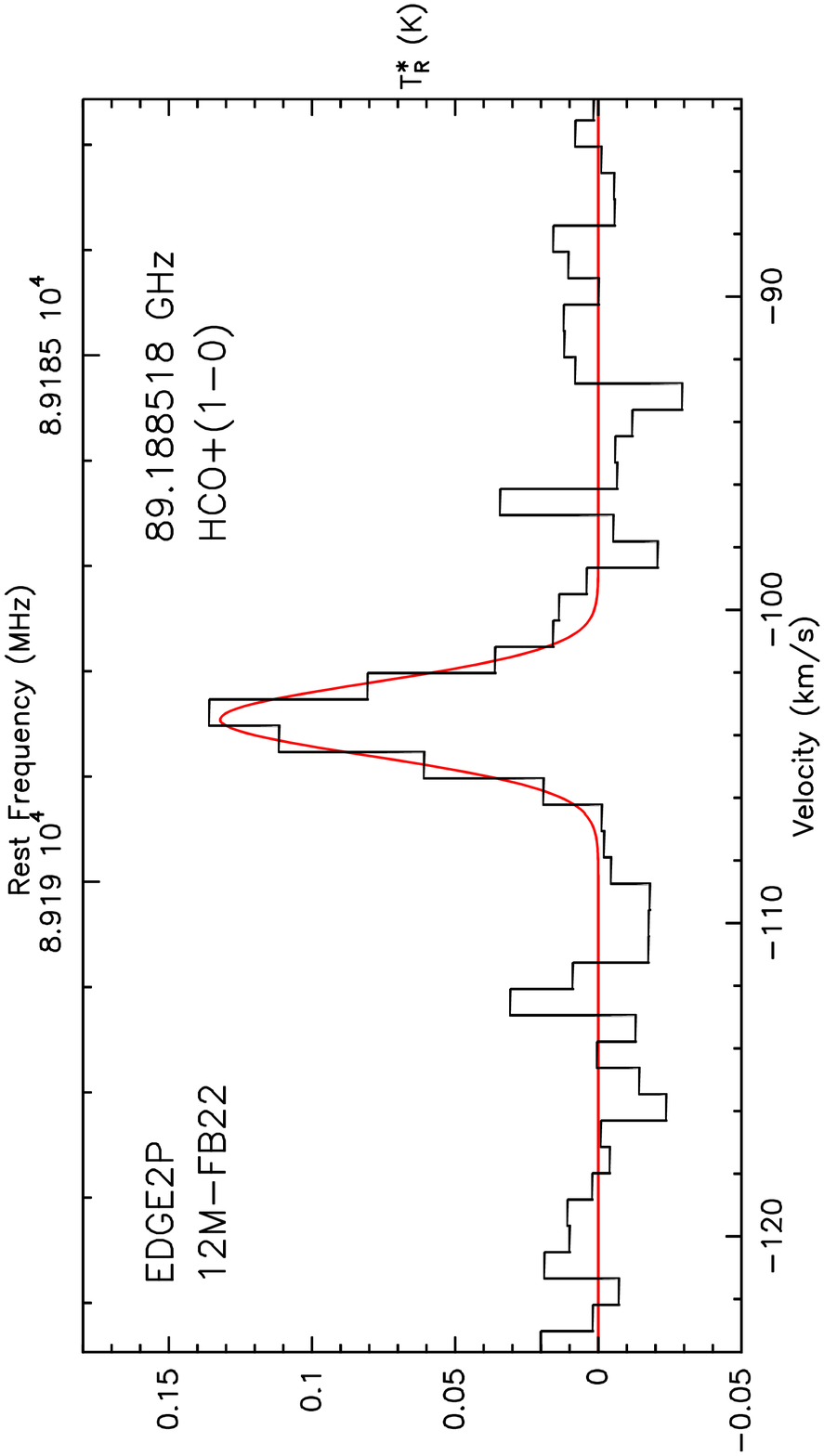}%
\hspace{3mm}%
\includegraphics[scale=0.20,angle=-90]{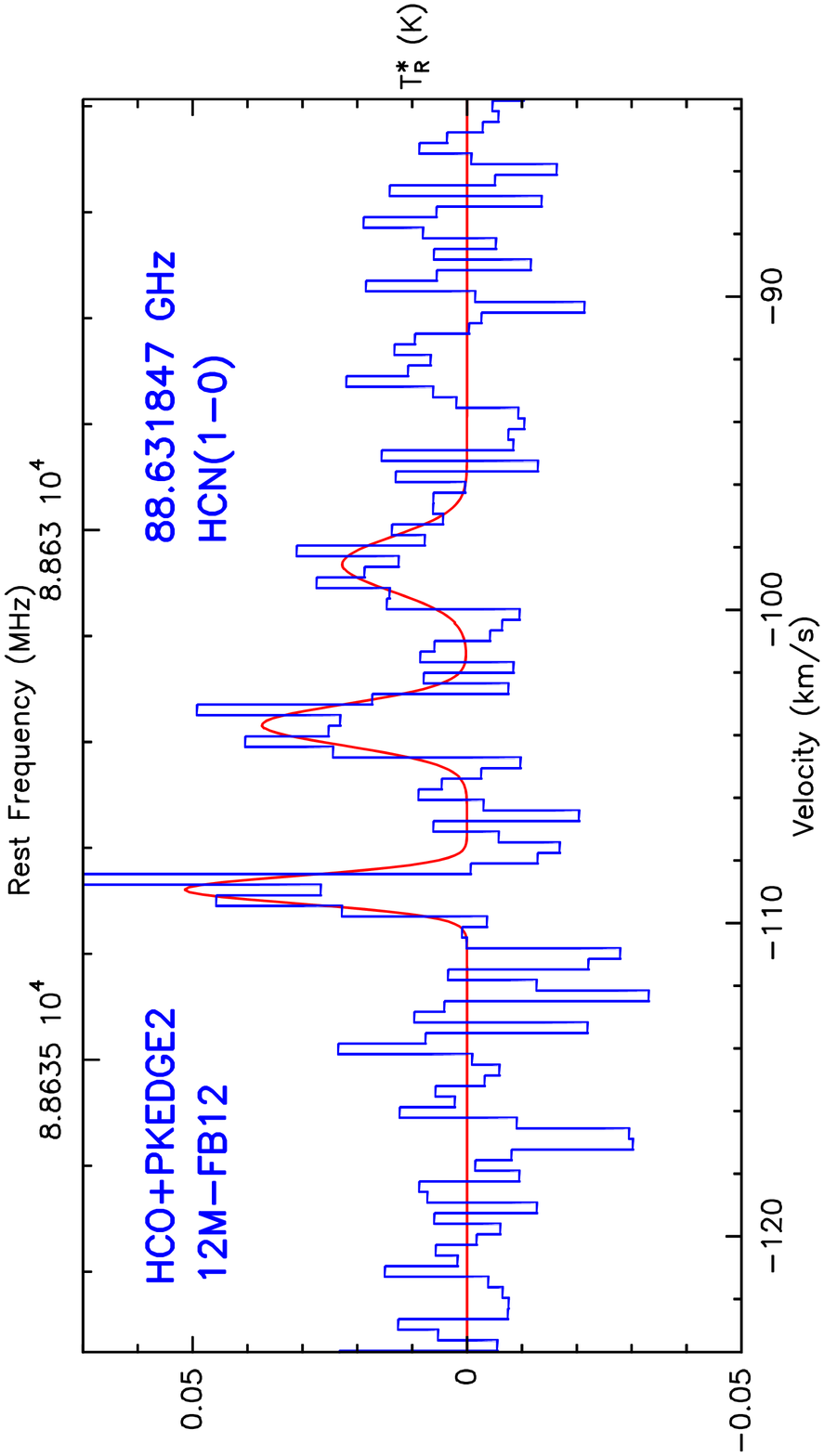}%
\hspace{2mm}%
\includegraphics[scale=0.20,angle=-90]{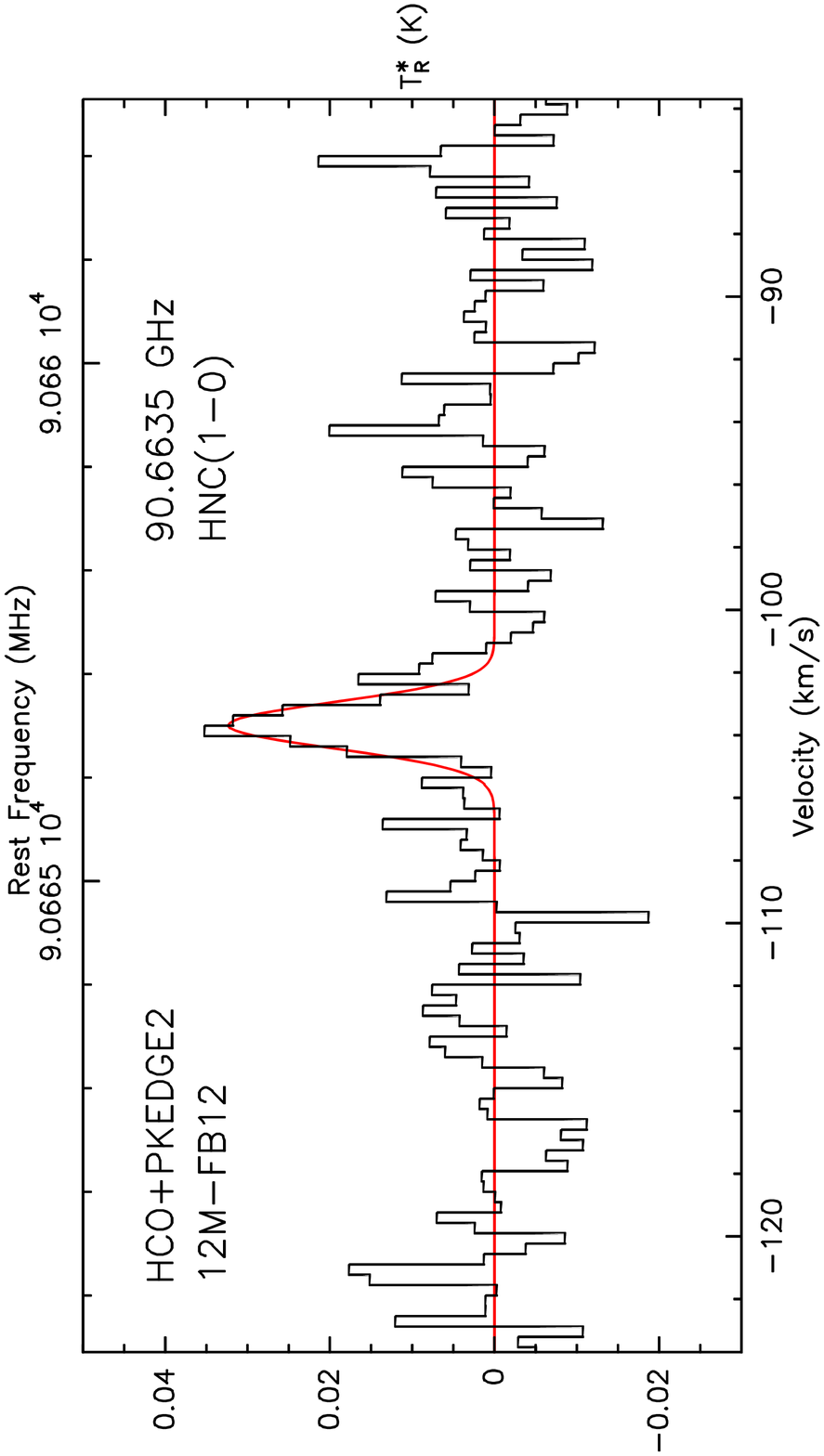}%
} 
\vspace{5mm} 
\hbox{ 
\includegraphics[scale=0.20,angle=-90]{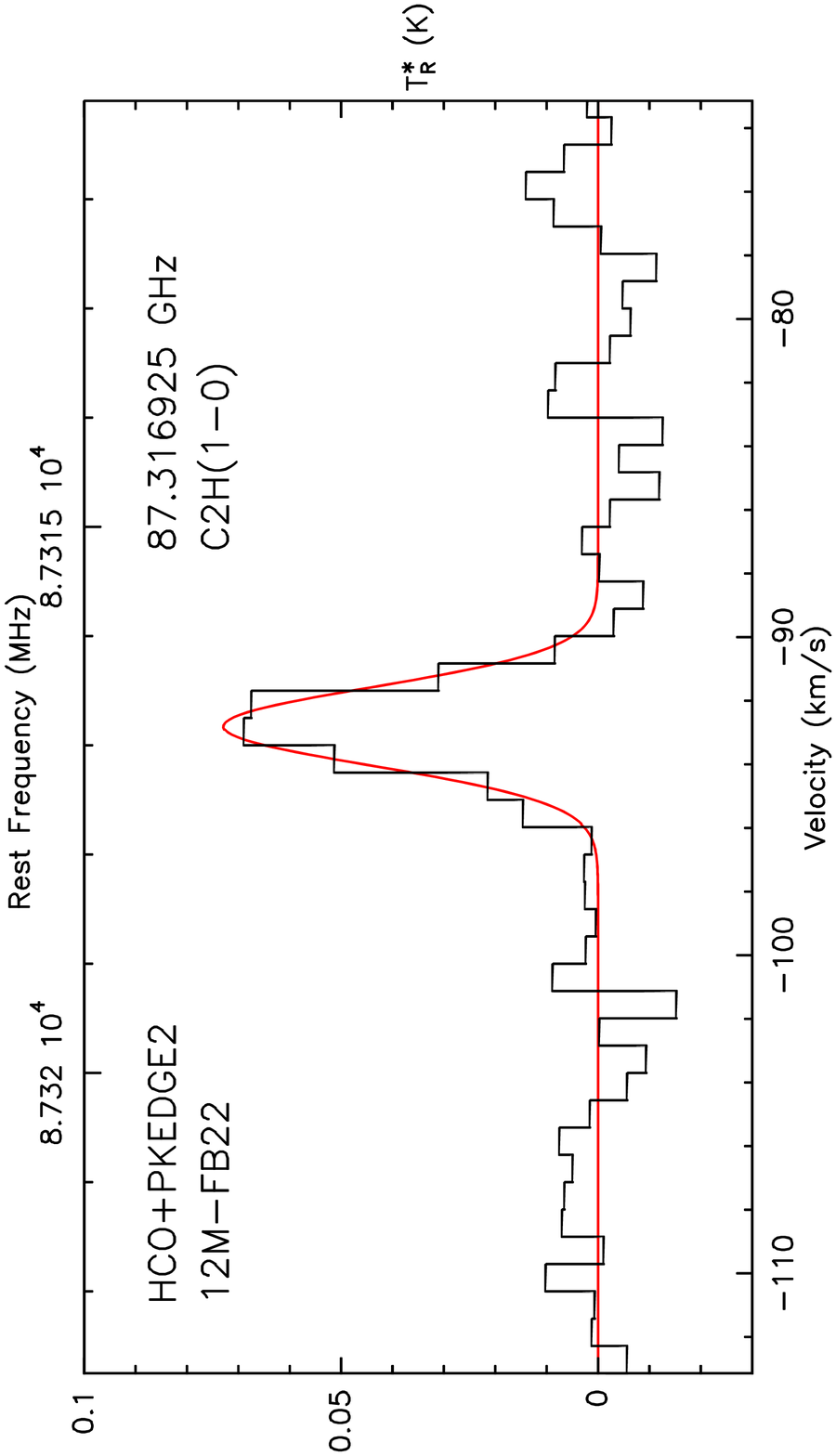}%
\hspace{3mm}%
\includegraphics[scale=0.20,angle=-90]{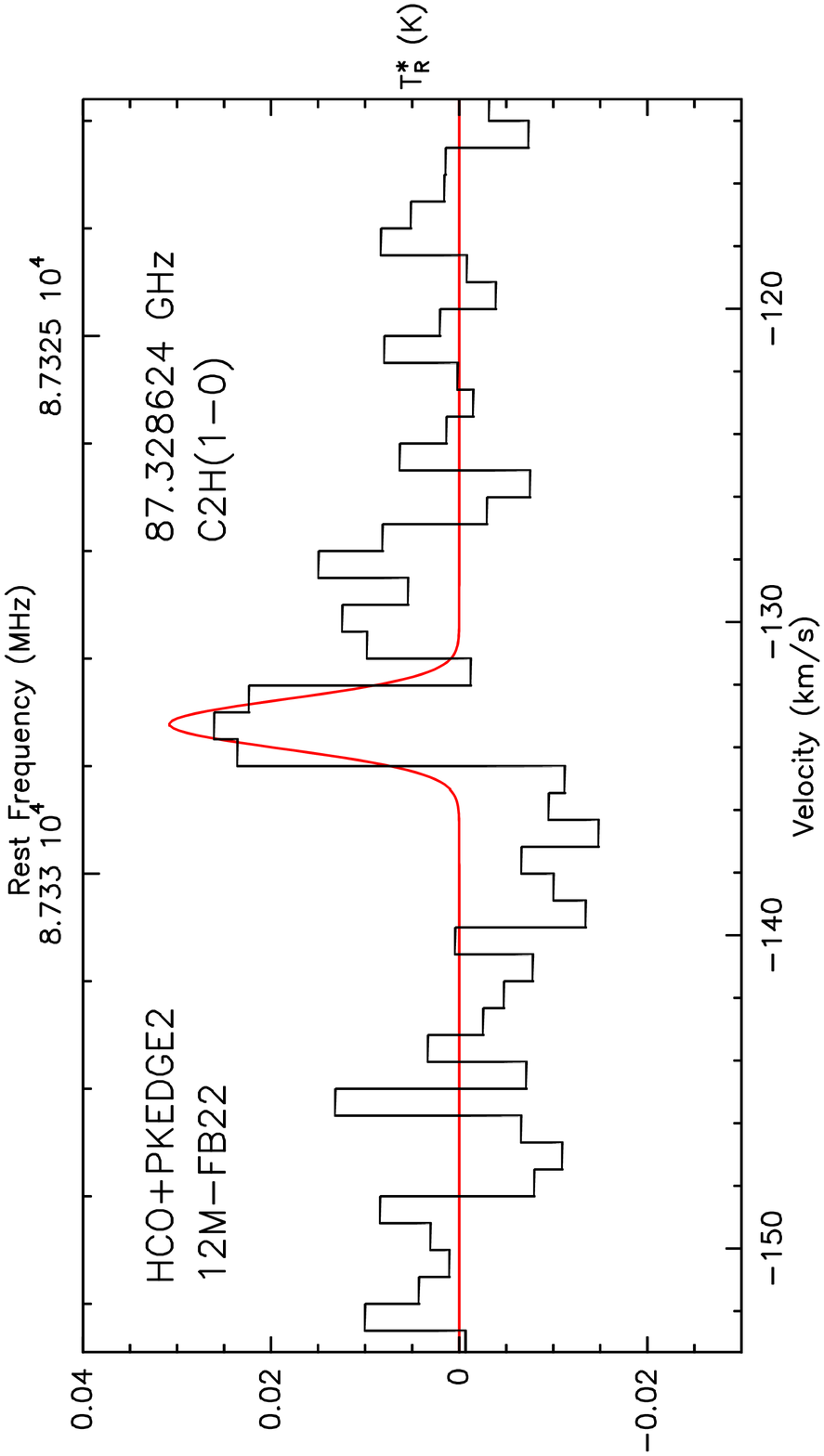}%
\hspace{3mm}%
\includegraphics[scale=0.20,angle=-90]{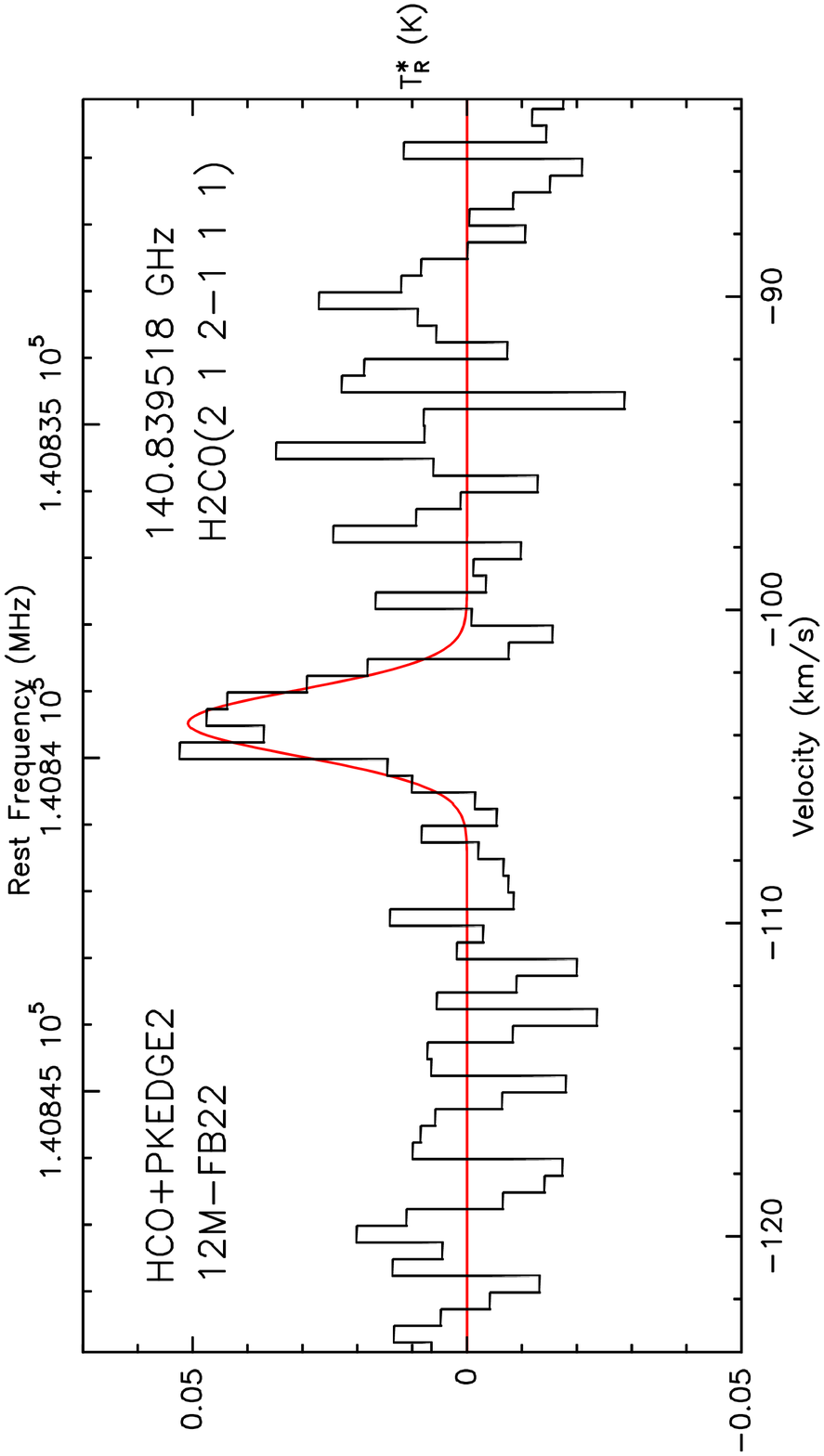}%
} 
\vspace{5mm} 
\hbox{ 
\hspace{-1mm}%
\includegraphics[scale=0.20,angle=-90]{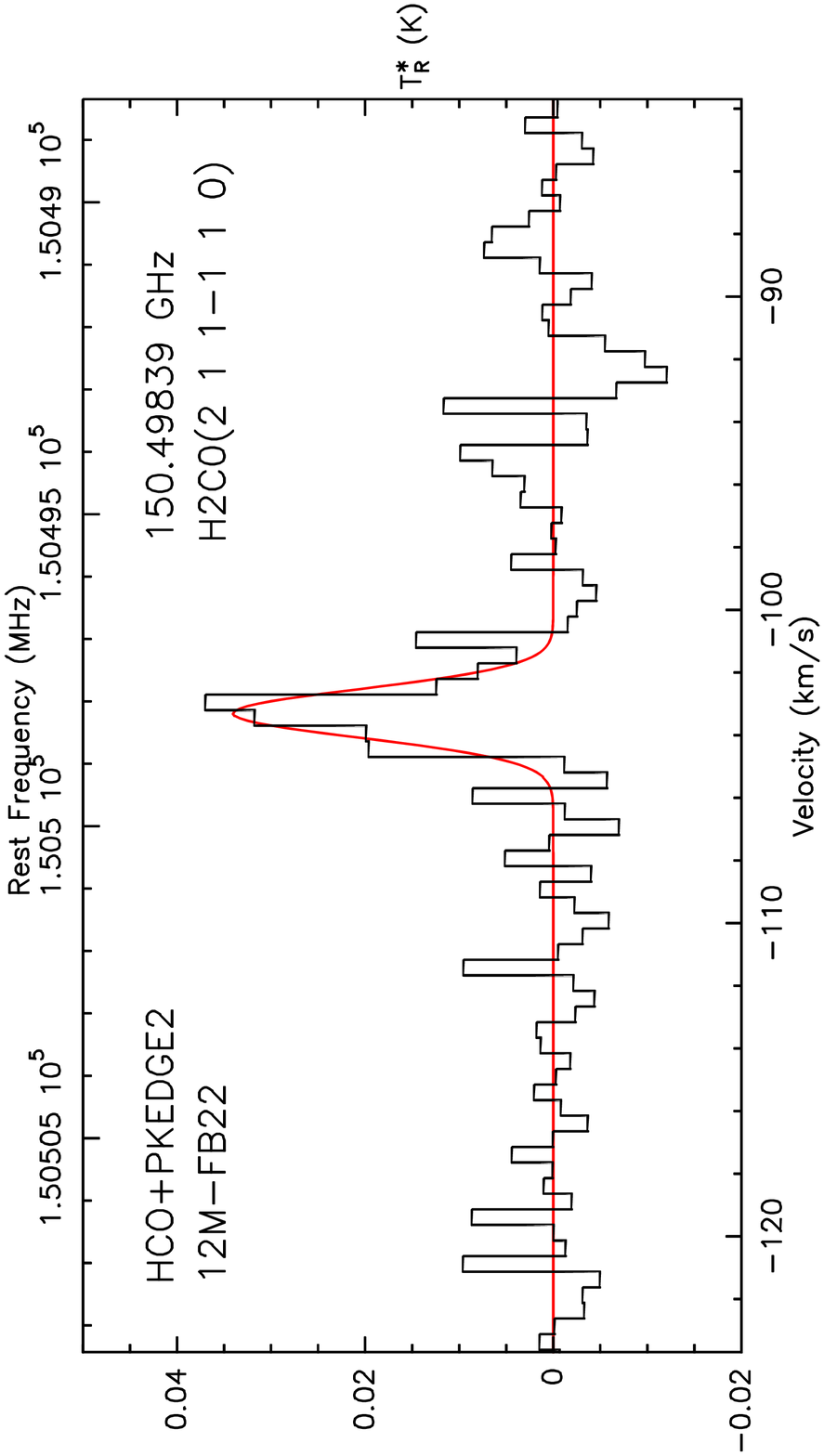}%
\hspace{3mm}%
\includegraphics[scale=0.20,angle=-90]{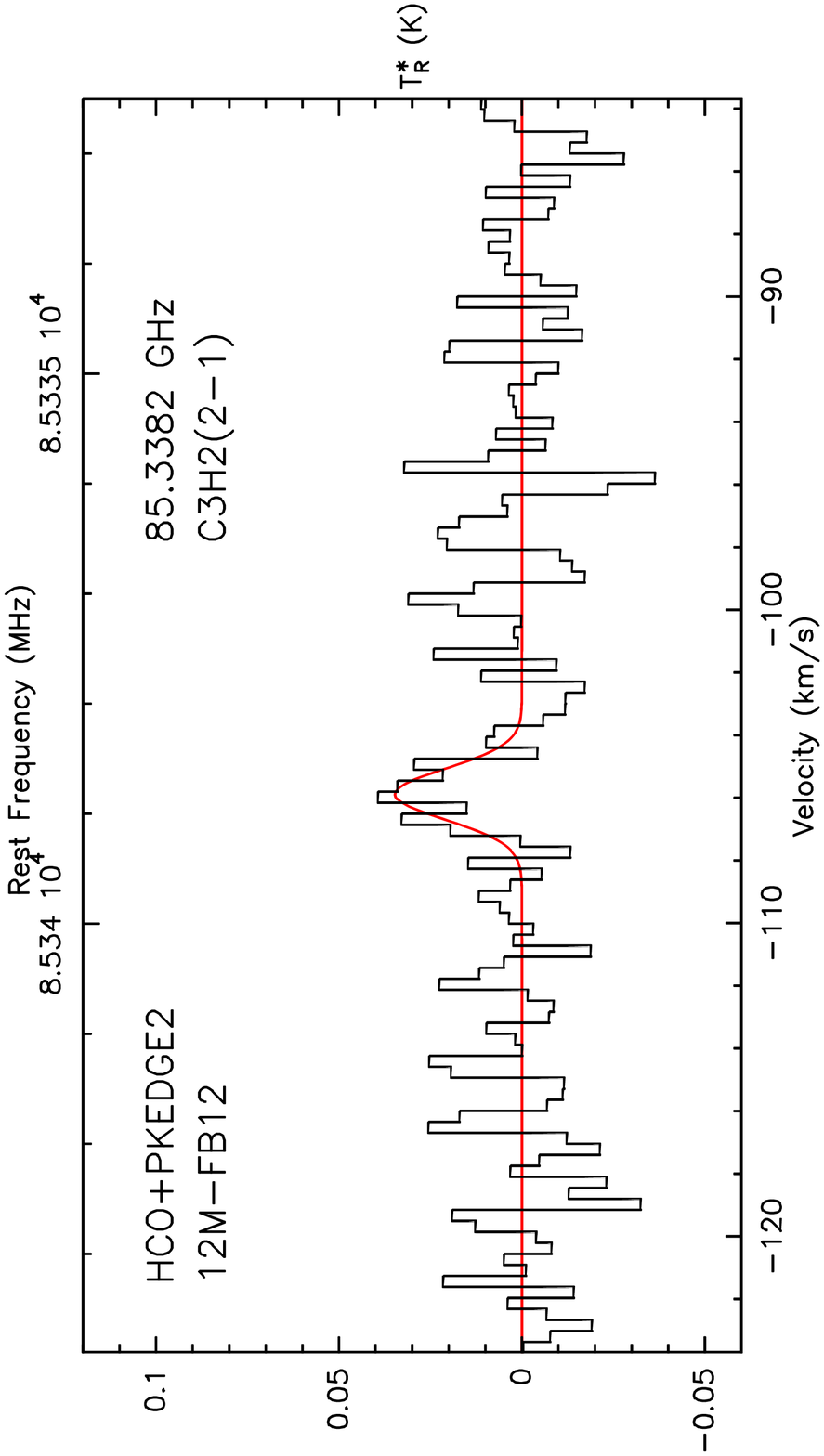}%
\hspace{2mm} 
\includegraphics[scale=0.20,angle=-90]{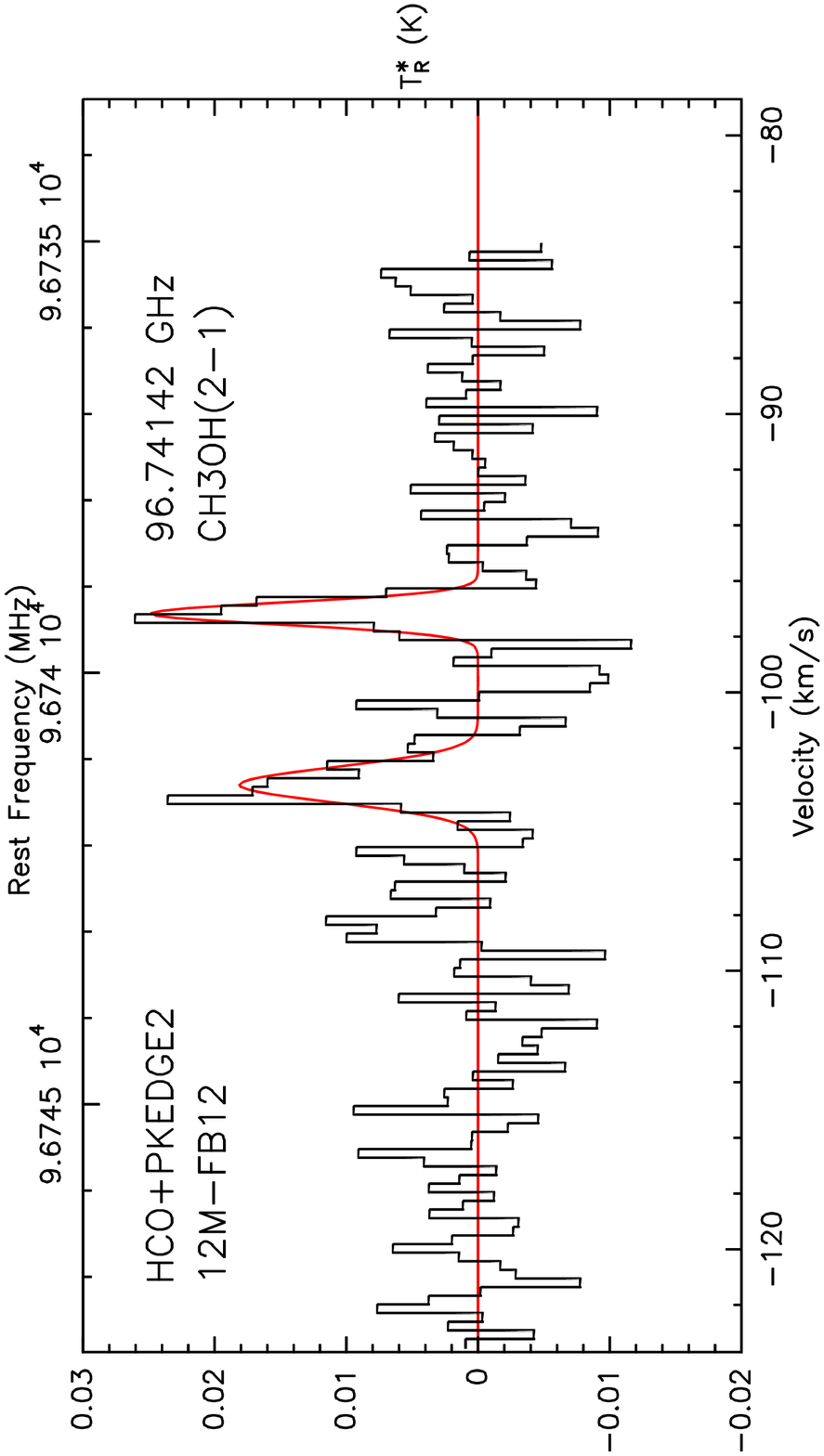}%
} 
\caption{EC2 spectra observed at the ARO 12\,m Feb-June 2002 position A ($\alpha_{2000}=$ 02:48:38.5, \newline 
$\delta=$ 58:28:28.1, $v_\mathrm{rad} =  -103.70$\,km\,s$^{-1}$). 
The ordinate is T$_R^*$(K), uncorrected for beam-filling.
\label{12m_jun02_sum}} 
\end{figure*}

\begin{figure*} 
\hbox{ 
\includegraphics[scale=0.20,angle=-90]{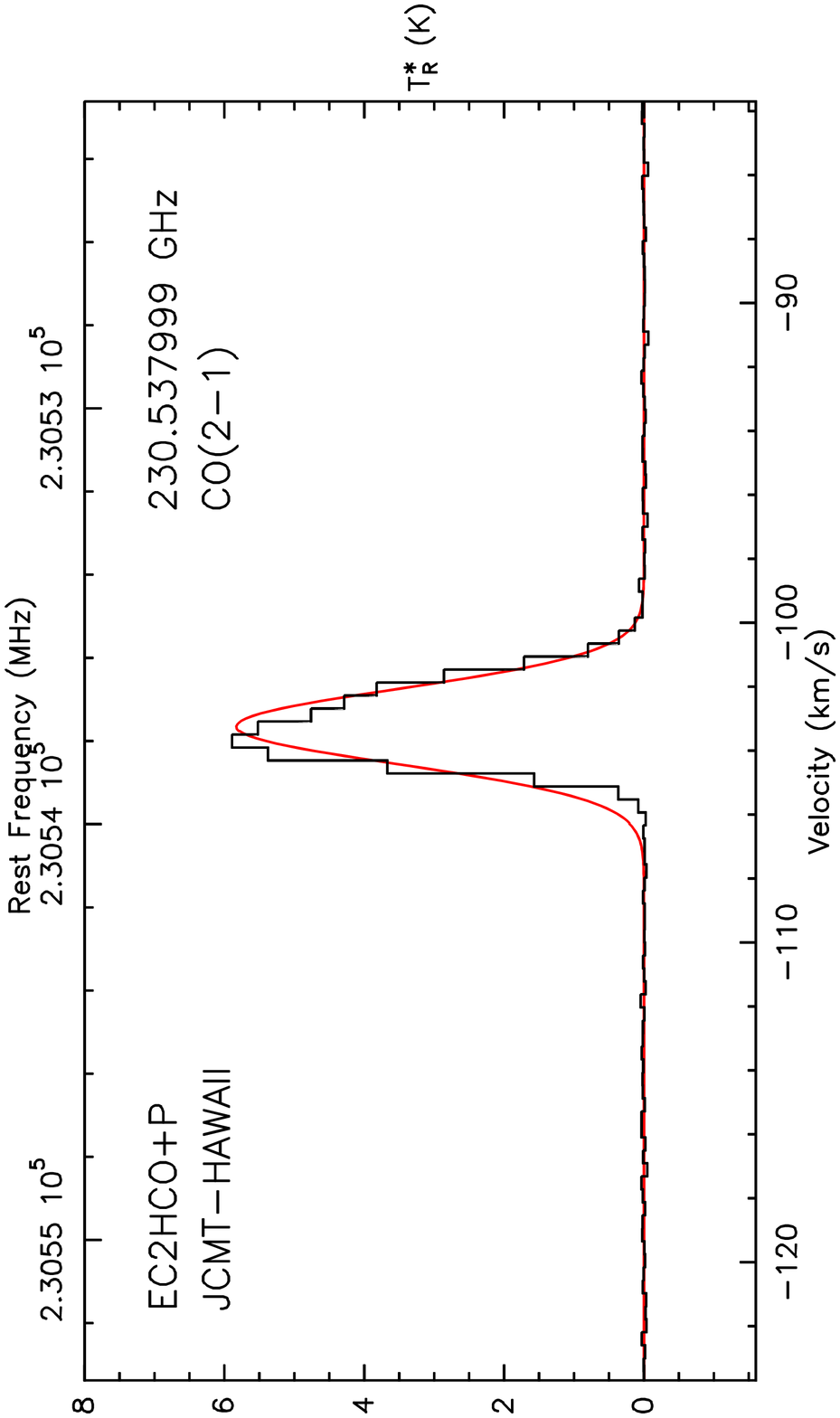}%
\hspace{3mm}%
\includegraphics[scale=0.20,angle=-90]{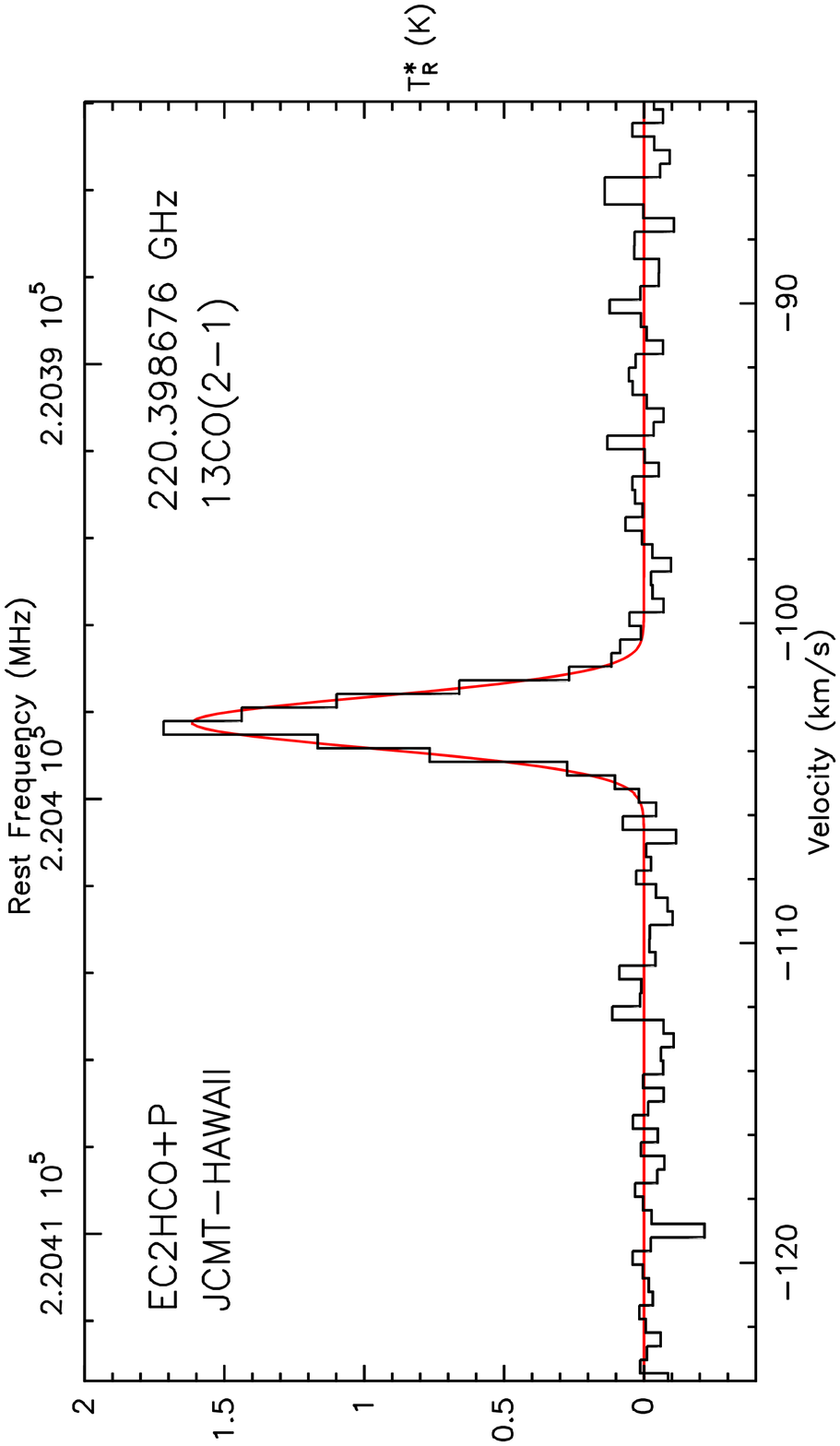}%
\hspace{3mm}%
\includegraphics[scale=0.20,angle=-90]{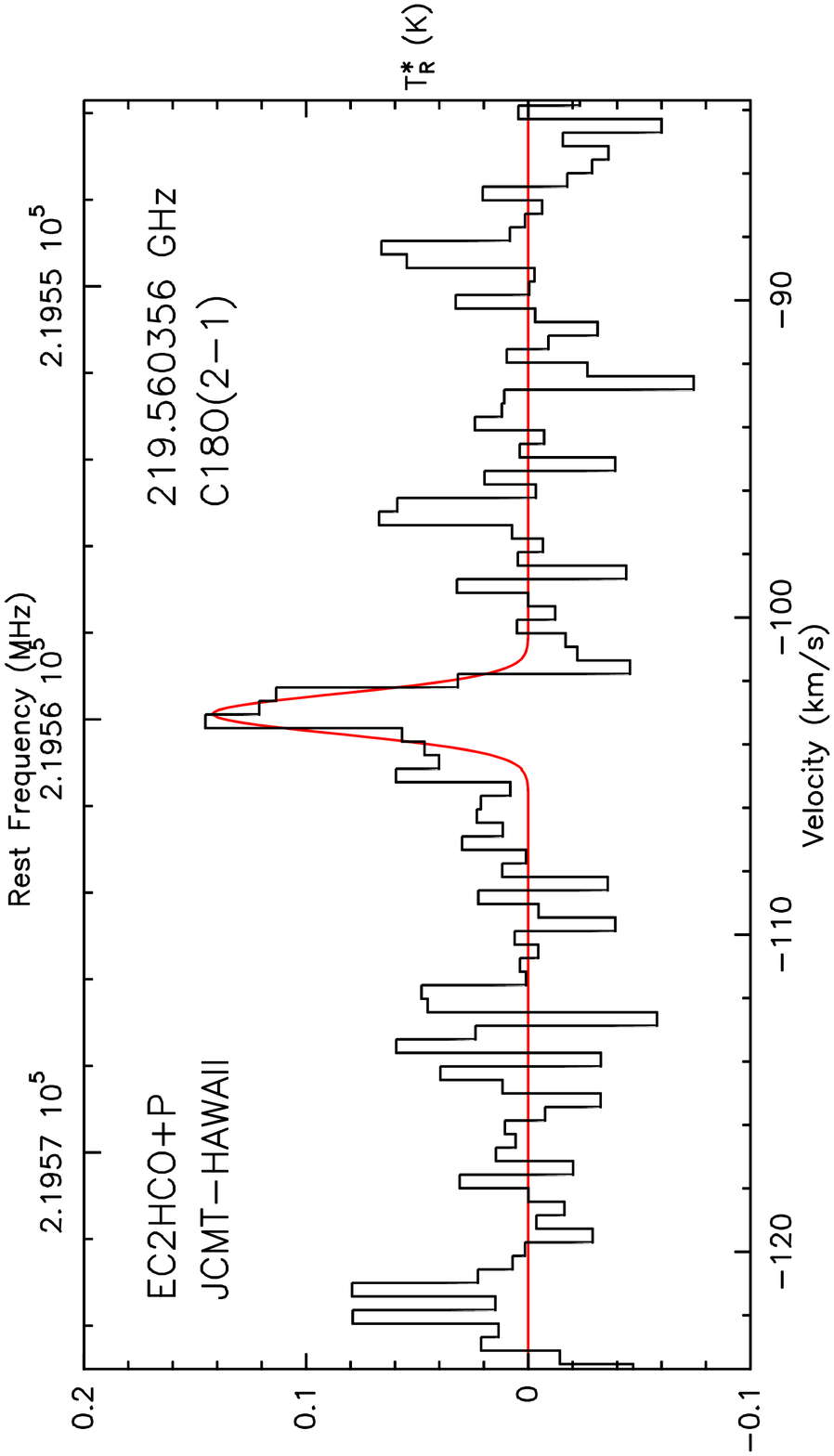}%
} 
\vspace{5mm} 
\hbox{ 
\includegraphics[scale=0.20,angle=-90]{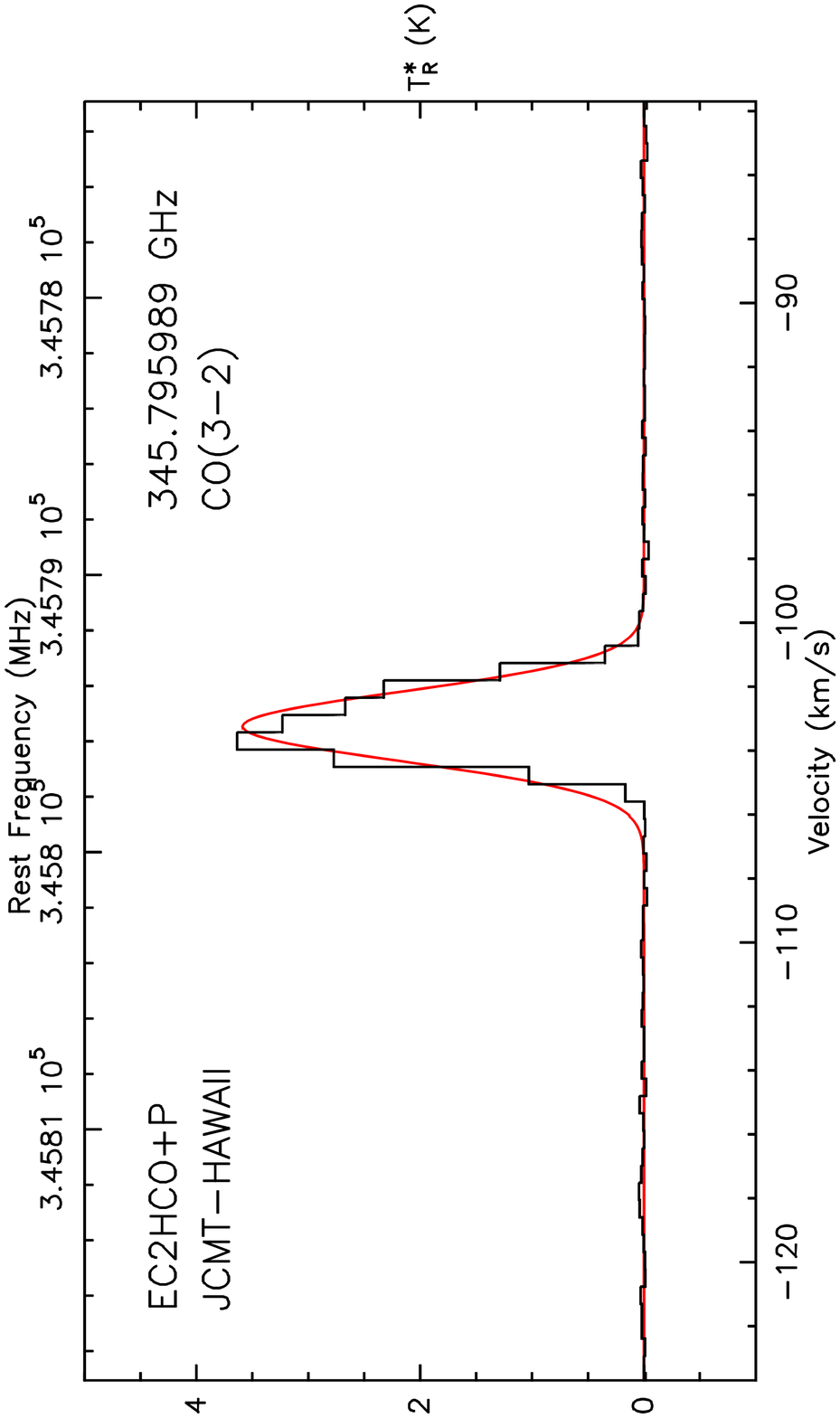}%
\hspace{2mm}%
\includegraphics[scale=0.20,angle=-90]{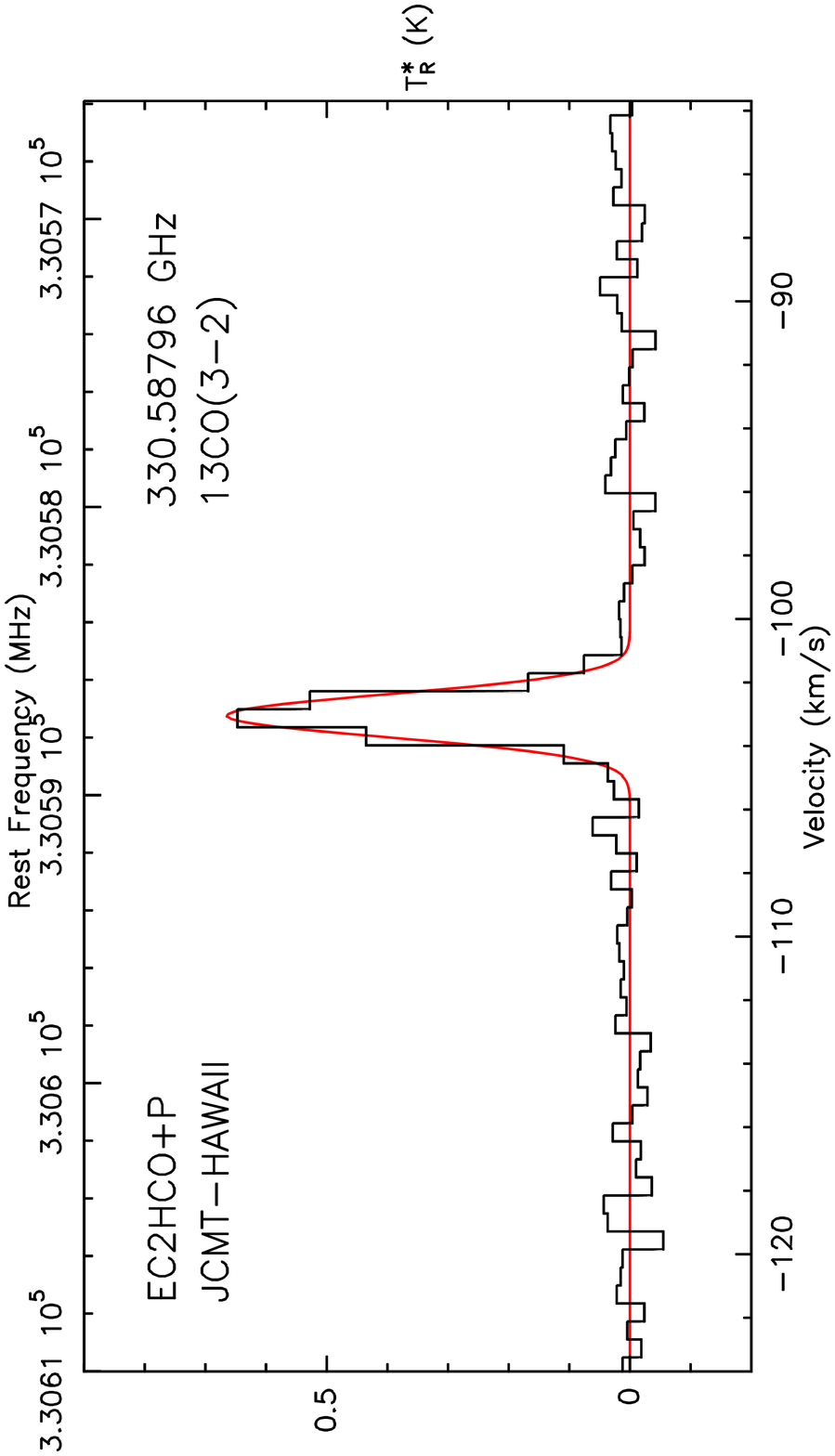}%
\hspace{2mm}%
\includegraphics[scale=0.20,angle=-90]{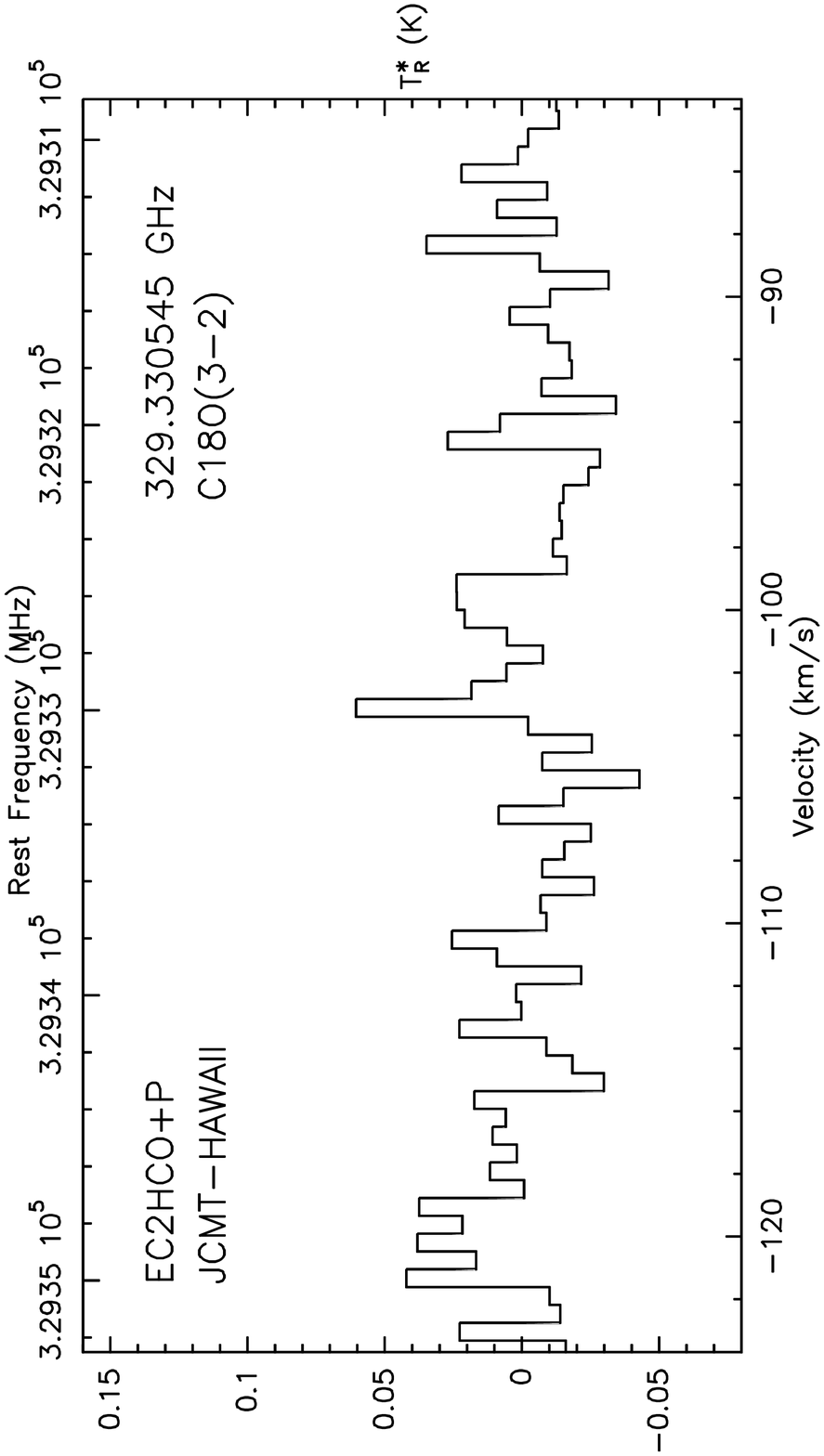}%
} 
\vspace{5mm} 
\hbox{ 
\hspace{-1mm}%
\includegraphics[scale=0.20,angle=-90]{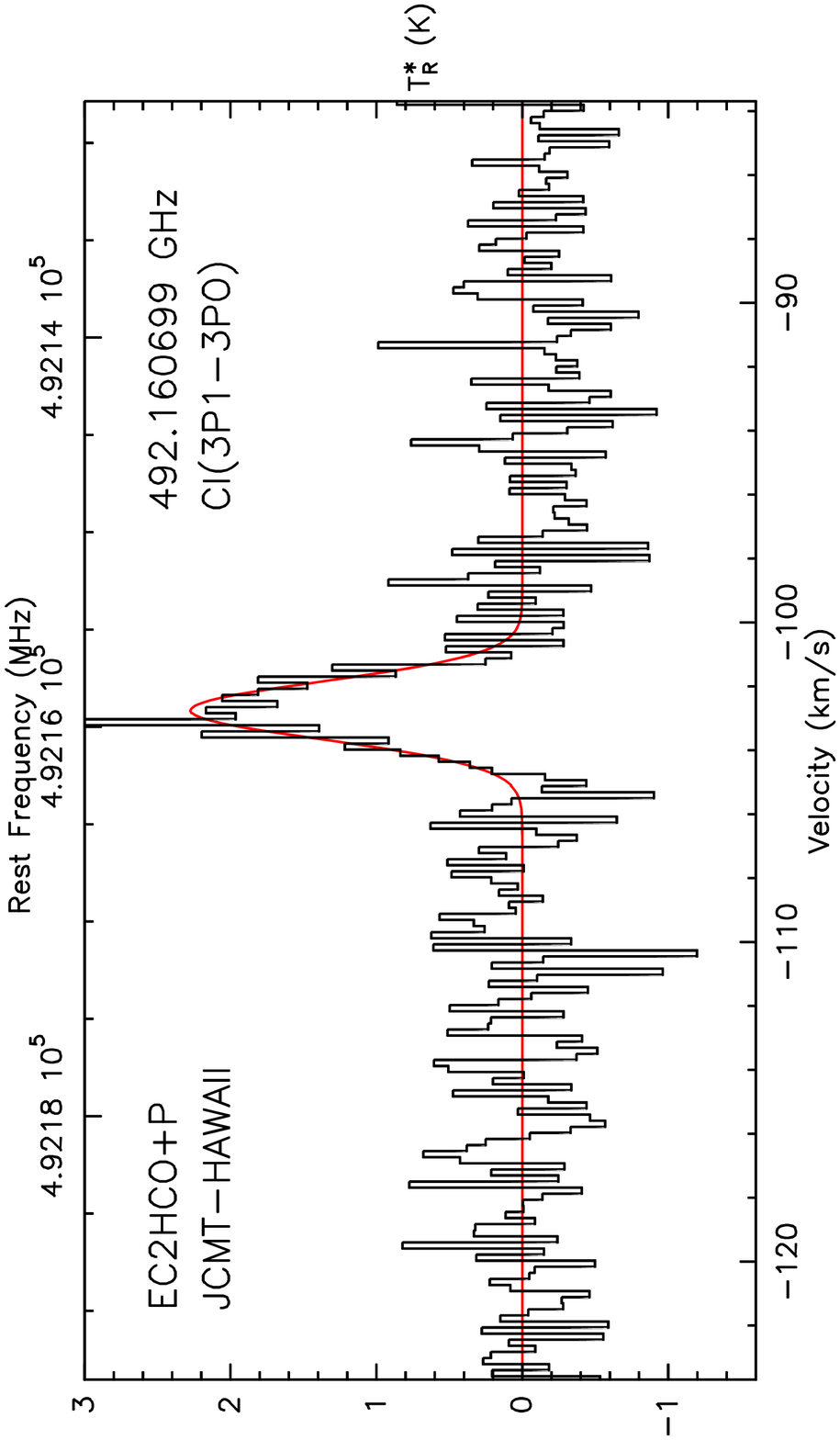}%
\hspace{2mm}%
\includegraphics[scale=0.20,angle=-90]{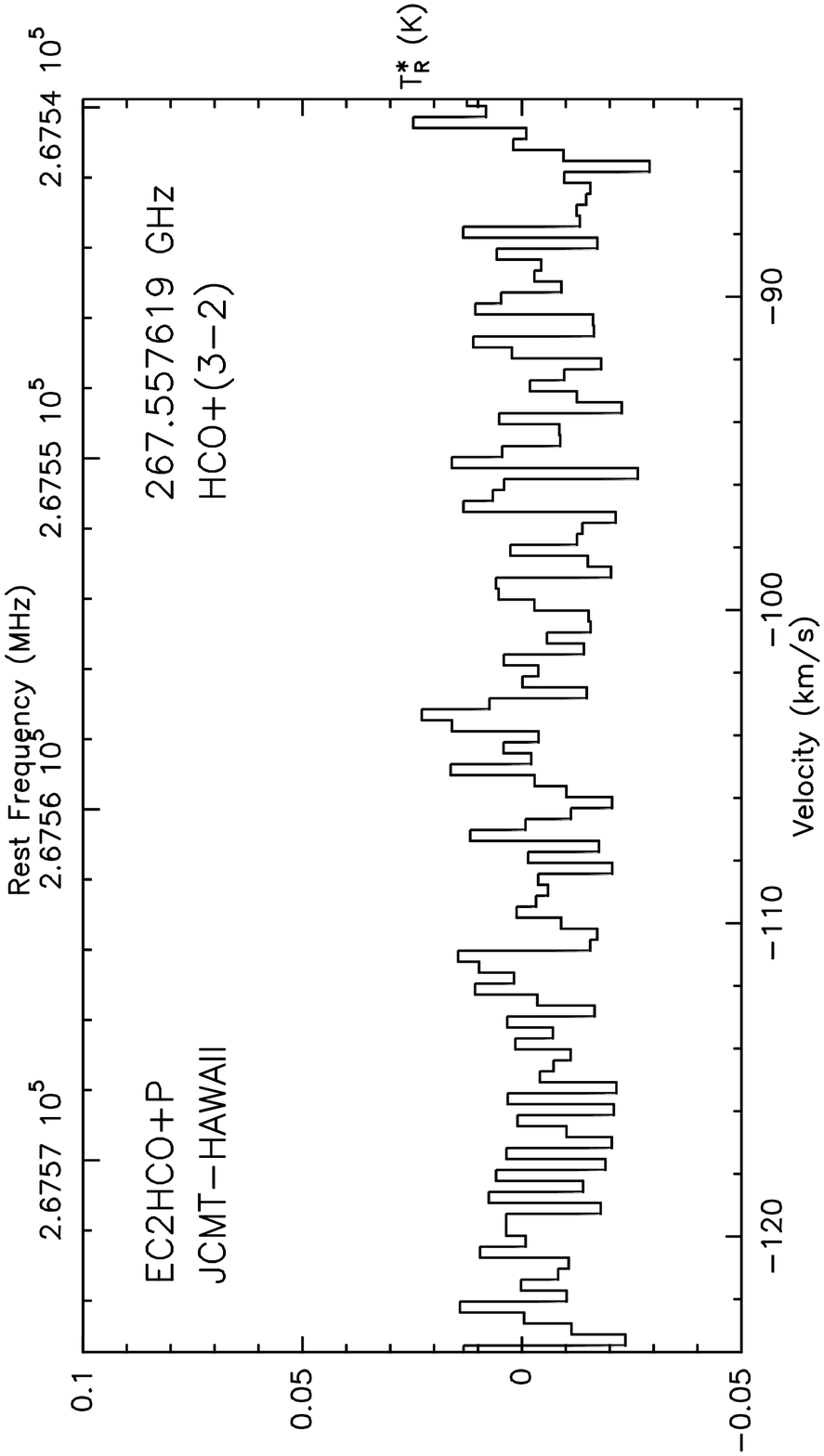}%
} 
\caption{EC2 spectra observed at the JCMT 15\,m June 2004 position A ($\alpha_{2000}=$ 02:48:38.5, \newline 
$\delta=$ 58:28:28.1, $v_\mathrm{rad} =  -103.70$\,km\,s$^{-1}$). 
The ordinate is T$_R^*$(K), uncorrected for beam-filling. 
\label{jcmt}} 
\end{figure*}

\begin{figure*} 
\hbox{ 
\includegraphics[scale=0.40]{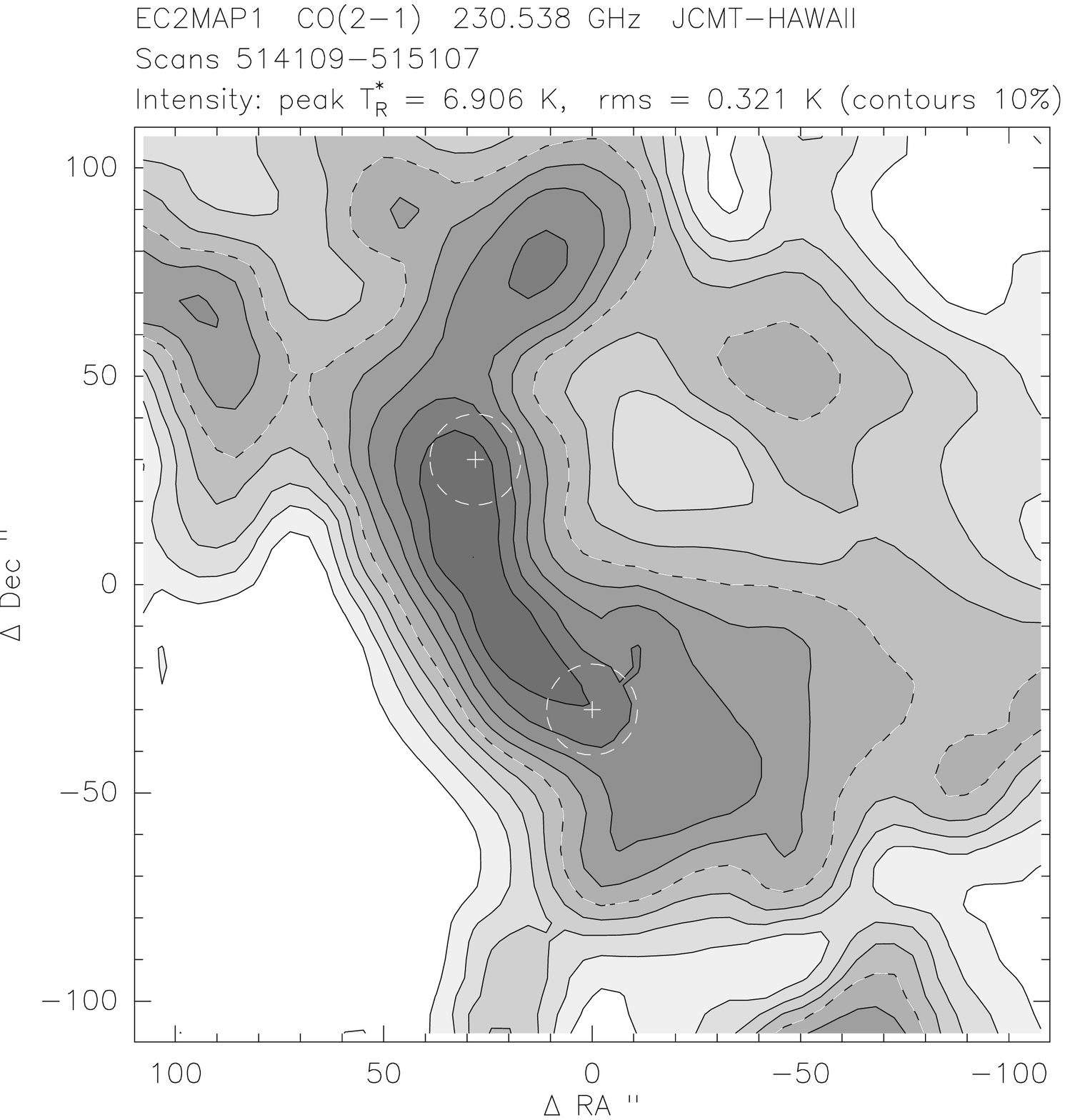} 
\hspace{3mm} 
\includegraphics[scale=0.40]{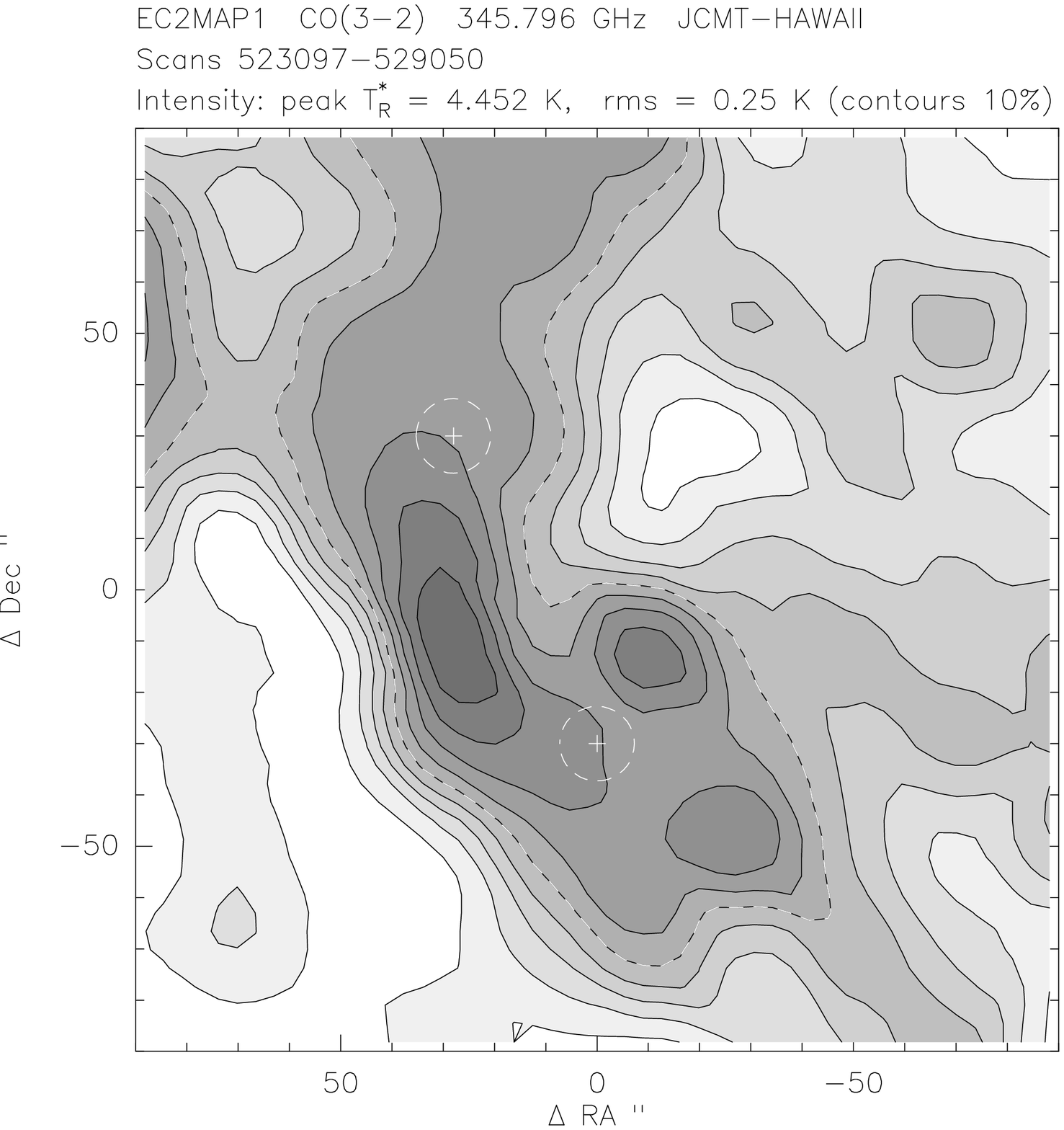} 
} 
\vspace{5mm} 
\hbox{ 
\includegraphics[scale=0.40]{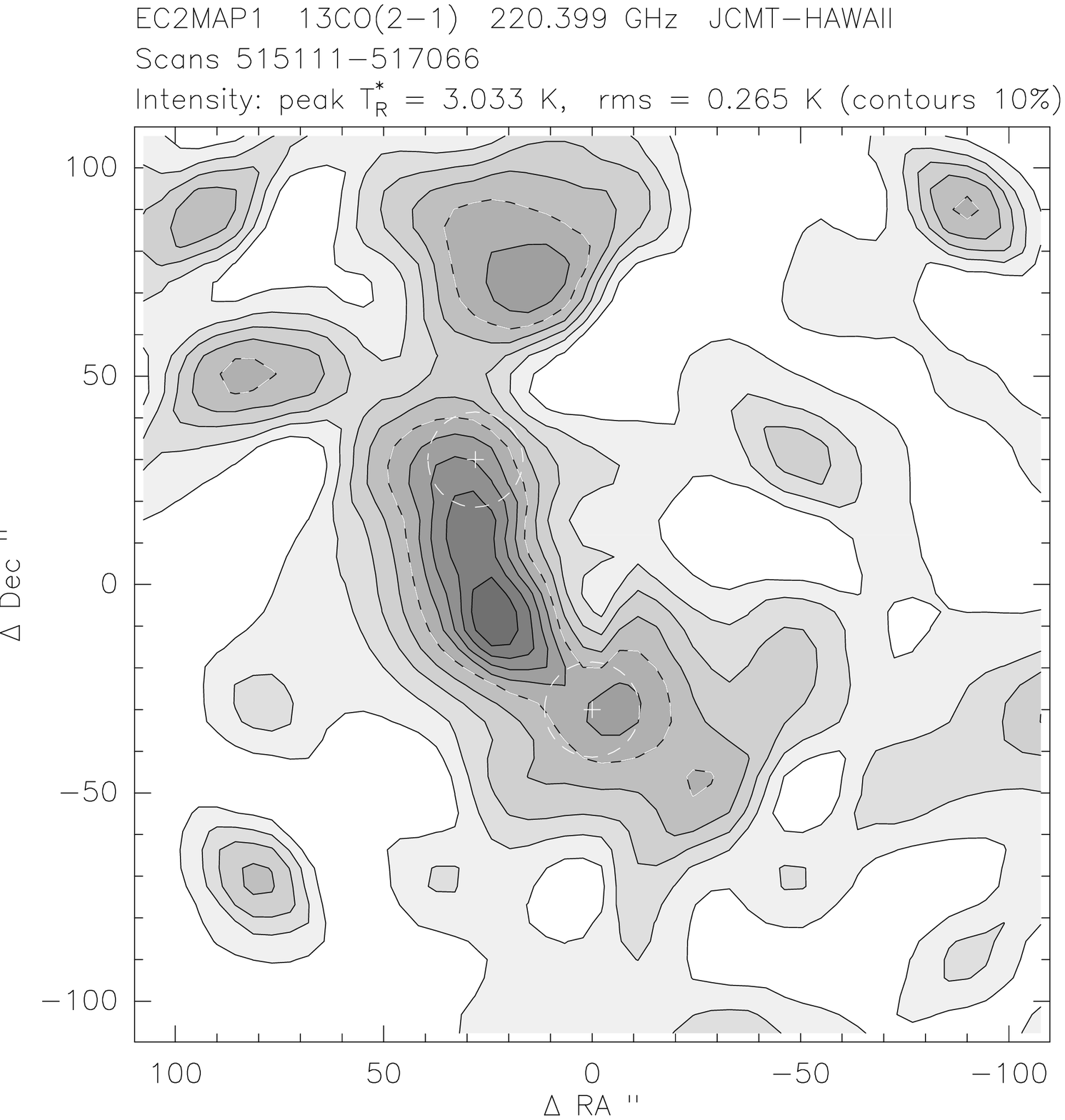} 
\hspace{3mm} 
\includegraphics[scale=0.40]{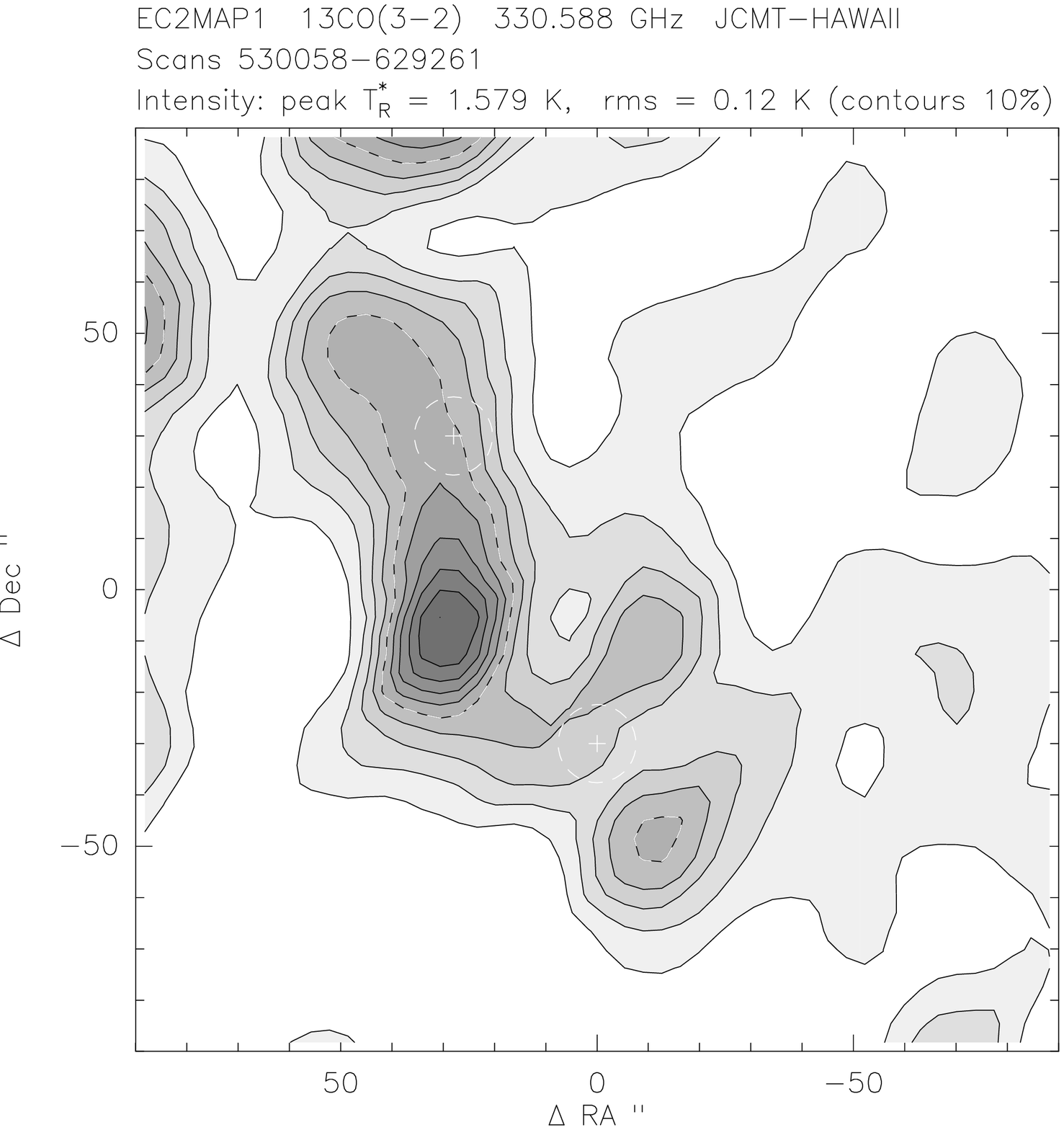} 
} 
\caption{JCMT 15\,m May-July 2005 maps of observed CO intensities centered on position E 
         ($\alpha_{2000}=$ 02:48:38.5, $\delta=$ 58:28:58.3, $v_\mathrm{rad} = -103.70 $ km\,s$^{-1}$) with position A 
         ($\alpha_{2000}=$ 02:48:38.5, $\delta=$ 58:28:28.1) 30\arcsec\ below map center. 
         Axis offsets in arcsec, contour levels 10\% of peak, dashed contours 50\% of peak (FWHP) 
         and white circle is half-power beam width (HPBW). 
         (a) CO 2--1: peak $T^*_\mathrm{R}$ = 6.906\,K, rms = 0.321\,K. 
         (b) CO 3--2: peak $T^*_\mathrm{R}$ = 4.452\,K, rms = 0.25\,K. 
         (c) $^{13}$CO 2--1: peak $T^*_\mathrm{R}$ = 3.033\,K, rms = 0.265\,K. 
         (d) $^{13}$CO 3--2: peak $T^*_\mathrm{R}$ = 1.579\,K, rms = 0.12\,K. 
\label{ec2_comap1_intensity}} 
\end{figure*}

\begin{figure} 
\includegraphics[width=75mm]{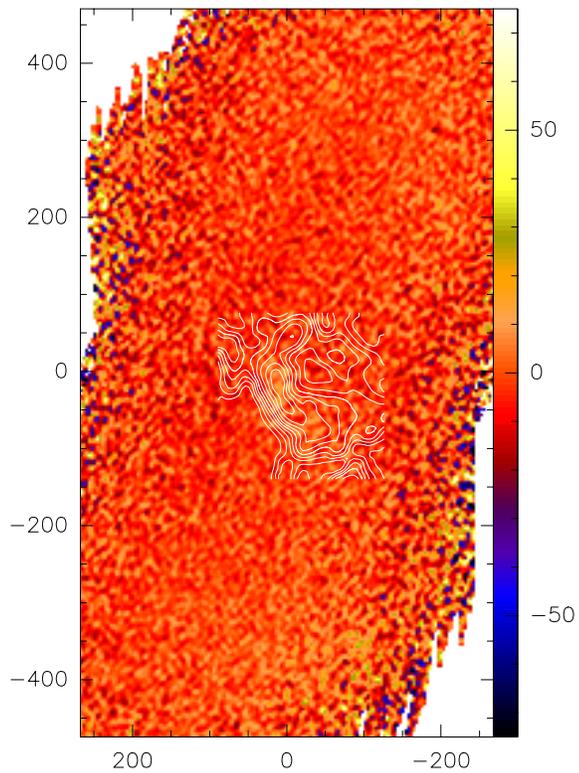} 
\caption{IRAM 30\,m Oct-Nov 2005 1.2\,mm dust map centered on position {\it EC2DUST1A} 
	 ($\alpha_{2000}=$ 02:48:41.0, $\delta=$ 58:29:27.9) with CO 2--1 intensity contours from 
          Fig. \ref{ec2_comap1_intensity}a superimposed. The rms noise for the map is 6.35\,mJy/beam 
          with a detected peak of 20.3\,mJy/beam. Axis offsets in arcsec. 
\label{EC2DUST1A_CO}} 
\end{figure}

\begin{figure} 
\includegraphics[width=75mm]{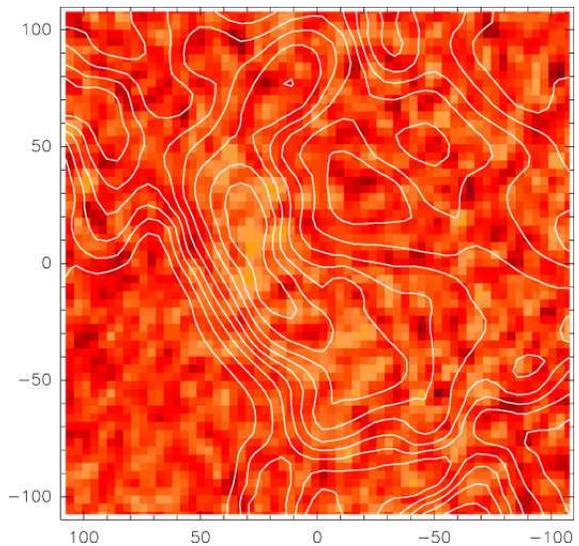} 
\caption{CO 2--1 intensity contours from {\it EC2MAP1} (Fig. \ref{ec2_comap1_intensity}a)  
	 overlaid on 1.2\,mm dust map (Fig. \ref{EC2DUST1A_CO} {\it EC2DUST1A}), showing 
	 the correspondence of peak CO 2--1 and 1.2\,mm dust emission. Axis offsets in arcsec. 
\label{EC2map1_dust}} 
\end{figure}

Detected lines, CO maps, dust maps and a summary of observations can be found in 
Figs.~\ref{ec2_map_positions} through \ref{EC2map1_dust} 
and Table~\ref{ec2_sumobs}. 

Before estimating abundances of the observed molecular species, we first have to determine the main physical parameters of the gas, i.e. its density and temperature, taking into account the beam-filling by deconvolving the detected intensities. To estimate the latter we use the $^{12}$CO and $^{13}$CO 2--1 maps in Fig. \ref{ec2_comap1_intensity} and assume that the source, which has an elongated shell-like structure, has a full-width at half power ($fwhp$) size of $150\arcsec \times 40\arcsec$ ($\theta_\mathrm{s} \equiv 77\arcsec$) for $^{12}$CO, and $100\arcsec \times 30\arcsec$ ($\theta_\mathrm{s} \equiv 55\arcsec$) for $^{13}$CO, C$^{18}$O and all other molecules. The peak intensities listed in Table~\ref{ec2_sumobs} are corrected for these source sizes using 
$T_\mathrm{mb} = T^*_\mathrm{R} / \eta_\mathrm{bf}$, with 
$\eta_\mathrm{bf} = \theta^{2}_\mathrm{s} / (\theta^{2}_\mathrm{s} + \theta^{2}_\mathrm{b})$, 
where $\eta_\mathrm{bf}$ is the beam filling correction factor.

There are several tracers sensitive to both density and temperature, though none that only traces density, but one that exclusively traces kinetic temperature. The species allowing us to determine $T_\mathrm{kin}$ is NH$_3$. 

We have detected at least four of the five groups of hyperfine components in the ($J,K$)=(1,1) line of NH$_3$ (see Fig.~\ref{bonn}, the satellite features being marked `HF'). Line intensity ratios are consistent with optically thin emission. Gaussian fits of the groups of hyperfine components are (with increasing velocity) 23.3, 17.2, 80.2, 5.7 and 11.1\,mJy\,km\,s$^{-1}$. For the (2,2) line we get 38.0$\pm$8.0\,mJy\,km\,s$^{-1}$. The (2,2)/(1,1) line intensity ratio is then 0.28$\pm$0.07, with the error being dominated by the low signal-to-noise ratio of the (2,2) line. To derive total beam-averaged column densities in an inversion doublet, we use 
  \begin{equation} 
    N(J,K) = 1.55 \times 10^{14} \frac{J(J+1)}{\nu \,K^2} \int{T_\mathrm{mb} \mathrm{d}V} 
  \end{equation} 
\citep[e.g. eq. 1 in][]{henkel0009} that is valid in the optically thin limit. The column density $N$, the frequency $\nu$, and the integral are in units of cm$^{-2}$, GHz, and K km\,s$^{-1}$, respectively. As already indicated in section~\ref{ec2obs}, the conversion factor between the Jy and the $T_\mathrm{mb}$ scale is 1.6. Thus we obtain total beam averaged column densities of 2.52 $\times$ 10$^{12}$ and 5.21 $\times$ 10$^{11}$\,cm$^{-2}$ in the (1,1) and (2,2) inversion doublets. The (2,2)/(1,1) column density ratio is then 0.21$\pm$0.05. With 
  \begin{equation} 
    N(2,2)/N(1,1) = (5/3) e^{-\Delta E/kT} 
  \end{equation} 
and $\Delta E/k = 40.957$\,K, the difference in energy between the (1,1) and (2,2) levels, we derive a rotational temperature $T = 20\pm3$\,K. 

With the also populated inversion singlet (0,0) level\footnote{The (0,0) level is ortho-ammonia, with a statistical weight of 2, but as there is only one state, this is factored by 0.5.} at 23.19\,K below the (1,1) state, the total NH$_{3}$ column density, summed over the (0,0), (1,1) and (2,2) levels is $N$(NH$_{3}$) = 5.77 $\times$ 10$^{12}$\,cm$^{-2}$, with an uncertainty of $\pm$15 per cent, averaged over the 40\arcsec\ beam. $J>K$ levels require extremely high densities \citep{mauersberger8505} and are therefore not relevant. 

Using this rotational temperature, which should be close to the kinetic temperature \citep{walmsley8306,danby8811}, the density of EC2 was estimated using Large Velocity Gradient (LVG) models for a number of molecular species, for a spherical cloud with constant density \citep[see][and references therein]{henkel8002,mauersberger9009}. The results from these models are included in Table \ref{ec2_lvgcomps}. 
First, we use a LVG model involving 40 ortho-H$_{2}$CO states up to $K_\mathrm{a}=3$ and the collision cross sections of \citet{green9107}. We find that the observed deconvolved line intensities for both the cm and mm lines are well reproduced with a density $n(\mathrm{H}_{2}) = 1.2 \times 10^{4}$\,cm$^{-3}$.
Second we fit the $^{13}$CO and C$^{18}$O spectra to an LVG model using collisional rates from \citet{flower0107} and an adopted ortho-para ratio of 3 for H$_2$, although the results are only weakly dependent on this. 
We first fitted the three $^{13}$CO deconvolved lines to derive a density of $n(\mathrm{H}_{2}) = 3.2 \times 10^{3}$\,cm$^{-3}$. With this density, C$^{18}$O (only two lines) and $^{12}$CO could then also be fitted. This gives isotopical abundances of approximately 100:10:1 for $^{12}$CO:$^{13}$CO:C$^{18}$O (see below). This can not reflect the actual isotopic abundance ratios, as the optically thin $^{12}$C/$^{13}$C ratio is $\sim$89 in the solar system and 60 to 70 in local diffuse clouds \citep{wilson94}. 
Using $n_\mathrm{lvg}$(X)/$n$(H$_{2}$) from Table \ref{ec2_lvgcomps}, the best fit also gives a fractional abundance with regard to H$_{2}$ of $\sim$10$^{-7}$ for $^{13}$CO, well below the value of $2 \times 10^{-6}$ typically found in local metal-rich molecular clouds. For the C$^{18}$O/H$_{2}$ ratio the model gives $\sim$10$^{-8}$ compared to the local value of $1.7 \times 10^{-7}$ with $^{16}$O/$^{18}$O = 500 \citep{frerking8211}. 

For the rotational HCN 1--0 transition, our detected hyperfine structure\footnote{The HCN spectra in Fig.~\ref{12m_jun02_sum} shows spikes near --110\,km\,s$^{-1}$, increasing the strength of what should be the weakest hyperfine component. Spectra taken 30\arcsec\ from position A in 2005 Dec. confirm the hyperfine structure for HCN.} 
is compatible with optically thin emission, implying that other lines should also be optically thin (HCN 1--0 is among the optically thickest 3\,mm lines after CO). Therefore, we have computed LVG models for several other species, and start with those for which we have measured more than one transition. 
For SO, using the collisional rates of \citet{green9410}, we can fit the $1_{0}$--$0_{1}$ and $3_{2}$--$2_{1}$ lines simultaneously, but only with $n(\mathrm{H}_{2}) = 1.4 \times 10^{5}$\,cm$^{-3}$, an order of magnitude larger than that derived from H$_2$CO and two orders of magnitude higher than that obtained from $^{13}$CO, implying the presence of small-scale clumping. 
The CS 2--1 and 3--2 deconvolved line temperatures can be roughly fitted \citep[using collisional rates from][]{turner9211} with the density of $1.2 \times 10^{4}$\,cm$^{-3}$ obtained for H$_{2}$CO. 
The model reproduces HCO$^{+}$ 1--0 deconvolved line temperatures for both the H$_{2}$CO and the SO densities \citep[using collisional rates from][]{flower9905}. 
The HCO$^{+}$ model line intensities appear to be consistent with the observed $J$ = 3--2 upper limit, at least in the case of the lower density of $1.2 \times 10^{4}$\,cm$^{-3}$. 
For HCN 1--0 we summed the deconvolved integrated intensities of the three hyperfine components and then divided by the average of the line widths to get a peak intensity. Using collisional rates from \citet{schoier0503}, this peak deconvolved line intensity was fitted for both the H$_{2}$CO and the SO densities ($1.2 \times 10^{4}$\,cm$^{-3}$ and $1.4 \times 10^{5}$\,cm$^{-3}$ respectively). The higher density model is likely to be the correct one; as in the case of HCN, radio observations are mainly sensitive to high density gas.
The model reproduces HNC deconvolved line temperatures for both the H$_{2}$CO and SO densities \citep[using rates from][]{schoier0503}. 
N$_{2}$H$^{+}$ was not detected, which is highly unusual and presumably a sign of N depletion. 

There are several ways to explain the discrepancies between the expected and derived from observations $^{12}$CO:$^{13}$CO:C$^{18}$O abundance ratios, and the surprisingly low fractional abundances of $^{13}$CO and C$^{18}$O:
(i) optical thickness of $^{12}$CO; (ii) $^{13}$CO enhancement by fractionation in cloud edges due to charge exchange reactions with C$^{+}$ \citep[e.g.][]{watson7605}; or (iii) small scale clumping, suggested by the H$_{2}$ density derived from the SO line intensity ratio. 
The energy barrier for the fractionation reaction $^{13}\mathrm{C}^{+} + \,^{12}\mathrm{CO} \,\longleftrightarrow\, ^{13}\mathrm{CO} + \,^{12}\mathrm{C}^{+}$ is $\sim$35\,K \citep{watson7605}, which requires low temperatures for efficient fractionation, although at 20\,K the enhancement would not be large. Low $A_\mathrm{V}$ and a high UV field are also required for the production of C$^{+}$. However, for higher UV fields, since $^{12}$CO self-shields more effectively than $^{13}$CO, selective photodissociation destroys $^{13}$CO faster than $^{12}$CO. So depending on the respective rates of the fractionation reaction and photodissociation, $^{13}$CO fractionation may or may not be taking place. Given the range of densities derived, small-scale structure and optical depths effects are the most likely explanations for the discrepancies (section \ref{chemods} examines the effect of high UV fields and low $A_\mathrm{V}$ in greater detail). 


In order to derive accurate molecular abundances one needs multiple transitions of a variety of isotopes to account for optical depth and excitation effects. Furthermore, the use of optically thin transitions of isotopes to derive accurate abundances of the main species is also fraught with difficulty since the underlying isotopic ratios may be different in this gas at the edge of the Galaxy than in the more local molecular clouds. To derive molecular abundances for species other than CO, H$_2$CO and NH$_3$, we have assumed LTE and an excitation temperature of 20\,K. Most of the derived abundances are not very sensitive to temperature, typically varying by less than $\pm$10 per cent for $T = 20 \pm 3$\,K, except for the abundances derived for CO 3--2 ($\pm$22\%), C{\sc i} ($\pm$18\%), SO ($\pm$13\%), H$_{2}$CO ($\pm$12\%) and NH$_{3}$($\pm$12\%). 

For optically thin line emission from linear molecules, we find that the total column density, $N$ (cm$^{-2}$), of a molecule with normalised electric dipole moment $\tilde{\mu}$ (in units of 10$^{-18}$ esu) with an excitation temperature $T_\mathrm{ex}$ emitting at frequency $\nu$ in a transition J$\rightarrow$J--1 can be written in the form: 
  \begin{equation} 
    N = \frac{1.603 \times 10^{13}}{\tilde{\mu}^{2}} \frac{T_\mathrm{ex}}{T_{0}^{2}} 
    e^{(J+1)T_0/2T_\mathrm{ex}} \int{T_\mathrm{mb}\mathrm{d}V}, 
  \end{equation} 
where $T_{0} = h\nu/k$ and the deconvolved integrated intensity for a Gaussian line profile is given by $1.06\,T_\mathrm{mb} \Delta V$ in units of K\,km\,s$^{-1}$. For non-detections we use the 3$\sigma$ upper limits for the integrated intensity: 
  \begin{equation} 
    \Delta (\mathrm{K\,km\,s}^{-1}) = 3\,rms_\mathrm{mb}(1 + \alpha) \sqrt{\Delta \nu \Delta V}, 
  \end{equation} 
where $\alpha$ is the calibration uncertainty taken to be 0.15, $\Delta \nu$ and $\Delta V$ are the spectral resolution and expected line-width (2\,km\,s$^{-1}$ is adopted here), both measured in km\,s$^{-1}$. 
For optically thin emission, the fractional abundance varies proportionally to the integrated line intensity. For CN we summed the intensities of our two observed hyperfine components and adopted a relative intensity of 0.456 to determine the integrated intensity and, thus, $N$(CN).  For HCN the sum of the three hyperfine components was used, while for C$_2$H we used the strongest hyperfine component at 87.317 GHz and adopted a relative intensity of 0.4167. 

We take a slightly different approach for SO, which has a $^3\Sigma$ ground state. We calculate $N$(SO) directly using the formula given by \citet{takano9512} with the partition function from the JPL Molecular Spectroscopy Catalogue \citep{pickett98}. For C{\sc i} we use a similar approach to that of \citet{tauber9505}. To estimate $N$(H$_{2}$CO) we use the LVG modelled density:
$N(\mathrm{H}_{2}\mathrm{CO}) = n(\mathrm{H}_{2}\mathrm{CO}) \times 1 \mathrm{pc} \times \Delta V$, 
where $n$(H$_{2}$CO) is the best fit abundance to the LVG calculation ($1.4 \times 10^{-6}$\,cm$^{-3}$) and $\Delta V = 1.3$ km\,s$^{-1}$ (the $1_{1,0}-1_{1,1}$ line is widened by hyperfine structure). 
Table~\ref{ec2_sumobs} lists the derived column densities. 

Table \ref{ec2_lvgcomps} compares line temperatures and column densities for EC2 derived from our LVG models and observed deconvolved integrated intensities, and it can be seen that in most cases there is good agreement between $N_\mathrm{mb}$(X) and $N_\mathrm{lvg}$(X). The LVG column densities were calculated from the best fit model densities with $N_\mathrm{lvg}$(X) = $n_\mathrm{lvg}$(X) $\times$ 1\,pc $\times$ $\Delta V$.
It should be pointed out that our assumption that $T_\mathrm{ex} = T_\mathrm{kin} = 20$\,K could be considered unrealistic, since excitation temperatures are crucial for abundance determination. CO is almost certainly thermalised, but other molecular species may not be thermalised with similar certainty. At the low densities we have derived from the observations, molecules like HCO$^{+}$, HCN and HNC would not be thermalised. However, our LVG results provide a realistic estimate of the real excitation temperature ($T_\mathrm{ex}$ in Table \ref{ec2_lvgcomps}) for a range of our detected molecules. Recalculating with these specific values (instead of $T_\mathrm{ex} = 20$\,K), gives column densities that only differ by factors of between 0.5 and 3.7 (except CS 3--2 by 8.2). 

To aid comparison with local molecular clouds, we have derived molecular abundances relative to that of HCO$^+$, assuming that the 1--0 transition is optically thin. Our upper limit to H$^{13}$CO$^+$ gives a lower limit to the $^{12}$C/$^{13}$C ratio of around 17, much less than the lower limit of 201$\pm$15 derived from CO 1--0 by \citet{wouterloot9611} towards WB 89-437 at a galactocentric distance of 16.4\,kpc, although we note that the value they derive from the 2--1 transition is much smaller, 104$\pm$60. Table~\ref{ec2_sumdens} shows the result and also the equivalent abundance ratio for the nearby dark clouds L134N (center position), TMC-1 (average), and the range of values derived in L134N \citep{dickens0010,pratap9709}. Note that these authors use isotopic species to determine the column densities of optically thick lines. In particular, they adopt $^{12}$C/$^{13}$C = 64 for determining the column densities of HCO$^+$, HCN and HNC and $^{16}$O/$^{18}$O = 500 to determine the CO and $^{13}$CO column densities. Finally we note that they used a dipole moment of 4.07 D for HCO$^+$ compared to the value of 3.9 D used here. Use of our value would increase their abundance ratios by about 10\%. Table~\ref{ec2_sumdens} also gives fractional abundances relative to molecular hydrogen using $N$(H$_{2}$) = $7.4 \times 10^{22}$\,cm$^{-2}$ and compares these values with those for the translucent clouds observed by \citet{turner0010}. 


Dust emission can be used to estimate the mass of cold molecular clouds and we use our detected peak continuum emission at 1.2\,mm to calculate the opacity and dust-to-gas ratio in EC2. First, we estimate $\tau_{\nu}$, the dust opacity at frequency $\nu$, using 
  \begin{equation}
    \tau_{\nu} = B(\nu) / \pi(\theta_\mathrm{b}/2)^{2} B_{\nu}(T_\mathrm{dust}), \: (\tau \ll 1)
  \end{equation}
from \citet{hildebrand8309}, where $B_{\nu}(T_\mathrm{dust})$ is the Planck function, $T_\mathrm{dust} = 20$\,K, the observed dust emission $B(\nu) = 20.3 \times 10^{-3}$ Jy, and the beam size $\theta_\mathrm{b} = 11\arcsec$, which gives a value of $\tau_{\nu} = 3.25 \times 10^{-4}$. 
The dust opacity per unit mass column density at frequency $\nu$, is given by
$\kappa_{\nu} = 3Q_{\nu} / 4a\rho$, 
where $Q_{\nu}$ is the dust emissivity which varies as $\nu^{\beta}$ [$B(\nu) \propto \nu^{\beta} B_{\nu}(T_\mathrm{dust})$], the average dust grain radius $a = 0.1\,\mu$m, and the density of the dust $\rho = 3$\,g\,cm$^{-3}$. Values for $Q_{\nu}$, $a$ and $\rho$ are taken from \citet{hildebrand8309}.
From \citet[p150]{whittet92} we note the relation $\rho_\mathrm{dust} = \tau_{\nu} / \kappa_{\nu} L$, 
and it therefore follows that the dust to gas ratio is:
  \begin{equation}
    \frac{M_\mathrm{dust}}{M_\mathrm{gas}} = \frac{\tau_{\nu}}{\kappa_{\nu} L\, n(\mathrm{H}_{2}) \mu m_\mathrm{H}}, 
  \end{equation}
where $L=D\theta_\mathrm{b}$ is the size of the source and $D$ is the heliocentric distance (14--20\,kpc). Table \ref{ec2_mass} lists derived dust to gas ratios for $D$ = 14 and 20\,kpc and a range of values for $\beta$ from 1 to 2 (taken to characterise dust grains that range from amorphous to crystalline), as well as appropriate values for $Q_{1200}$ and $\kappa_{\nu}$. 

It now remains to relate the dust to gas ratio to possible values of extinction in EC2. For the standard dust-to-gas mass ratio of 0.01 in the local ISM, it is found that $N(\mathrm{H})/E(B-V) = 5.8 \times 10^{21}$ atoms cm$^{-2}$ mag$^{-1}$ \citep{bohlin7808} and with $R_\mathrm{V} = A_\mathrm{V}/E(B-V)$, we can write:
  \begin{equation}
     A_\mathrm{V} = \frac{2 n(\mathrm{H}_{2}) LR_\mathrm{V}}{5.8 \times 10^{21}} \: \frac{M_\mathrm{dust}/M_\mathrm{gas}}{0.01}. 
  \end{equation}
As discussed in \citet{ruffle0409}, the value of $R_\mathrm{V}$ may be lower in certain regions of the ISM, so we calculate values of $A_\mathrm{V}$ for $R_\mathrm{V} = 3.1$ and $R_\mathrm{V} = 2.0$ at the position of maximum continuum emission. As our CO maps show, EC2 is clumpy in structure and the peak dust detection of 20.3\,mJy/beam ($\theta_\mathrm{b} = 11\arcsec$) traces the peak CO emission of one such clump. Over the map, the rms value of 6.35\,mJy/beam is $\sim$3 times less than the peak, so extinction in EC2 as a whole is likely to be correspondingly lower.

Finally, we estimate upper and lower limits for the mass of EC2 based on a density of $n(\mathrm{H}_{2}) = 1.2 \times 10^{4}$\,cm$^{-3}$. From our maps of the two CO 2--1 peaks in EC2 (Fig. \ref{ec2_map_positions}) we estimate that the cloud consists of two dense clumps $\sim$75\arcsec\ in diameter, which yield a mass range of $9.3 \times 10^{4}$ to $2.7 \times 10^{5}\,M_{\sun}$ for the distance range 14--20\,kpc. Taking the clumpy structure of EC2 into account, the average density may be of the order $n(\mathrm{H}_{2}) = 10^{3}$\,cm$^{-3}$, so our mass estimate is $M_\mathrm{EC2} \sim 10^{4}\,M_{\sun}$ (based on their CO detections \citet{digel94} estimated $M_\mathrm{EC2} = 3.7 \times 10^{4}\,M_{\sun}$). This value can be compared with a virial mass estimate using the method of \citet[equation 3]{maclaren8810}, $M_\mathrm{vir}/M_{\sun} = k_{2} r \Delta V^{2}$, where $k_{2} = 190$ (for $\rho \propto 1/r$), $r$ is the radius of the clumps in pc, and $\Delta V$ is the $^{13}$CO line width (FWHM) in km\,s$^{-1}$. 
\citeauthor{maclaren8810} argue that the virial theorem will only yield a meaningful result if the estimate of $\Delta V$ is found from the average throughout the whole volume of the cloud, including central core regions. An optically thick line such as $^{12}$CO will only probe the cloud to the surface where the optical depth is unity. On the other hand, an optically thin line, such as $^{13}$CO should allow sampling of velocities throughout the central, denser regions of the cloud. However, taking the ``radius'' of EC2 from the $^{12}$CO observations should avoid the effect of excluding the lower density outer regions, i.e. an overestimation of the total mass.
For the distance range 14--20\,kpc, the radius of each of the two clumps, $r$ is 2.5--3.7\,pc and the line width for the strongest $^{13}$CO 2--1 transition is 1.87 km\,s$^{-1}$, which yields a total virial mass $M_\mathrm{vir}$ = 3.3--4.9 $\times 10^{3}\,M_{\sun}$, somewhat less than our direct estimate above. The lower virial mass suggests that the cloud is gravitationally unstable and a likely site of star formation as suggested by \citet{yasui0610}.

\section{Discussion} 

The data in Table~\ref{ec2_sumdens} indicate some interesting differences 
between molecular abundances relative to HCO$^+$ in EC2 and local dark 
molecular clouds. In particular sulfur-bearing molecules, (CS, SO), appear 
to be very over-abundant; the nitrogen-bearing species HCN is comparable while 
HNC appears to be marginally under-abundant -- the HNC/HCN ratio is less than 
one, whereas in dark clouds it is usually close to or slightly greater than one -- and the radicals CN 
and C$_2$H are very much over-abundant. The latter result is typical of 
photon-dominated regions in which photoprocesses ensure that radicals exist in 
high abundance. At first glance it would appear that the molecular clouds at 
the edge of the Galaxy would be less likely to show the effects of PDR 
chemistry; certainly the flux of UV photons must be much less at the Galactic 
edge than in the local ISM. However, the critical parameter for determining 
whether or not photons dominate chemistry is the ratio of UV flux to grain 
surface area, particularly the area of the small grains which extinguish UV 
photons most efficiently. At large galactocentric radii, the metal abundances 
relative to hydrogen are expected to be much reduced, consistent with a reduced dust-to-gas ratio, as discussed previously. 
In addition, although the region of EC2 appears to contain young T Tauri 
stars \citep{yasui0610}, there is no evidence of the late-type stars which 
produce dust grains.

\section{Chemical Modelling}\label{chemods} 

To investigate the properties of EC2 and to see whether it is typical of 
material which has been less processed, we have made a pseudo-time-dependent 
chemical kinetic model. This uses observationally derived temperatures and 
densities and varies elemental abundances, photon flux, cosmic ray ionisation 
(CRI) rate and gas-to-dust ratio in an attempt to fit the observed results. 
Chemical reaction rates were taken from Rate99: The UMIST database for 
astrochemistry \citep{rate99}. 

Unless otherwise stated, the model assumes $T_\mathrm{kin}$ = 20\,K, $n$(H$_{2}$) = $1.2 \times 10^4$\,cm$^{-3}$, based on NH$_{3}$, H$_{2}$CO, and CO observations, and uses the initial elemental abundances listed in Table~\ref{ec2_inabund}. For  the metals (C, N, O, S, Si and Fe) these are reduced by a factor of five from those typically used for local ISM clouds; we find that without this reduction in elemental abundances, agreement with observations is worsened for $A_\mathrm{V} > 1$\,mag. We have investigated varying the CRI rate and the visual extinction ($A_\mathrm{V}$) in order to try and find the best fit to the observations. We also varied the UV photon field, initial abundances, and, specifically, the abundances of sulfur or nitrogen. 

Our JCMT CO maps probed at higher resolution (15-23\arcsec), but the bulk of our detections (with the ARO 12\,m) covered beams of 42-70\arcsec. Our Effelsberg detections had beams of 30\arcsec\ for SO, 40\arcsec\ for NH$_{3}$, and 55\arcsec\ and 160\arcsec\ for H$_2$CO. As described in section \ref{analysis}, our detected peak intensities were deconvolved for these beam sizes ($\theta_\mathrm{b}$) and our estimated $fwhp$ clump size ($\theta_\mathrm{s}$).
The peak CO 2--1 emission seen in our maps is within $\sim$30\arcsec, or half a typical beamwidth, of position A. Our ARO 12m detections of CO, CS, SO, HCO$^+$, HCN, C$_2$H and C$_3$H$_2$ at nearby positions B, C, D, E, G, H and J \citep{ruffle06phd} are commensurate with their position A detections, given that the intensity contours seen in Fig. \ref{ec2_map_positions} would be `smeared' by the relevant beam sizes (55-70\arcsec).
At a distance of 20\,kpc, a 60\arcsec\ beam subtends a linear distance of about 6\,pc or a column density of about $2 \times 10^{23}$\,cm$^{-2}$ for $n$(H$_{2}$) = $1.2 \times 10^4$\,cm$^{-3}$. 
Since our CO maps show that EC2 is extended and clumpy, the `one-point' model adopted is simplistic; nevertheless, given the lack of data on very small scales, one might hope that the model helps constrain some of the global parameters in the cloud in a `beam-averaged' sense. 321 models with varying parameters were run and the results of testing each one against our observations are detailed in \citet{ruffle06phd}. 

To test the agreement between each model and our observations, we looked at the ratio of the observed abundance relative to H$_{2}$ from Table~\ref{ec2_sumdens} to the model prediction (or the inverse of that ratio if it was greater than 1) for that species. For each molecule/transition we then square its agreement factor and sum these squares for each model, normalising to get a `Fit' number between 0 and 1. This should be closest to 1 for the model which agrees best. 
In the case of our multiple transition detections of $^{12}$CO, $^{13}$CO and C$^{18}$O, we have used a $\sigma$ weighted mean of each isotopomer's observed column densities, where $\sigma = T_\mathrm{mb} / rms_\mathrm{mb}$.
However, as the models only consider the main isotope of an atom (with the exception of H and D) we have assumed that the isotopic ratios have their local interstellar values ($^{12}$C/$^{13}$C = 60, C$^{16}$O/C$^{18}$O = 500) and factored our observed (mean) abundances relative to $N$(H$_{2}$) accordingly. For the $^{13}$CO and C$^{18}$O transitions, this gives us a reasonably consistent value of $N$(CO)/$N$(H$_{2}$) $\sim 1.4 \times 10^{-6}$, although the main $^{12}$CO lines give a value closer to $10^{-7}$. Since the main lines of $^{12}$CO are optically thick (0.6-2.2 from Table \ref{ec2_lvgcomps}), the ratios derived from observation are lower limits. 

We have selected 10 models for detailed examination, which can be compared 
in Tables \ref{ec2_chemod_fits}, \ref{ec2_abundModels1-2-3-4}, \ref{ec2_abundModels2-5-6-7} and 
\ref{ec2_abundModels2-8-9-10}. We investigated the agreement factor at 
steady state which for our choice of parameters is reached, in general, at 
times less than than 10$^5$ yr. 
Fig. \ref{ec2_3Dplots} plots 3D parameter surfaces of these test results for 
various combinations of the models, where two parameters are varied and the 
others are kept constant. For ease of comparison, numbered disks, marked on 
the 3D plots, indicate the examined models relative to the parameter surfaces. 
Fig. \ref{frac_Models} 
shows fractional abundances varying over time for five of the models. 

\subsubsection*{Varying CRI and $A_\mathrm{V}$}

We investigated the agreement tests for values of $A_\mathrm{V}$ between 0.5 and 10~mag and a CRI rate between 0.5 and 30 times the standard interstellar rate of $1.3 \times 10^{-17}$\,s$^{-1}$. The UV field was set to the local ISM value.
 Although many of the models give very similar results, the agreement factor is highest for a CRI rate above 10 $\times$ the ISM value and an $A_\mathrm{V}$ of 1~mag. Fig.~\ref{ec2_3Dplots}a shows the general trend of varying $A_\mathrm{V}$ and CRI. It can be seen that increasing the CRI rate beyond 20 $\times$ the ISM value extends agreement in the model to $A_\mathrm{V}$ = 2~mag. What is striking however, is the failure of the model for values of $A_\mathrm{V}$ above 2~mag. It should be noted that model results for $A_\mathrm{V} \le 0.5$~mag should be treated with caution, as the model may fail for such low values of $A_\mathrm{V}$, since it does not treat the self-shielding of H$_{2}$ and CO accurately. 

Table~\ref{ec2_abundModels1-2-3-4} shows how well these models actually agree with the observations for four selected models (1--4). Model~1 uses `standard' ISM values for the $A_\mathrm{V}$ and CRI rate, Model~2 is the best fit model (CRI = $2.6 \times 10^{-16}$ s$^{-1}$, $A_\mathrm{V}$ = 1~mag), while Model~3 uses CRI = $1.3 \times 10^{-16}$ s$^{-1}$ and $A_\mathrm{V}$ = 2~mag, to illustrate some effects which arise from changing these parameters. Model~4 is the same as Model~3, but with a UV field 10 times the local rate. 

For Model~1 the predicted CO/H$_{2}$ ratio is 1 order of magnitude higher than 
all the observations. Models~2 and 4 agree within a factor of 2 for $^{12}$CO 
but underestimate $^{13}$CO and C$^{18}$O by $\sim$1 order of magnitude, 
whereas Model~3 reverses those agreements. 
Model~1 does badly at reproducing the observed value for atomic C, since when 
the $A_\mathrm{V}$ is 10~mag the cloud is shielded from UV radiation and the 
bulk of the carbon becomes incorporated into CO.  When the $A_\mathrm{V}$ is 
lowered and/or the CRI rate increased, however, CO can be broken down, so that 
the abundance of C{\sc i} remains significant even at steady-state. We find 
that an $A_\mathrm{V}$ of 1 or 2~mag gives a better agreement with the EC2 
observation of $N$(C)/$N$(H$_{2}$). 

The predicted value of C$_{2}$H/H$_{2}$ in Model~1 agrees with the observations
to within a factor of 2, while the other three Models are 10 to 40 times too 
high. Conversely, Models~1 and 3 over-predict the CN/H$_{2}$ ratio by over 2 
orders of magnitude, with Models~2 and 4 over-estimating it by 40 and 50 times 
respectively. The time-dependent results show that Models 2 and 4 have good 
agreement overall once $t$ is greater than about 2000 yr, whereas neither 
Model~1 or 3 show good agreement beyond the first 500 years (see Fig. 
\ref{frac_Models}f).

\subsubsection*{Varying UV photon field, CRI and $A_\mathrm{V}$}

\begin{figure*} 
\hbox{
\includegraphics[width=75mm]{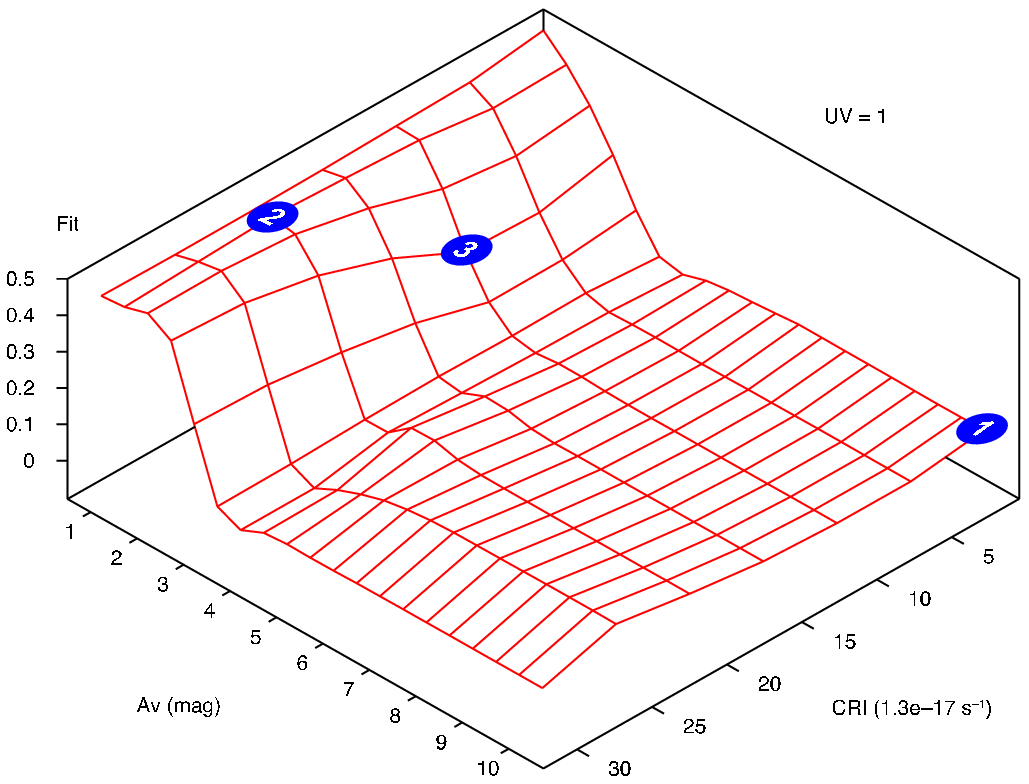}%
\hspace{10mm}
\includegraphics[width=75mm]{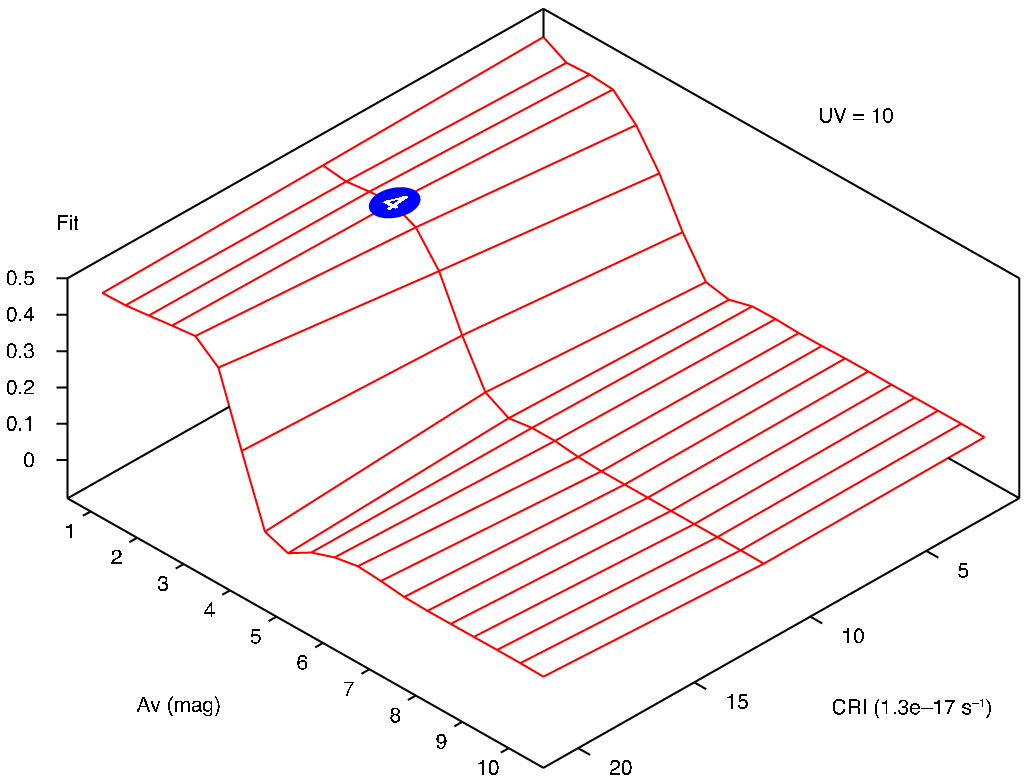}}
\vspace{2mm}
\hbox{
\includegraphics[width=75mm]{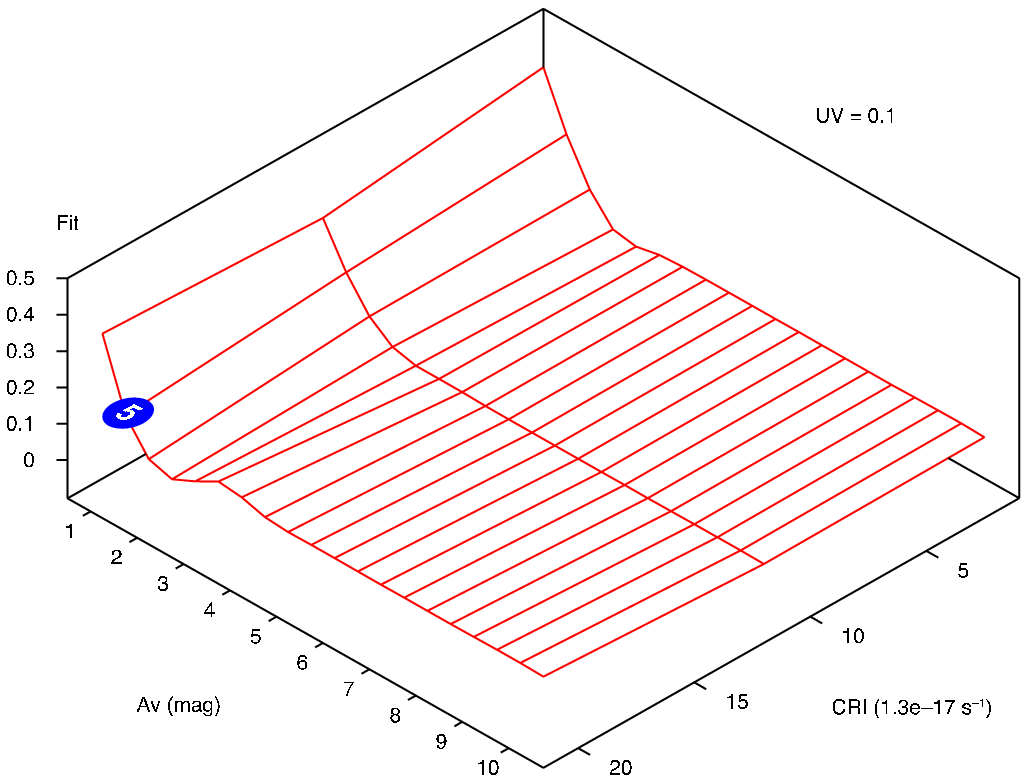}%
\hspace{10mm}
\includegraphics[width=75mm]{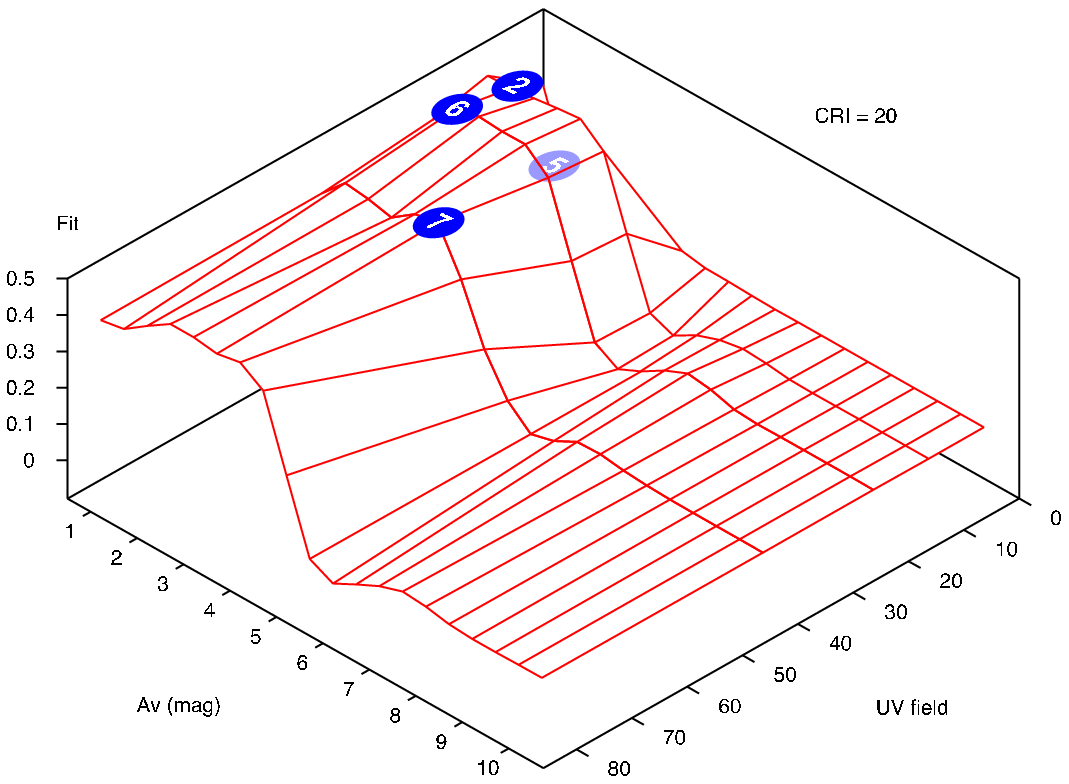}}
\vspace{2mm}
\hbox{
\includegraphics[width=75mm]{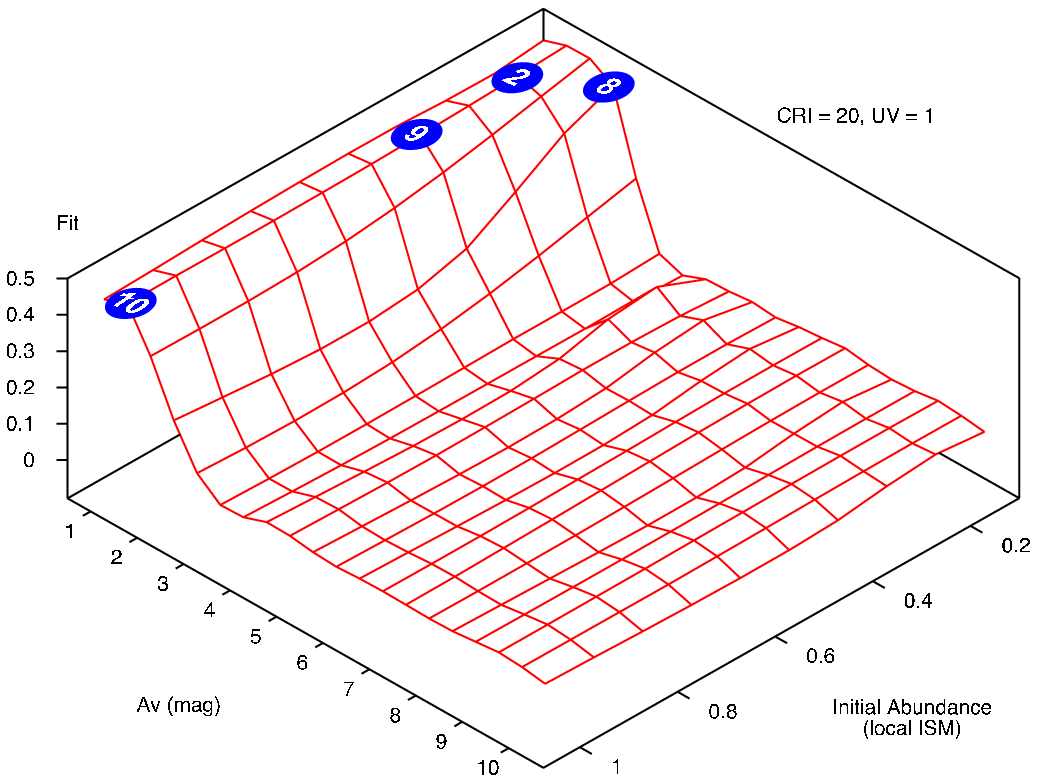}%
\hspace{10mm}
\includegraphics[width=75mm]{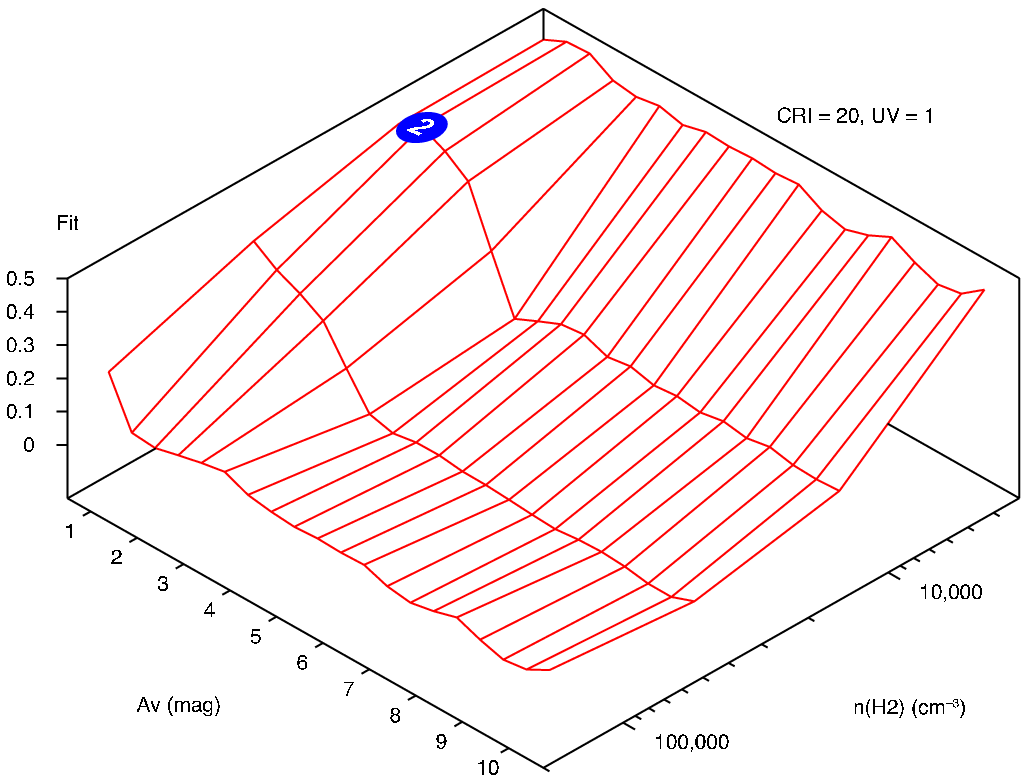}}
\caption{Agreement factor for models with varying: 
         (a) $A_\mathrm{V}$ (mag) and cosmic ray ionisation (CRI); 
	 (b) $A_\mathrm{V}$ and CRI with UV field increased by a factor of 10 compared to local ISM values; 
	 (c) $A_\mathrm{V}$ and CRI with UV field reduced by a factor of 10; 
	 (d) $A_\mathrm{V}$ and an increasing UV field; 
	 (e) $A_\mathrm{V}$ and most initial abundances reducing relative to typical local ISM values; 
         (f) $A_\mathrm{V}$ and $n(\mathrm{H}_{2})$.
         Unless indicated above, parameters were fixed at: 
	 $T$ = 20\,K; $n(\mathrm{H}_{2}) = 1.2 \times 10^{4}$\,cm$^{-3}$; 
         CRI rate = $20 \times$ the standard ISM rate of $1.3 \times 10^{-17}$\,s$^{-1}$; 
         UV field set to local ISM values; 
	 most initial abundances reduced by a factor of 5 from typical local ISM values (see Table \ref{ec2_inabund}). 
	 Numbered disks indicate specific models relative to the parameter surface. 
\label{ec2_3Dplots}} 
\end{figure*}

\begin{figure*} 
\hbox{
\includegraphics[width=81mm]{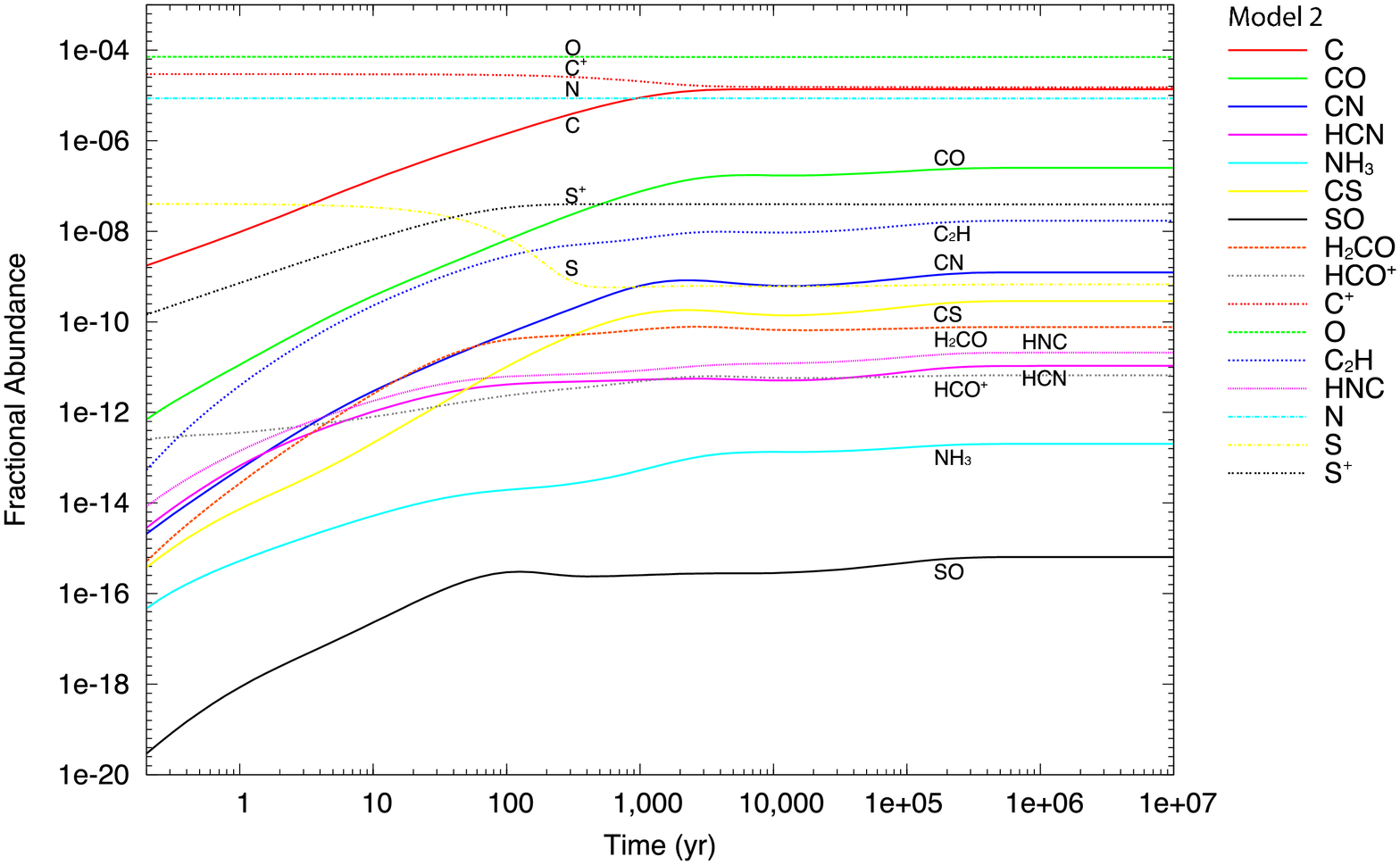}%
\hspace{2mm}
\includegraphics[width=81mm]{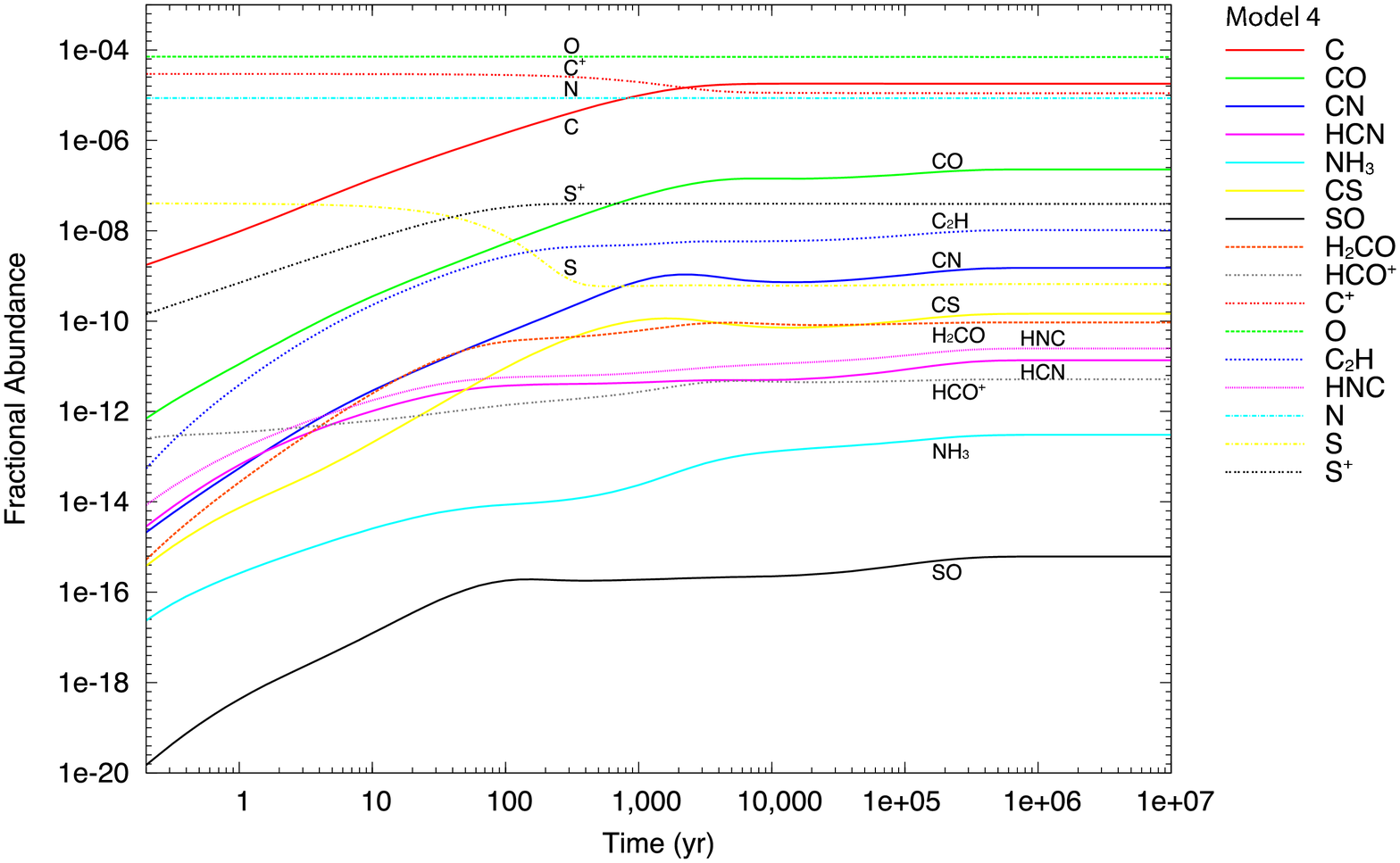}} 
\vspace{2mm}
\hbox{
\includegraphics[width=81mm]{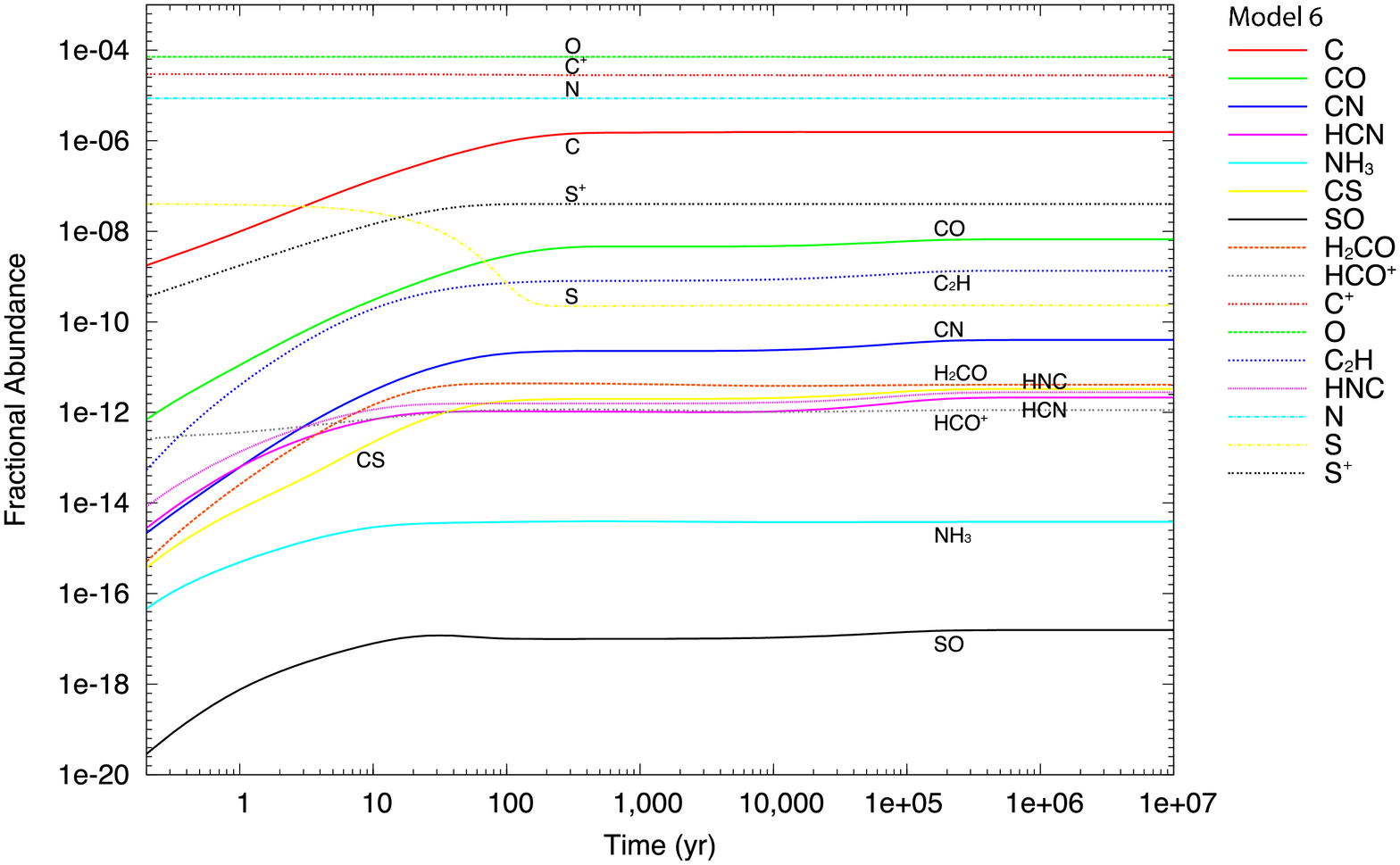}%
\hspace{2mm}
\includegraphics[width=81mm]{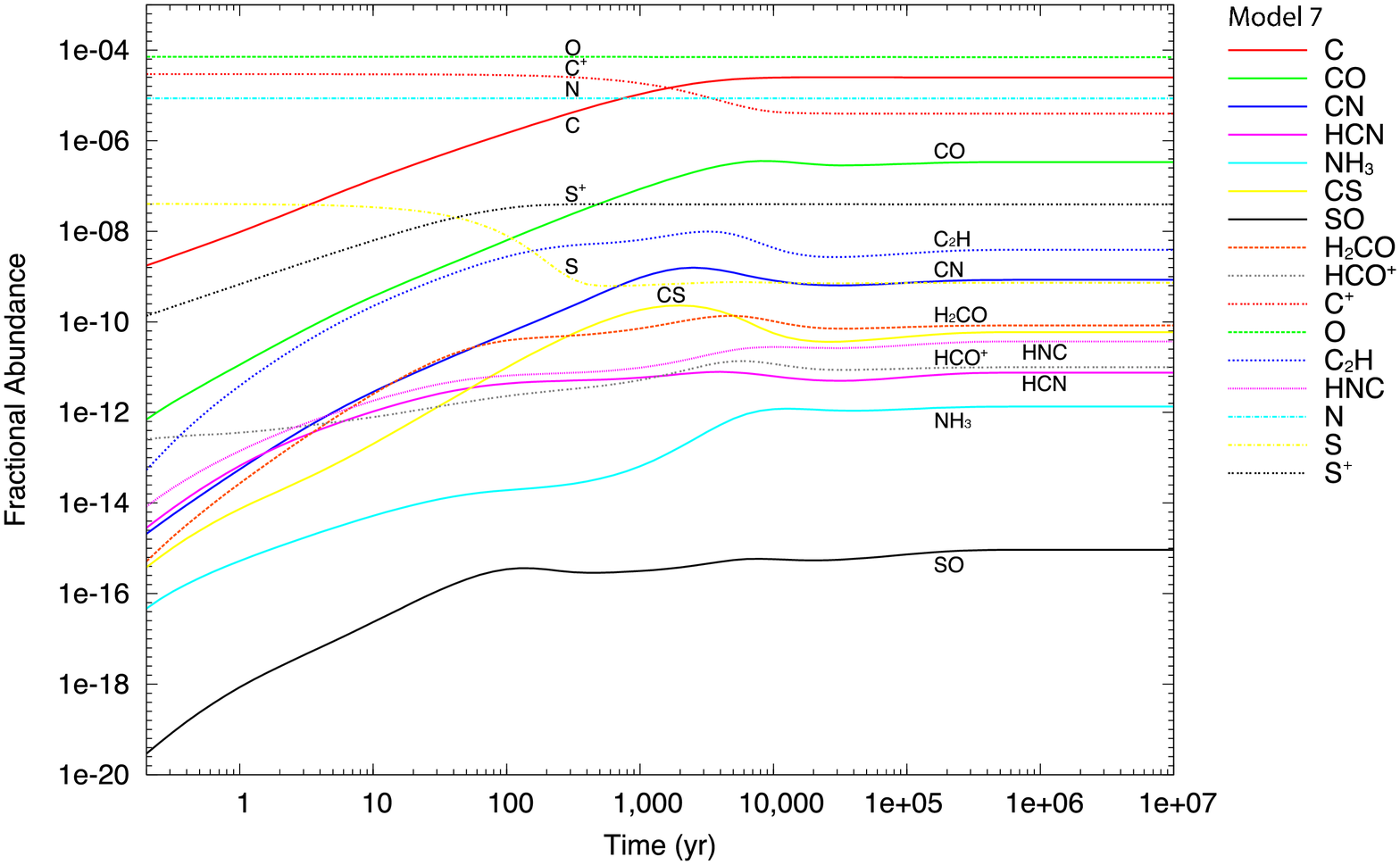}} 
\vspace{2mm}
\hbox{
\includegraphics[width=81mm]{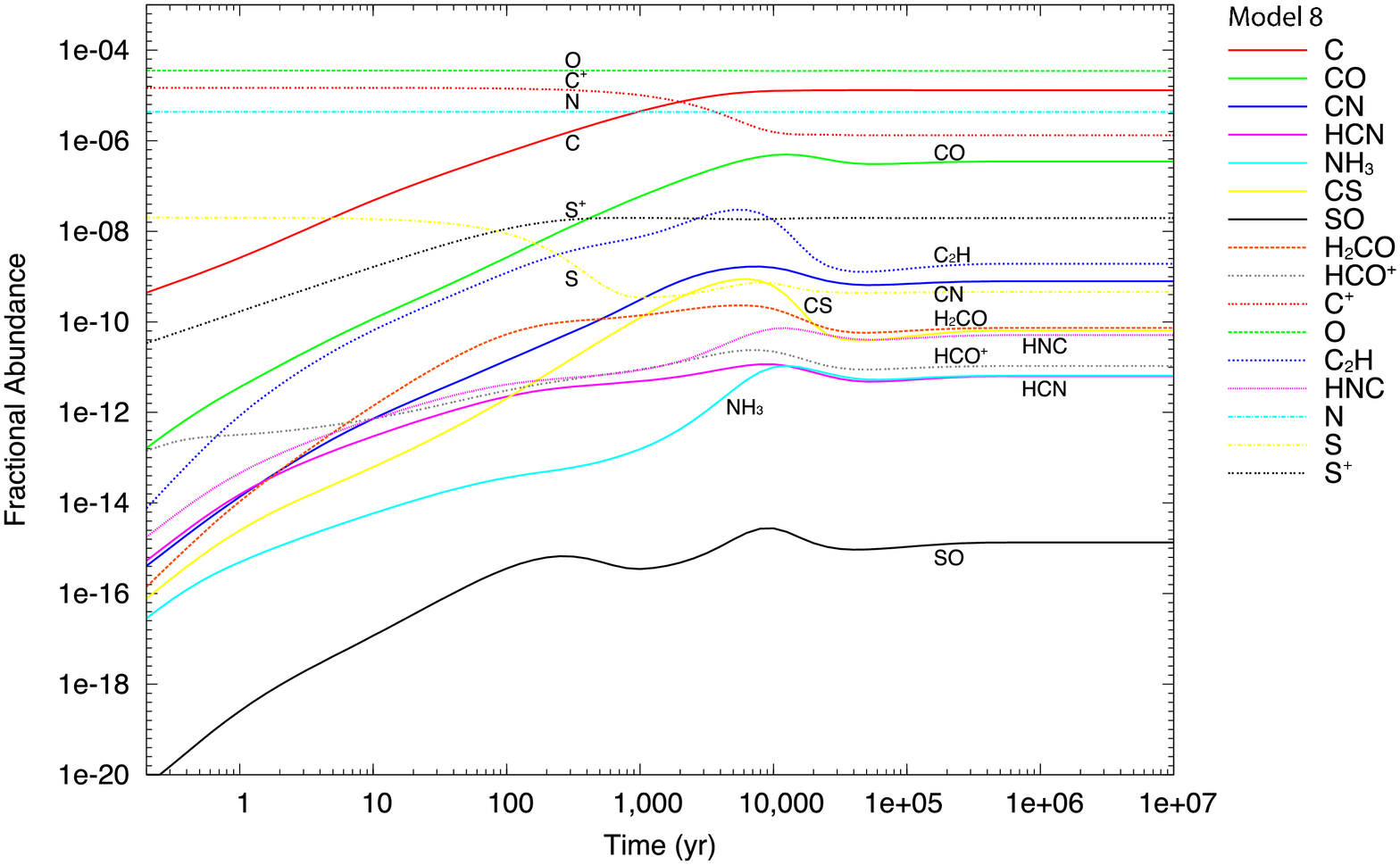}%
\hspace{2mm}
\includegraphics[width=71mm]{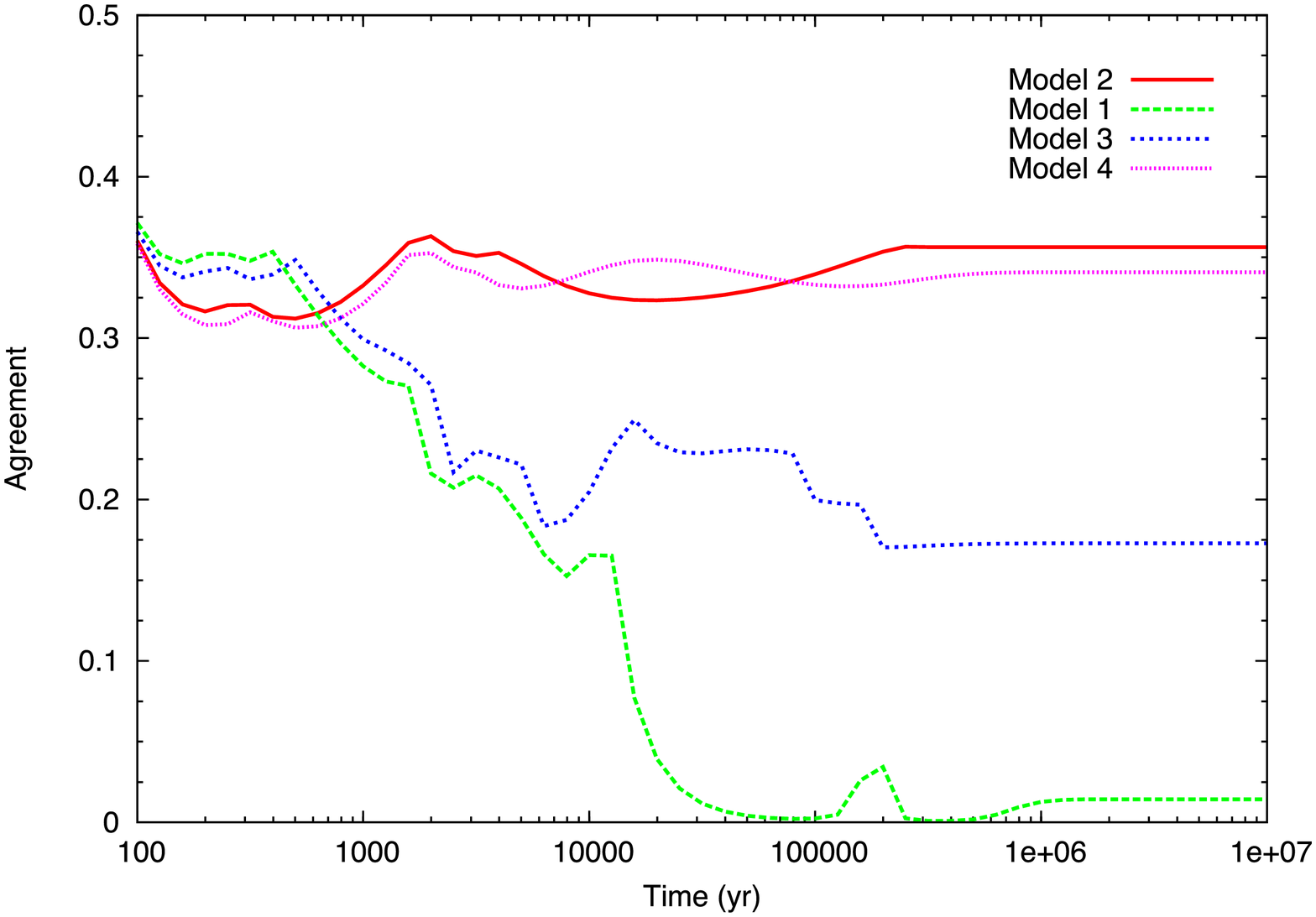}} 
\caption{Fractional abundances varying over time (see Tables \ref{ec2_abundModels1-2-3-4}, 
         \ref{ec2_abundModels2-5-6-7} and \ref{ec2_abundModels2-8-9-10}) for: \newline
	 (a) Model 2, where $A_\mathrm{V}$ = 1 mag, CRI = 20$\times$, UV = 1$\times$, 
	 IA = 0.2$\times$, $n$(H$_{2}$) = $1.2 \times 10^{4}$\,cm$^{-3}$, $T$ = 20\,K; \newline
	 (b) Model 4, where $A_\mathrm{V}$ = 2 mag, CRI = 10$\times$, UV = 10$\times$, 
	 IA = 0.2$\times$, $n$(H$_{2}$) = $1.2 \times 10^{4}$\,cm$^{-3}$, $T$ = 20\,K; \newline
	 (c) Model 6, where $A_\mathrm{V}$ = 1 mag, CRI = 20$\times$, UV = 20$\times$, 
	 IA = 0.2$\times$, $n$(H$_{2}$) = $1.2 \times 10^{4}$\,cm$^{-3}$, $T$ = 20\,K; \newline
	 (d) Model 7, where $A_\mathrm{V}$ = 3 mag, CRI = 20$\times$, UV = 40$\times$, 
	 IA = 0.2$\times$, $n$(H$_{2}$) = $1.2 \times 10^{4}$\,cm$^{-3}$, $T$ = 20\,K; \newline
	 (e) Model 8, where $A_\mathrm{V}$ = 2 mag, CRI = 20$\times$, UV = 1$\times$, 
	 IA = 0.1$\times$, $n$(H$_{2}$) = $1.2 \times 10^{4}$\,cm$^{-3}$, $T$ = 20\,K. \newline
         (f) Agreement factor varying over time for Models 1--4, where $A_\mathrm{V}$ (mag) 
	 and cosmic ray ionisation (CRI) rates are varied 
	 (see Figs. \ref{ec2_3Dplots}a,b and Table \ref{ec2_abundModels1-2-3-4}).
\label{frac_Models}} 
\end{figure*}

We have also looked at the effect of varying the photon field in the models. Fig. \ref{ec2_3Dplots}b,c shows agreement factors for models with varying $A_\mathrm{V}$ and CRI for (b) a UV field increased by a factor of 10 compared to the local ISM value, and (c) a UV field reduced by a factor of 10. It can be seen that a higher UV field improves agreement up to an $A_\mathrm{V}$ of 2~mag regardless of CRI rate (although for CRI 20 $\times$ this is extended to 3~mag), with Model~4 the best fit. A lower UV field produces very poor agreement (e.g. Model~5).

Fig. \ref{ec2_3Dplots}d plots fit results from models where the CRI rate is fixed at 20 times the standard value, $A_\mathrm{V}$ is varied, and the UV field is varied from one-tenth to 80 times the local interstellar field. As expected, an increase in the UV photon field extends agreement to higher values of $A_\mathrm{V}$ (4~mag for 80 $\times$ UV). Table \ref{ec2_abundModels2-5-6-7} shows the agreement for three of these models. Interestingly, 20 $\times$ UV produces a good fit for atomic carbon at an $A_\mathrm{V}$ of 1~mag (Model~6), whereas 40 $\times$ UV at an $A_\mathrm{V}$ of 3~mag gives a good fit for CO (Model~7). At steady-state CO is primarily destroyed by photons (only if the UV field is strong enough, or the $A_\mathrm{V}$ low enough) or by He$^{+}$ and H$_3^+$, whose abundances increase with increasing CRI rate. Again a UV field below the local value produces very poor general agreement, but apart from very high UV fields ($\sim$70 $\times$) at an $A_\mathrm{V}$ of $\sim$4~mag, there is poor agreement for any $A_\mathrm{V}$ above 3~mag. Interestingly the high UV (20 $\times$) Model~6 shows remarkable agreement from the earliest times and Model~7 (40 $\times$ UV) agrees well from $t = 2 \times 10^4$ yr. 

The observed upper limit to the DCO$^{+}$/HCO$^{+}$ ratio, 0.08, is not a 
useful limit since all models with the exception of Model~1, satisfy this 
constraint.
For C$_{2}$H/H$_{2}$ and CN/H$_{2}$, the ratios are overestimated by 2 orders 
of magnitude in the low UV Model~5, but the fit improves with increasing UV 
field, with good agreement reached at 20 $\times$ UV (Model~6). Finally, we 
note that there is a much better fit to the ammonia abundance in low UV 
Model~5 (which is also achieved for NH$_{3}$ in Model~3 with UV = 1 and CRI = 
10 times the local ISM rates).

\subsubsection*{Varying initial abundances and $A_\mathrm{V}$} 

Initial elemental abundances (see Table~\ref{ec2_inabund}) were varied between 
local ISM values and 10 per cent of those values. The parameter surface in 
Fig. \ref{ec2_3Dplots}e shows that for $A_\mathrm{V}$ above 1~mag, the 
agreement worsens appreciably for initial abundances (IA) above 20 per cent of 
local cloud values. Table \ref{ec2_abundModels2-8-9-10} compares  
Models~8--10, and for IA = 0.1 (Model~8) there is good agreement, except for 
atomic C and CN ($\sim$1 order of magnitude too great), NH$_{3}$ ($\sim$1 order 
to little), and SO (underestimated by 5 orders). Model~8's agreement varies 
considerably in the first 10,000 years or so, before settling down at a 
steady-state below Model~2 (IA = 0.2). 


The observed CS(2--1)/H$_{2}$ ratio is fairly well matched by our best fit 
Models~2 and 4 and very well matched by Models~7 and 8 (UV = 40 and IA = 0.1 
respectively). Except for Model~1, though, the predictions for SO are too low 
by several orders of magnitude. Therefore, we tested the effect of increasing 
the initial abundance of sulfur relative to other initial abundances (which 
are set at 20 per cent of local cloud values). As expected the abundances of 
sulfur-bearing molecules scale linearly with any change in the initial 
abundance of S. For the N-bearing species, all models except 1 give good 
agreement for HC$_{3}$N and N$_{2}$H$^{+}$, but the derived abundances 
are upper limits, so 
this result must be treated with some caution. For HCN and HNC the picture is 
less clear, but abundances are again well matched for our best fit Models~2 
and 4. Models~3 and 5 are in good agreement with the NH$_{3}$/H$_{2}$ ratio, 
while all other models (except Model~1) predict ratios 2 to 5 orders of 
magnitude too low. For CN the high UV Model~6 agrees with the observed 
abundance. 

\subsubsection*{Varying $n(\mathrm{H}_{2})$ and $A_\mathrm{V}$}

Fig. \ref{ec2_3Dplots}f shows the effect of varying $n(\mathrm{H}_{2})$ and $A_\mathrm{V}$ with a UV field set to the local ISM value and a CRI rate 20 times the local ISM value.  Densities above $n(\mathrm{H}_{2}) = 1.2 \times 10^{4}$\,cm$^{-3}$ produce a marked fall-off in agreement. We also found that, even for a UV field 10 times the local interstellar value, the visual extinction must be below $A_\mathrm{V} = 3$~mag for reasonable agreement.

\subsubsection*{Deuteration}

We failed to detect any deuterated species in this survey, with the upper 
limits being: DCO$^{+}$/HCO$^{+}$ $<$ 0.08; DCN/HCN $<$ 1.5; 
C$_{2}$D/C$_{2}$H $<$ 0.06 and HDCO/H$_{2}$CO $<$ 0.01.  The models presented 
here use the underlying D/H ratio from the local ISM.  Model~1 predicts 
DCO$^{+}$/HCO$^{+}$ $\sim$0.25, which is higher than that seen in local dark 
clouds such as TMC-1, a result of the lower abundance of heavy elements. 
Models~2 and 4, however, predict very low DCO$^{+}$/HCO$^{+}$ ratios 
($\sim$10$^{-3}$), whereas Model~3 predicts $\sim$10$^{-2}$. The large upper 
limits on DCN/HCN and C$_{2}$D/C$_{2}$H mean that these are not useful in 
constraining the models. Likewise for HDCO and DCO$^{+}$ the fractionation 
ratios agree except for Model~1.

\subsubsection*{Fractional abundances over time} 

To this point our model results have been discussed in terms of steady-state 
abundances and it is pertinent to ask whether there might be any significant
difference if we consider the time-dependent results. Fig. \ref{frac_Models} 
plots the fractional abundances over time for Models 2, 4, 6, 7 and 8. 
For our best-fit model, Model~2, steady-state is reached very quickly, after around 5,000 years (apart from some slight changes in HCN and HCO$^{+}$). 
Depending on the choice of the UV field and visual extinction, other models
reach steady state on somewhat shorter ($\sim$500 yr, Model~6) or somewhat 
longer ($\sim10^4$ yr, Model~7) times. The longest time to reach steady state 
is $\sim$5 10$^4$ yr (Model~8) which has the lowest initial elemental 
abundances. Given that the likely age of the
EC2 molecular cloud is greater than a few thousand years, we expect that the
steady-state results can be used to compare with the observed abundances.

\section{Conclusions} 

In order to deduce its physical and chemical properties, we have observed 
continuum emission and a large number of molecular transitions in EC2 at the 
Galactic edge. We have also made CO maps of EC2 and used these to calculate 
deconvolved line intensities. A temperature of 20\,K was estimated from 
hyperfine detections of ammonia and a gas density of $n(\mathrm{H}_{2}) 
\sim 10^{4}$\,cm$^{-3}$ was determined by comparing LVG models of a number of 
species to their deconvolved line detections. Molecular abundances were also 
determined from the LVG models and found to be in good agreement with 
abundances calculated directly from the deconvolved line intensities. From the 
peak continuum emission we calculated a dust mass for EC2 $\ge 150\,M_{\sun}$ 
and a dust-to-gas mass ratio $\ge 0.001$.

Through the use of chemical models we have been able to establish the most 
likely chemical and physical properties of EC2 (although we note that not all 
observed abundances can be reproduced in a self-consistent manner - see Table \ref{ec2_chemod_fits} 
where our chemical model agreement results are ranked by fit). There is 
an indication that heavy elements may be depleted by about a factor of five 
relative to local molecular clouds (similar to those in dwarf irregular 
galaxies and damped Lyman alpha systems). Such reduced abundances may be 
attributed to the low level of star formation in this region and are probably 
also related to the continuing infall of primordial (or low metallicity) halo 
gas since the Milky Way formed. 
The models also suggest a high UV photon field in EC2 (10--20 $\times$ local 
values), where an increased UV field allows for values of $A_\mathrm{V}$ up to 
4~mag, especially if this increased field is combined with an increase in CRI 
(10--20 $\times$), although the models are less sensitive to increases in the CRI 
rate. High CRI rates ($>$20 $\times$ the ISM value) without an increase in UV 
field only allow for extinction up to 2~mag. Gas densities much above 
$n(\mathrm{H}_{2}) = 1.2 \times 10^{4}$\,cm$^{-3}$ are excluded by the models, 
even if the UV field is increased. Some of our models indicate that 
steady-state is reached very quickly after around 5,000 years and that a high 
UV field can reduce this time to just $\sim$500\,yr. For a very high UV 
photon field (40 $\times$) and an extinction of $A_\mathrm{V}$ = 3~mag, however,
steady-state is not reached until $10^{4}$\,yr.

In the context of ratios relative to HCO$^{+}$, sulfur-bearing molecules 
appear to be very over-abundant by at least an order of magnitude compared to 
local dark clouds. The observed 
high abundances (again relative to HCO$^{+}$) of the radicals C$_{2}$H and CN 
are typical of photon-dominated regions (PDRs). This may be related to a large
value of the UV flux to grain surface area when compared to local clouds. In
particular, we find that our best-fit models are consistent with reduced 
elemental abundances and a low dust-to-gas mass ratio. 
In addition, although EC2 does contain young stars, there is no evidence of 
the late-type stars which produce dust grains, thereby justifying the 
assumption of a high ratio of UV flux to grain surface area. We conclude 
therefore, that despite the position of EC2 in the Galaxy, UV photons (rather 
than cosmic rays) play an important role in establishing its detailed chemical 
composition. The observed clumpy structure also leads to an enhanced role
for PDR-like chemistry in EC2. 

Given that EC2 is in a region of extremely low gas pressure and very small 
spiral arm perturbation, the question remains as to the origin of the 
structure and chemistry in EC2.  \citet{stil0112} speculate that old SN 
shells may be a source of dense clouds in low density environments such as 
the outer Galaxy. They go on to show that the SNR associated with EC2, 
GSH 138--01--94, is the largest and 
oldest SNR known to exist in the Milky Way. It consists of a H\,{\sc i} shell 
at a kinematic galactocentric distance of 23.6\,kpc, with an expansion 
velocity of 11.8$\pm$0.9\,km\,s$^{-1}$, an expansion age of 4.3\,Myr and a 
timescale for dissolving into the ISM of 18\,Myr. 
\citeauthor{stil0112} associate EC2 with the approaching side of the 
H\,{\sc i} shell, reducing the distance of EC2 to $R \sim 23.6$\,kpc, 
compared with 28\,kpc \citep{digel94}, and in closer 
agreement with the photometric distance range of 15-19\,kpc for the B star 
MR1 \citep{smartt96}. EC2 could even have been formed by GSH 138--01--94 from swept-up 
interstellar gas, through Rayleigh-Taylor instabilities. It could, therefore, be as 
young as the ages derived from the time-dependent calculations, discussed 
above. Although shock chemistry may be driven by the SNR, the short 
time-scales to reach steady-state imply that UV-driven chemistry can reset
shock abundances. One test of the importance of shock chemistry would be to 
search for SiO emission, a result of an enhanced Si gas-phase abundance due to
sputtering of interstellar grains.

We conclude that the formation, structure and subsequent chemistry 
of EC2 may be the direct result of shock fronts from GSH 138--01--94 
propagating through the medium sometime between 1,000 and 10,000 years ago. 
The role of GSH 138--01--94 in relation to 
EC2 is reinforced by \citet{yasui0610}, who conclude that the 
three-dimensional geometry of GSH 138--01--94, EC2 and the embedded star 
cluster, as well as the cluster's age ($\sim$1\,Myr, much less than the SNR's 
expansion age of 4.3\,Myr), strongly suggest that that the cloud collapse and 
subsequent star formation observed in EC2 was triggered by the SNR H\,{\sc i} shell.

\acknowledgements 

Astrophysics at the University of Manchester and Queen's University Belfast is supported by PPARC. 
PMER acknowledges receipt of a PPARC studentship. 
DAL is sponsored by an American Astronomical Society Small Research Grant and a Hofstra research grant.
Astrophysics at Williams College is supported in part by grants from NASA, the National Science Foundation, 
and the Committee for Research and Exploration of the National Geographic Society. 
We thank the referee for valuable suggestions that have improved and clarified this work. 
We are grateful to the telescope operators and staff of the ARO 12\,m, 
Effelsberg, IRAM 30\,m and JCMT telescopes for their help in 
securing these data. The James Clerk Maxwell Telescope (JCMT) is operated by the 
Joint Astronomy Centre on behalf of the United Kingdom Particle Physics and 
Astronomy Research Council, the Netherlands Organization for Scientific 
Research, and the National Research Council of Canada. The Kitt Peak 12\,m 
telescope is operated by the Arizona Radio Observatory (ARO), Steward 
Observatory, University of Arizona, with additional funding from the Academia 
Sinica Institute of Astronomy \& Astrophysics. 
The 100\,m telescope at Effelsberg is operated by the MPIfR (Max-Planck-Institut 
f{\"u}r Radioastronomie) and is supported by the MPG (Max-Planck-Gesellschaft). 
The 30\,m at IRAM (Institute de Radioastronomie Millimetrique) is supported by 
INSU/CNRS (France), MPG (Germany) and IGN (Spain).

{\it Facilities:} \facility{ARO:12\,m}, \facility{Effelsberg}, \facility{IRAM:30\,m}, \facility{JCMT}.


\clearpage

\begin{deluxetable}{lccccccc} 
\tabletypesize{\footnotesize} 
\tablecaption{Summary of observations and detections toward EC2 position A. Peak intensities corrected for source size using 
              $T_\mathrm{mb} = T^*_\mathrm{R} / \eta_\mathrm{bf}$, where 
              $\eta_\mathrm{bf} = \theta^{2}_\mathrm{s} / (\theta^{2}_\mathrm{s} + \theta^{2}_\mathrm{b})$; 
              column densities and abundances estimated for an excitation temperature of 20\,K 
              (see section \ref{analysis}).\label{ec2_sumobs}} 
\tablewidth{0pt} 
\tablehead{ 
Molecule	&  Transition	&  Freq.	&  $T_\mathrm{mb}$	&  $\Delta V$	&  $\Delta\nu$	&  $rms$	&  $N$  	\\
		&  		&  (GHz)	&  			& (km\,s$^{-1}$)	& (km\,s$^{-1}$)	&  (K)		&  (cm$^{-2}$) 
} 
\startdata 
CO\tablenotemark{a}	&  1--0		&  115.271	&  7.863	&  2.70	&  0.650	&  0.151	&  $2.48 \pm 0.06 \times 10^{16}$	\\ 
$^{13}$CO	&  1--0		&  110.201	&  1.499	&  2.24	&  0.272	&  0.131	&  $4.24 \pm 0.45 \times 10^{15}$	\\ 
C$^{18}$O	&  1--0		&  109.782	&  0.119	&  2.05	&  0.683	&  0.027	&  $3.11 \pm 0.94 \times 10^{14}$	\\ 
C$^{17}$O	&  1--0		&  112.359	&  ---		&  ---	&  0.667	&  0.012	&  	   $<5.62 \times 10^{13}$	\\ 
CO		&  2--1		&  230.538	&  6.289	&  2.76	&  0.406	&  0.026	&  $8.81 \pm 0.05 \times 10^{15}$	\\ 
$^{13}$CO	&  2--1		&  220.399	&  1.897	&  1.87	&  0.425	&  0.073	&  $1.90 \pm 0.09 \times 10^{15}$	\\ 
C$^{18}$O	&  2--1		&  219.560	&  0.167	&  1.48	&  0.427	&  0.036	&  $1.33 \pm 0.41 \times 10^{14}$	\\ 
CO		&  3--2		&  345.796	&  3.715	&  2.56	&  0.542	&  0.020	&  $4.92 \pm 0.03 \times 10^{15}$	\\ 
$^{13}$CO	&  3--2		&  330.588	&  0.716	&  1.69	&  0.567	&  0.027	&  $6.37 \pm 0.33 \times 10^{14}$	\\ 
C$^{18}$O	&  3--2		&  329.330	&  ---		&  ---	&  0.569	&  0.022	&  	   $<3.94 \times 10^{13}$	\\ 
C{\sc i}	&  $^3P_1$--$^3P_0$ &  492.161	&  2.354	&  2.23	&  0.190	&  0.412	&  $7.69 \pm 1.45 \times 10^{16}$	\\ 
CS		&  2--1		&  97.981	&  0.294	&  2.02	&  0.306	&  0.031	&  $3.38 \pm 0.45 \times 10^{12}$	\\ 
CS		&  3--2		&  146.969	&  0.071	&  1.47	&  0.510	&  0.013	&  $3.75 \pm 0.88 \times 10^{11}$	\\ 
C$^{34}$S	&  3--2		&  144.617	&  ---		&  ---	&  0.207	&  0.015	&  	   $<1.13 \times 10^{11}$	\\ 
CN		&  1,$\frac{3}{2},\frac{3}{2}$--0,$\frac{1}{2},\frac{1}{2}$	
				&  113.488	&  0.032	&  1.64	&  0.660	&  0.008	&  $2.29 \pm 0.80 \times 10^{12}$	\\ 
CN		&  1,$\frac{3}{2},\frac{5}{2}$--0,$\frac{1}{2},\frac{3}{2}$	                                                   
				&  113.491	&  0.042	&  2.18	&  0.660	&  0.008	&					\\ 
SO		&  1$_0$--0$_1$	&  30.002	&  0.095	&  1.80	&  0.390	&  0.018	&  $1.88 \pm 0.45 \times 10^{13}$	\\ 
SO		&  3$_2$--2$_1$	&  99.300	&  0.133	&  1.95	&  0.302	&  0.016	&  $4.17 \pm 0.63 \times 10^{12}$	\\ 
DCO$^+$		&  1--0		&  72.039	&  ---		&  ---	&  0.416	&  0.018	&  	   $<1.17 \times 10^{11}$	\\ 
H$^{13}$CO$^+$\tablenotemark{b}	& 1--0 & 86.754 &  ---		&  ---	&  0.346	&  0.019	&  	   $<8.27 \times 10^{10}$	\\ 
HCO$^+$		&  1--0		&  89.189	&  0.351	&  2.74	&  0.840	&  0.035	&  $1.45 \pm 0.20 \times 10^{12}$	\\ 
HCO$^+$		&  3--2		&  267.558	&  ---		&  ---	&  0.350	&  0.013	&  	   $<1.79 \times 10^{10}$	\\ 
DCN\tablenotemark{b} &  1--0	&  72.415	&  ---		&  ---	&  0.414	&  0.148	&  	   $<1.65 \times 10^{12}$	\\ 
H$^{13}$CN	&  1--0		&  86.340	&  ---		&  ---	&  0.347	&  0.014	&  	   $<1.03 \times 10^{11}$	\\ 
HCN		&  1--0		&  88.632	&  0.274	&  1.54	&  0.338	&  0.032	&  $1.10 \pm 0.30 \times 10^{12}$	\\ 
HNC		&  1--0		&  90.664	&  0.083	&  1.74	&  0.331	&  0.018	&  $3.48 \pm 1.13 \times 10^{11}$	\\ 
C$_2$D\tablenotemark{b}	&  1--0	&  72.108	&  ---		&  ---	&  0.416	&  0.021	&  	   $<3.32 \times 10^{12}$	\\ 
C$_2$H		&  1--0		&  87.317	&  0.199	&  2.89	&  0.858	&  0.019	&  $5.15 \pm 0.64 \times 10^{13}$	\\ 
C$_2$H		&  		&  87.329	&  0.085	&  1.86	&  0.858	&  0.019	&					\\ 
N$_2$H$^+$	&  1--0		&  93.174	&  ---		&  ---	&  0.322	&  0.020	&  	   $<9.67 \times 10^{10}$	\\ 
H$_2$CO		&  1$_{1,0}$--1$_{1,1}$	&    4.830  &  -0.256	&  3.2	&  0.303	&  0.055	&  $5.61 \pm 1.20 \times 10^{12}$	\\ 
H$_2$CO		&  2$_{1,1}$--2$_{1,2}$	&   14.488  &  -0.021	&  1.3	&  1.515	&  0.011	&					\\ 
H$_2$CO		&  2$_{1,2}$--1$_{1,1}$	&  140.840  &   0.085	&  2.48	&  0.532	&  0.022	&					\\ 
H$_2$CO		&  2$_{1,1}$--1$_{1,0}$	&  150.498  &   0.054	&  1.83	&  0.498	&  0.008	&					\\ 
HDCO\tablenotemark{b} &  2$_{1,1}$--1$_{1,0}$ &  134.285 & ---	&  ---	&  0.223	&  0.012 	&	   $<6.44 \times 10^{10}$	\\ 
NH$_3$		&  1--1		&  23.694	&  0.040	&  2.0	&  0.500	&  0.006	&  $5.77 \pm 0.87 \times 10^{12}$	\\ 
NH$_3$		&  2-2		&  23.723	&  0.015	&  2.6	&  0.500	&  0.005	&					\\ 
HC$_3$N		&  9--8		&  81.881	&  ---		&  ---	&  0.366	&  0.021	&  	   $<2.62 \times 10^{11}$	\\ 
C$_3$H$_2$	&  2$_{1,2}$--1$_{0,1}$	&  85.338 &  0.098	&  1.92	&  0.351	&  0.039	&					\\ 
CH$_3$OH	&  2$_{-1}$--1$_{-1}$\,E&  96.739 &  0.060	&  0.93	&  0.310	&  0.012	&					\\  
CH$_3$OH	&  2$_0$--1$_0$\,A$^+$	&  96.741 &  0.043	&  1.49	&  0.310	&  0.012	&					\\ 
\enddata
\tablenotetext{a}{Data taken with the ARO 12\,m in Jan. 2006.}
\tablenotetext{b}{Data taken with the ARO 12\,m in Feb. 2005.}
\end{deluxetable} 

\clearpage

\begin{deluxetable}{llcccccccccccc}  
\tabletypesize{\footnotesize} 
\tablecaption{Comparison of line temperatures and column densities for EC2 derived from Large Velocity Gradient (LVG) models 
              and observed deconvolved integrated intensities (see section \ref{analysis}, where total column densities are 
              calculated from individual observed transitions).\label{ec2_lvgcomps}} 
\tablewidth{0pt} 
\small
\tablehead{ 
\multicolumn{2}{l}{Molecule and}	&  $T_\mathrm{mb}$	&  $T_\mathrm{lvg}$	&  $N_\mathrm{mb}$(X)	&  $N_\mathrm{lvg}$/	&  $N_\mathrm{lvg}$(X)	&  $n_\mathrm{lvg}$(X)	&  $n$(H$_{2}$)	&  $\tau$	&  $T_\mathrm{ex}$	\\
\multicolumn{2}{l}{Transition}		&  (K)			&  (K)			&  			&  $N_\mathrm{mb}$	&  			&  			&  		&  		&  (K)			
} 
\startdata  
CO		&  1--0			&  7.863	&  7.850	&  $2.48 \times 10^{16}$	&  1.0	&  $2.50 \times 10^{16}$	&  $3.00 \times 10^{-3}$	&  $3.20 \times 10^{3}$	&  0.622	&  20.40	\\
$^{13}$CO	&  1--0			&  1.499	&  1.500	&  $4.24 \times 10^{15}$	&  0.7	&  $2.76 \times 10^{15}$	&  $4.00 \times 10^{-4}$	&  $3.20 \times 10^{3}$	&  0.056	&  30.90	\\
C$^{18}$O	&  1--0			&  0.119	&  0.124	&  $3.11 \times 10^{14}$	&  0.6	&  $1.96 \times 10^{14}$	&  $3.10 \times 10^{-5}$	&  $3.20 \times 10^{3}$	&  0.004	&  37.50	\\
CO		&  2--1			&  6.289	&  7.720	&  $8.81 \times 10^{15}$	&  2.9	&  $2.55 \times 10^{16}$	&  $3.00 \times 10^{-3}$	&  $3.20 \times 10^{3}$	&  2.220	&  13.60	\\
$^{13}$CO	&  2--1			&  1.897	&  1.810	&  $1.90 \times 10^{15}$	&  1.2	&  $2.31 \times 10^{15}$	&  $4.00 \times 10^{-4}$	&  $3.20 \times 10^{3}$	&  0.388	&  10.20	\\
C$^{18}$O	&  2--1			&  0.167	&  0.157	&  $1.33 \times 10^{14}$	&  1.1	&  $1.42 \times 10^{14}$	&  $3.10 \times 10^{-5}$	&  $3.20 \times 10^{3}$	&  0.033	&   9.42	\\
CO		&  3--2			&  3.715	&  3.760	&  $4.92 \times 10^{15}$	&  4.8	&  $2.37 \times 10^{16}$	&  $3.00 \times 10^{-3}$	&  $3.20 \times 10^{3}$	&  2.120	&  10.50	\\
$^{13}$CO	&  3--2			&  0.716	&  0.699	&  $6.37 \times 10^{14}$	&  3.3	&  $2.09 \times 10^{15}$	&  $4.00 \times 10^{-4}$	&  $3.20 \times 10^{3}$	&  0.271	&   8.62	\\
C$^{18}$O	&  3--2			&   --- 	&  0.056	&  $3.94 \times 10^{13}$	&  4.9	&  $1.91 \times 10^{14}$	&  $3.10 \times 10^{-5}$	&  $3.20 \times 10^{3}$	&  0.020	&   8.44	\\
CS		&  2--1			&  0.294	&  0.297	&  $3.38 \times 10^{12}$	&  6.8	&  $2.31 \times 10^{13}$	&  $3.70 \times 10^{-6}$	&  $1.20 \times 10^{4}$	&  1.370	&   3.22	\\
CS		&  3--2			&  0.071	&  0.071	&  $3.75 \times 10^{11}$	&  45	&  $1.68 \times 10^{13}$	&  $3.70 \times 10^{-6}$	&  $1.20 \times 10^{4}$	&  5.620	&   3.00	\\
SO		&  1$_0$--0$_1$		&  0.095	&  0.096	&  $1.88 \times 10^{13}$	&  0.2	&  $3.11 \times 10^{12}$	&  $5.60 \times 10^{-7}$	&  $1.40 \times 10^{5}$	&  -0.017	&  -2.83	\\
SO		&  3$_2$--2$_1$		&  0.133	&  0.129	&  $4.17 \times 10^{12}$	&  0.8	&  $3.37 \times 10^{12}$	&  $5.60 \times 10^{-7}$	&  $1.40 \times 10^{5}$	&  0.045	&   6.62	\\
HCO$^{+}$	&  1--0			&  0.351	&  0.346	&  $1.45 \times 10^{12}$	&  1.9	&  $2.70 \times 10^{12}$	&  $3.20 \times 10^{-7}$	&  $1.20 \times 10^{4}$	&  0.659	&   3.57	\\
HCN		&  1--0			&  0.274	&  0.274	&  $1.10 \times 10^{12}$	&  0.9	&  $9.50 \times 10^{11}$	&  $2.00 \times 10^{-7}$	&  $1.40 \times 10^{5}$	&  0.079	&   4.50	\\
HNC		&  1--0			&  0.083	&  0.080	&  $3.48 \times 10^{11}$	&  7.7	&  $2.68 \times 10^{12}$	&  $5.00 \times 10^{-7}$	&  $1.20 \times 10^{4}$	&  0.656	&   2.93	\\
N$_2$H$^{+}$	&  1--0			&   --- 	&  0.028	&  $9.67 \times 10^{10}$	&  1.8	&  $1.70 \times 10^{11}$	&  $2.75 \times 10^{-8}$	&  $1.20 \times 10^{4}$	&  0.142	&   3.51	\\
H$_2$CO		&  1$_{1,0}$--1$_{1,1}$	&  -0.256	&  -0.250	&  $5.61 \times 10^{12}$	&  2.5	&  $1.38 \times 10^{13}$	&  $1.40 \times 10^{-6}$	&  $1.20 \times 10^{4}$	&  0.152	&   0.95	\\
H$_2$CO		&  2$_{1,1}$--2$_{1,2}$	&  -0.021	&  -0.021	&  				&  1.0	&  $5.61 \times 10^{12}$	&  $1.40 \times 10^{-6}$	&  $1.20 \times 10^{4}$	&  0.016	&   1.43	\\
H$_2$CO		&  2$_{1,2}$--1$_{1,1}$	&  0.085	&  0.096	&  				&  1.9	&  $1.07 \times 10^{13}$	&  $1.40 \times 10^{-6}$	&  $1.20 \times 10^{4}$	&  0.623	&   3.05	\\
H$_2$CO		&  2$_{1,1}$--1$_{1,0}$	&  0.054	&  0.049	&  				&  1.4	&  $7.90 \times 10^{12}$	&  $1.40 \times 10^{-6}$	&  $1.20 \times 10^{4}$	&  0.501	&   2.94	\\
\enddata  
\tablecomments{{LVG} column densities calculated with $N_\mathrm{lvg}$(X) = $n_\mathrm{lvg}$(X) $\times$ 1\,pc $\times$ $\Delta V$.}
\end{deluxetable}

\clearpage

\begin{deluxetable}{llccccrr} 
\tablecaption{Comparison of molecular abundances ratios relative to HCO$^{+}$ and H$_{2}$ in EC2 (position A), 
	 L134N (center), TMC-1 (average) and L134N (range) \citep{dickens0010,pratap9709}, 
         Trans(lucent) cloud observations by \citep{turner0010}.
	 For EC2 $T_\mathrm{ex}$ = 20\,K, $n$(H$_{2}$) = $1.2 \times 10^{4}$\,cm$^{-3}$ and $N$(H$_{2}$) = $7.4 \times 10^{22}$\,cm$^{-2}$ 
	 assumed in calculation of fractional abundances from individual observed transitions (see section \ref{analysis}).\label{ec2_sumdens}} 
\tablewidth{0pt} 
\small
\tablehead{ 
Molecule	&  	&  X/HCO$^{+}$	&  L134N	&  TMC-1	&  L134N range	&  \multicolumn{1}{c}{Trans}    & \multicolumn{1}{c}{X/H$_{2}$}	
} 
\startdata  
CO		&  1--0	&  17078	&  11000\tablenotemark{b}	&  7800\tablenotemark{b}	&  		&		&     $3.3 \times 10^{-07}$	\\
$^{13}$CO	&  1--0	&   2919	&  172\tablenotemark{a}	&  122\tablenotemark{a}	&  111-188	&		&     $5.7 \times 10^{-08}$	\\
C$^{18}$O	&  1--0	&    214	&  22.00		&  15.60		&  14.2-24	&		&     $4.2 \times 10^{-09}$	\\
C$^{17}$O	&  1--0	&  $<$38.8	&  ---			&  ---			&  		&		&  $<$$7.6 \times 10^{-10}$	\\
CO		&  2--1	&   6070	&  11000\tablenotemark{b}	&  7800\tablenotemark{b}	&  		&		&     $1.2 \times 10^{-07}$	\\
$^{13}$CO	&  2--1	&   1308	&  172\tablenotemark{a}	&  122\tablenotemark{a}	&  		&		&     $2.6 \times 10^{-08}$	\\
C$^{18}$O	&  2--1	&     92	&  22.00		&  15.60		&  		&		&     $1.8 \times 10^{-09}$	\\
CO		&  3--2	&   3389	&  11000\tablenotemark{b}	&  7800\tablenotemark{b}	&  		&		&     $6.6 \times 10^{-08}$	\\
$^{13}$CO	&  3--2	&    439	&  172\tablenotemark{a}	&  122\tablenotemark{a}	&  		&		&     $8.6 \times 10^{-09}$	\\
C$^{18}$O	&  3--2	&  $<$27.2	&  22.00	&  15.60	&  		&				&  $<$$5.3 \times 10^{-10}$	\\
C{\sc i}	&  	&  53020	&  ---		&  ---		&  		&				&     $1.0 \times 10^{-06}$	\\
CS		&  2--1	&      2.33	&  0.124	&  0.320	&  0.069-0.138	&  $1.1 \times 10^{-08}$	&     $4.6 \times 10^{-11}$	\\
CS		&  3--2	&      0.26	&  0.124	&  0.320	&  		&				&     $5.1 \times 10^{-12}$	\\
C$^{34}$S	&  3--2	&   $<$0.08	&  ---		&  ---		&  		&				&  $<$$1.5 \times 10^{-12}$	\\
CN\tablenotemark{d}	&  	&      1.58	&  0.061	&  0.070      & $<$0.045-0.069	&				&     $3.1 \times 10^{-11}$	\\
SO		&  1--0	&     13.0	&  0.719	&  0.130	&  0.264-0.738	&  $3.2 \times 10^{-08}$	&     $2.5 \times 10^{-10}$	\\
SO		&  3--2	&      2.87	&  0.719	&  0.130	&  0.264-0.738	&				&     $5.6 \times 10^{-11}$	\\
DCO$^{+}$	&  1--0	&   $<$0.08	&  ---		&  ---		&  		&				&  $<$$1.6 \times 10^{-12}$	\\
H$^{13}$CO$^{+}$&  1--0	&   $<$0.06	& \tablenotemark{c}	& \tablenotemark{c}	&  		&				&  $<$$1.1 \times 10^{-12}$	\\
HCO$^{+}$	&  1--0	&  		&  		&  		&  		&  $2   \times 10^{-09}$	&     $2.0 \times 10^{-11}$	\\
DCN		&  1--0	&   $<$1.13     &  		&  		&  		&				&  $<$$2.2 \times 10^{-11}$	\\
H$^{13}$CN	&  1--0	&   $<$0.07	& \tablenotemark{c}	& \tablenotemark{c}	&  		&				&  $<$$1.4 \times 10^{-12}$	\\
HCN		&  1--0	&      0.76	&  0.925	&  0.490	&  0.555-0.968	&  $3.6 \times 10^{-08}$	&     $1.5 \times 10^{-11}$	\\
HNC		&  1--0	&      0.24	&  3.251	&  1.680	&  1.324-3.963	&  $2.5 \times 10^{-09}$	&     $4.7 \times 10^{-12}$	\\
C$_2$H\tablenotemark{e} &  1--0 & 35.5	&  0.288	&  0.300	&  0.171-0.333	&  $6.6 \times 10^{-08}$	&     $7.0 \times 10^{-10}$	\\
N$_2$H$^{+}$	&  1--0	&   $<$0.07	&  0.077	&  0.013	&  0.031-0.077	&  $\sim 1 \times 10^{-09}$	&  $<$$1.3 \times 10^{-12}$	\\
H$_2$CO		&  	&      3.87	&  		&  		&  		&  $6.3 \times 10^{-09}$	&     $7.6 \times 10^{-11}$	\\
NH$_3$		&  	&      4.0	&  7.635	&  2.770	&  4.284-9.127	&  $2.1 \times 10^{-08}$	&     $7.8 \times 10^{-11}$	\\
HC$_3$N		&  9--8	&   $<$0.18	&  0.054	&  0.150	&  0.030-0.073	&  $5   \times 10^{-10}$	&  $<$$3.5 \times 10^{-12}$	\\
CH$_3$OH	&  2--1	&	  	&  0.641	&  0.099	&  0.311-0.641	&  $1.8 \times 10^{-08}$	&				\\
\enddata  
\small
\tablenotetext{a}{A $^{13}$CO/C$^{18}$O ratio of 7.81 was assumed in L134N and TMC-1.}
\tablenotetext{b}{A $^{12}$CO/C$^{18}$O ratio of 500 was assumed in L134N and TMC-1.}
\tablenotetext{c}{A $^{12}$C/$^{13}$C ratio of 64 was assumed in L134N and TMC-1.}
\tablenotetext{d}{From the sum of the two 113.49 GHz components assuming a relative intensity for this line of 0.456.}
\tablenotetext{e}{From the 87.317 GHz component assuming a relative intensity for this line of 0.4167.}
\end{deluxetable}

\clearpage

\begin{deluxetable}{lccccc}  
\tablewidth{0pt}
\tablecaption{Range of dust to gas ratios and extinction for EC2 based on $D$ and $\beta$.\label{ec2_mass}} 
\tablehead{ 
$\beta$						&  1		&  1.25		&  1.5		&  1.75		&  2 
} 
\startdata 
$Q_{1200}$					&  7.81e-5	&  4.44e-5	&  2.52e-5	&  1.43e-5	&  8.14e-6	\\
$\kappa_{\nu}$ (cm$^{2}$\,g$^{-1}$)		&  1.953	&  1.110	&  0.630	&  0.358	&  0.203	\\
$M_\mathrm{dust}/M_\mathrm{gas}$ ($D$ = 14\,kpc)&  0.0015	&  0.0027	&  0.0048	&  0.0084	&  0.0148	\\
$M_\mathrm{dust}/M_\mathrm{gas}$ ($D$ = 20\,kpc)&  0.0011	&  0.0019	&  0.0033	&  0.0059	&  0.0104	\\
$A_\mathrm{V}$ (mag) ($R_\mathrm{V}=3.1$)	&  4.6		&  8.0		&  14.1		&  24.9		&  43.8		\\
$A_\mathrm{V}$ (mag) ($R_\mathrm{V}=2.0$)	&  2.9		&  5.2		&   9.1		&  16.1		&  28.3		\\
\enddata 
\tablecomments{$Q_{\nu}$ and $\kappa_{\nu}$ calculated from the {\it 1983 Chicago Assumptions} of \citet{hildebrand8309}.} 
\end{deluxetable}

\begin{deluxetable}{lrclr} 
\tablecaption{Initial abundances relative to H. Most (C$^{+}$, N, O, S, Si and Fe$^{+}$) have been reduced 
	      by a factor of 5 from those typically used to model the local ISM.\label{ec2_inabund}} 
\tablewidth{0pt} 
\tablehead{ 
Species		&  Abundance			&  $\;$	&  Species	&  Abundance 
} 
\startdata  
He		&  $1.4  \times 10^{-1}$	&	&  HD		&  $3.2  \times 10^{-5}$	\\ 
C$^{+}$		&  $1.46 \times 10^{-5}$	&	&  N		&  $4.28 \times 10^{-6}$	\\ 
O		&  $3.51 \times 10^{-5}$	&	&  S		&  $2.0  \times 10^{-8}$	\\ 
Si		&  $3.99 \times 10^{-9}$	&	&  Fe$^{+}$	&  $1.99 \times 10^{-9}$	\\ 
\enddata  
\end{deluxetable}

\begin{deluxetable}{lccccc}
\tablecaption{EC2 chemical model agreement results ranked by fit as described in section \ref{chemods}, 
where $n$(H$_{2}$) = $1.2 \times 10^{4}$\,cm$^{-3}$, $T$ = 20\,K, $A_\mathrm{V}$ in mag, 
CRI in terms of the standard ISM rate of $1.3 \times 10^{-17}$\,s$^{-1}$, UV field in terms of the local ISM, 
and most initial abundances (IA) reduced by factor indicated from typical local ISM values. \label{ec2_chemod_fits}}
\tablewidth{0pt} 
\tablehead{ 
Model	&  Fit 	&  $A_\mathrm{V}$	&  CRI $\times$	&  UV $\times$	&  IA $\times$ 
} 
\startdata  
1	&  0.01	&  10	&  1	&  1	&  0.2	\\
5	&  0.11	&  1	&  20	&  0.1	&  0.2	\\
3	&  0.16	&  2	&  10	&  1	&  0.2	\\
8	&  0.40	&  2	&  20	&  1	&  0.1	\\
10	&  0.41	&  1	&  20	&  1	&  1	\\
9	&  0.42	&  1	&  20	&  1	&  0.4	\\
2	&  0.43	&  1	&  20	&  1	&  0.2	\\
4	&  0.43	&  2	&  10	&  10	&  0.2	\\
6	&  0.43	&  1	&  20	&  20	&  0.2	\\
7	&  0.44	&  3	&  20	&  40	&  0.2	\\
\enddata  
\end{deluxetable}

\begin{deluxetable}{lrcccc} 
\tablecaption{Comparison of observations of EC2 with four different model predictions at steady-state. 
Abundances given relative to $N$(H$_{2}$) = $7.4 \times 10^{22}$\,cm$^{-2}$, with $n$(H$_{2}$) = $1.2 \times 10^{4}$\,cm$^{-3}$ and $T$ = 20\,K. Note that the abundances of the $^{13}$C species have been multiplied by 60 and that of the C$^{18}$O species by 500.
Most model initial abundances relative to H reduced by a factor of 5 from typical local ISM values. 
Bold type indicates agreement to within a factor of 5. \label{ec2_abundModels1-2-3-4}} 
\tablewidth{0pt} 
\tablehead{ 
Species &  Observed  &  Model~1              &  Model~2              &  Model~3              &  Model~4       \\ 
        &  in EC2    &  $A_\mathrm{V}$ = 10 mag     &  $A_\mathrm{V}$ = 1 mag      &  $A_\mathrm{V}$ = 2 mag      &  $A_\mathrm{V}$ = 2 mag\\ 
        &            &  CRI = 1 $\times$     &  CRI = 20 $\times$    &  CRI = 10 $\times$    &  CRI = 10 $\times$      \\ 
        &            &  UV = 1 $\times$      &  UV = 1 $\times$      &  UV = 1 $\times$      &  UV = 10 $\times$ 
} 
\startdata  
$^{13}$CO 	&    1.45e-06 &      2.90e-05    &      2.51e-07    & {\bf 2.22e-06 }  &      2.28e-07    \\ 
C 		&    1.04e-06 &      1.41e-08    &      1.37e-05    &      2.47e-05    &      1.79e-05    \\ 
C$^{18}$O 	&    1.36e-06 &      2.90e-05    &      2.51e-07    & {\bf 2.22e-06 }  &      2.28e-07    \\ 
C$_{2}$D 	& $<$4.49e-11 &  {\bf 1.14e-10}  & {\bf 1.48e-11 }  &      6.70e-11    & {\bf 1.18e-11 }  \\ 
C$_{2}$H 	&    6.96e-10 & {\bf 1.65e-09 }  &      1.71e-08    &      2.79e-08    &      1.04e-08    \\ 
CN 		&    3.09e-11 &      7.61e-09    &      1.24e-09    &      6.82e-09    &      1.51e-09    \\ 
CO 		&    1.22e-07 &      2.90e-05    & {\bf 2.51e-07 }  &      2.22e-06    & {\bf 2.28e-07 }  \\ 
CS (2--1) 	&    4.57e-11 &      6.97e-09    &      2.87e-10    &      2.11e-09    & {\bf 1.46e-10 }  \\ 
CS (3--2) 	&    5.07e-12 &      6.97e-09    &      2.87e-10    &      2.11e-09    &      1.46e-10    \\ 
DCN 		& $<$2.23e-11 &      9.37e-10    & {\bf 1.24e-14 }  & {\bf 2.98e-13 }  & {\bf 2.08e-14 }  \\ 
DCO$^{+}$ 	& $<$1.58e-12 &      1.78e-09    & {\bf 6.16e-15 }  & {\bf 5.12e-13 }  & {\bf 7.53e-15 }  \\ 
H$^{13}$CN 	& $<$8.35e-11 &      6.13e-09    & {\bf 1.07e-11 }  &      1.28e-10    & {\bf 1.36e-11 }  \\ 
H$^{13}$CO$^{+}$ & $<$6.71e-11 &      7.26e-09   & {\bf 6.60e-12 }  & {\bf 5.36e-11 }  & {\bf 5.16e-12 }  \\ 
H$_{2}$CO 	&    7.58e-11 &      1.60e-08    & {\bf 7.70e-11 }  &      1.19e-09    & {\bf 9.38e-11 }  \\ 
HC$_{3}$N 	& $<$3.54e-12 &      1.55e-10    & {\bf 7.26e-13 }  & {\bf 1.19e-11}   & {\bf 2.57e-13 }  \\ 
HCN 		&    1.49e-11 &      6.13e-09    & {\bf 1.07e-11 }  &      1.28e-10    & {\bf 1.36e-11 }  \\ 
HCO$^{+}$ 	&    1.96e-11 &      7.26e-09    & {\bf 6.60e-12 }  & {\bf 5.36e-11 }  & {\bf 5.16e-12 }  \\ 
HDCO 		& $<$8.70e-13 &      4.08e-09    & {\bf 2.68e-13 }  &      2.59e-11    & {\bf 4.44e-13 }  \\ 
HNC 		&    4.70e-12 &      1.58e-08    & {\bf 2.09e-11 }  &      4.05e-10    &      2.48e-11    \\ 
N$_{2}$H$^{+}$	& $<$1.31e-12 &      4.21e-10    & {\bf 1.88e-14 }  & {\bf 1.55e-12}   & {\bf 4.75e-14 }  \\ 
NH$_{3}$	&    7.80e-11 &      1.21e-07    &      2.02e-13    & {\bf 5.14e-11 }  &      3.04e-13    \\ 
SO		&    1.34e-10 &      7.16e-09    &      6.43e-16    &      1.47e-14    &      6.16e-16    \\ 
\enddata
\tablecomments{
For Tables \ref{ec2_abundModels1-2-3-4}, \ref{ec2_abundModels2-5-6-7} and \ref{ec2_abundModels2-8-9-10}: Models only consider main isotopes (with the exception of H and D), so for $^{13}$CO, H$^{13}$CN and H$^{13}$CO$^{+}$, models are compared to observed abundances $\times$ 60; and for C$^{18}$O, models are compared to observed abundance $\times$ 500. 
For multiple transition detections of $^{12}$CO, $^{13}$CO and C$^{18}$O, a $\sigma$ weighted mean of each isotopomer's observed column density is used, where $\sigma = T_\mathrm{mb} / rms_\mathrm{mb}$.}
\end{deluxetable}

\begin{deluxetable}{lrccc} 
\tablecaption{Comparison of observations of EC2 with three different model predictions at steady-state, but with varying UV field. 
Abundances given relative to $N$(H$_{2}$) = $7.4 \times 10^{22}$\,cm$^{-2}$, with $n$(H$_{2}$) = $1.2 \times 10^{4}$\,cm$^{-3}$ and $T$ = 20\,K.  Note that the abundances of the $^{13}$C species have been multiplied by 60 and that of the C$^{18}$O species by 500. 
Most model initial abundances relative to H reduced by a factor of 5 from typical local ISM values. 
Bold type indicates agreement to within a factor of 5. \label{ec2_abundModels2-5-6-7}} 
\tablewidth{0pt} 
\tablehead{ 
Species &  Observed  &  Model~5              &  Model~6              &  Model~7              \\ 
        &  in EC2    &  $A_\mathrm{V}$ = 1 mag    &  $A_\mathrm{V}$ = 1 mag      &  $A_\mathrm{V}$ = 3 mag      \\ 
        &            &  CRI = 20 $\times$    &  CRI = 20 $\times$    &  CRI = 20 $\times$    \\ 
        &            &  UV = 0.1 $\times$    &  UV = 20 $\times$     &  UV = 40 $\times$ 
} 
\startdata  
$^{13}$CO 	&    1.45e-06 & {\bf 2.79e-06 }  &      6.66e-09    & {\bf 3.39e-07 }  \\ 
C 		&    1.04e-06 &      2.23e-05    & {\bf 1.55e-06 }  &      2.50e-05    \\ 
C$^{18}$O 	&    1.36e-06 & {\bf 2.79e-06 }  &      6.66e-09    & {\bf 3.39e-07 }  \\ 
C$_{2}$D 	& $<$4.49e-11 & {\bf 1.17e-10}   & {\bf 1.03e-12 }  & {\bf 5.90e-12 }  \\ 
C$_{2}$H 	&    6.96e-10 &      6.97e-08    & {\bf 1.34e-09 }  &      3.90e-09    \\ 
CN 		&    3.09e-11 &      1.24e-08    & {\bf 3.98e-11 }  &      8.46e-10    \\ 
CO 		&    1.22e-07 &      2.79e-06    &      6.66e-09    & {\bf 3.39e-07 }  \\ 
CS (2--1) 	&    4.57e-11 &      7.17e-09    &      3.30e-12    & {\bf 5.92e-11 }  \\ 
CS (3--2) 	&    5.07e-12 &      7.17e-09    & {\bf 3.30e-12 }  &      5.92e-11    \\ 
DCN 		& $<$2.23e-11 & {\bf 4.51e-13 }  & {\bf 2.27e-15 }  & {\bf 1.13e-14 }  \\ 
DCO$^{+}$ 	& $<$1.58e-12 & {\bf 6.42e-13 }  & {\bf 8.48e-16 }  & {\bf 1.86e-14 }  \\ 
H$^{13}$CN 	& $<$8.35e-11 & {\bf 2.44e-10}   & {\bf 2.13e-12 }  & {\bf 7.56e-12 }  \\ 
H$^{13}$CO$^{+}$ & $<$6.71e-11 & {\bf 8.63e-11}  & {\bf 1.13e-12 }  & {\bf 9.94e-12 }  \\ 
H$_{2}$CO 	&    7.58e-11 &      1.14e-09    &      4.09e-12    & {\bf 8.30e-11 }  \\ 
HC$_{3}$N 	& $<$3.54e-12 &      5.24e-11    & {\bf 1.00e-15 }  & {\bf 1.09e-13 }  \\ 
HCN 		&    1.49e-11 &      2.44e-10    &      2.13e-12    & {\bf 7.56e-12 }  \\ 
HCO$^{+}$ 	&    1.96e-11 & {\bf 8.63e-11 }  &      1.13e-12    & {\bf 9.94e-12 }  \cr
HDCO 		& $<$8.70e-13 &      1.98e-11    & {\bf 8.56e-15 }  & {\bf 6.28e-13 }  \cr
HNC 		&    4.70e-12 &      5.57e-10    & {\bf 2.78e-12 }  &      3.68e-11    \cr
N$_{2}$H$^{+}$	& $<$1.31e-12 & {\bf 1.92e-12}   & {\bf 2.87e-17 }  & {\bf 6.52e-14 }  \cr
NH$_{3}$	&    7.80e-11 & {\bf 6.18e-11 }  &      3.83e-15    &      1.35e-12    \cr
SO		&    1.34e-10 &      2.96e-14    &      1.56e-17    &      9.25e-16    \cr
\enddata  
\end{deluxetable}

\begin{deluxetable}{lrccc} 
\tablecaption{Comparison of observations of EC2 with three different model predictions at steady-state, but with varying initial abundances. 
Abundances given relative to $N$(H$_{2}$) = $7.4 \times 10^{22}$\,cm$^{-2}$, with $n$(H$_{2}$) = $1.2 \times 10^{4}$\,cm$^{-3}$ and $T$ = 20\,K.  Note that the abundances of the $^{13}$C species have been multiplied by 60 and that of the C$^{18}$O species by 500.
Most model initial abundances relative to H reduced by a factor (IA shown below) from typical local ISM values. 
Bold type indicates agreement to within a factor of 5. \label{ec2_abundModels2-8-9-10}} 
\tablewidth{0pt} 
\tablehead{ 
Species &  Observed  &  Model~8              &  Model~9              &  Model~10              \\ 
        &  in EC2    &  $A_\mathrm{V}$ = 2 mag     &  $A_\mathrm{V}$ = 1 mag      &  $A_\mathrm{V}$ = 1 mag      \\ 
        &            &  CRI = 20 $\times$    &  CRI = 20 $\times$    &  CRI = 20 $\times$    \\ 
        &            &  UV = 1 $\times$      &  UV = 1 $\times$      &  UV = 1 $\times$      \\ 
        &            &  IA = 0.1 $\times$    &  IA = 0.4 $\times$    &  IA = 1 $\times$ 
} 
\startdata  
$^{13}$CO	&    1.45e-06 & {\bf 3.48e-07 }  & {\bf 6.36e-07 }  & {\bf 2.21e-06 }  \\ 
C 		&    1.04e-06 &      1.30e-05    &      3.05e-05    &      8.90e-05    \\ 
C$^{18}$O 	&    1.36e-06 & {\bf 3.48e-07 }  & {\bf 6.36e-07 }  & {\bf 2.21e-06 }  \\ 
C$_{2}$D 	& $<$4.49e-11 & {\bf 3.58e-12 }  & {\bf 1.22e-11 }  & {\bf 7.92e-12 }  \\ 
C$_{2}$H 	&    6.96e-10 & {\bf 1.92e-09 }  &      1.99e-08    &      1.98e-08    \\ 
CN 		&    3.09e-11 &      7.89e-10    &      3.69e-09    &      8.57e-09    \\ 
CO 		&    1.22e-07 & {\bf 3.48e-07 }  &      6.36e-07    &      2.21e-06    \\ 
CS (2--1) 	&    4.57e-11 & {\bf 6.44e-11 }  &      9.73e-10    &      3.46e-09    \\ 
CS (3--2) 	&    5.07e-12 &      6.44e-11    &      9.73e-10    &      3.46e-09    \\ 
DCN 		& $<$2.23e-11 & {\bf 8.97e-15 }  & {\bf 2.10e-14 }  & {\bf 3.19e-14 }  \\ 
DCO$^{+}$ 	& $<$1.58e-12 & {\bf 3.12e-14 }  & {\bf 4.19e-15 }  & {\bf 2.51e-15 }  \\ 
H$^{13}$CN 	& $<$8.35e-11 & {\bf 6.31e-12 }  & {\bf 3.04e-11 }  & {\bf 7.78e-11 }  \\ 
H$^{13}$CO$^{+}$ & $<$6.71e-11 & {\bf 1.06e-11 } & {\bf 6.18e-12 }  & {\bf 5.80e-12 }  \\ 
H$_{2}$CO 	&    7.58e-11 & {\bf 7.36e-11 }  & {\bf 7.73e-11 }  & {\bf 6.19e-11 }  \cr
HC$_{3}$N 	& $<$3.54e-12 & {\bf 1.85e-13 }  & {\bf 1.19e-12 }  & {\bf 1.24e-12 }  \cr
HCN 		&    1.49e-11 & {\bf 6.31e-12 }  & {\bf 3.04e-11 }  &      7.78e-11    \cr
HCO$^{+}$ 	&    1.96e-11 & {\bf 1.06e-11 }  & {\bf 6.18e-12 }  & {\bf 5.80e-12 }  \cr
HDCO 		& $<$8.70e-13 & {\bf 9.07e-13}   & {\bf 1.68e-13 }  & {\bf 7.19e-14 }  \cr
HNC 		&    4.70e-12 &      5.08e-11    &      4.23e-11    &      8.88e-11    \cr
N$_{2}$H$^{+}$	& $<$1.31e-12 & {\bf 4.78e-14 }  & {\bf 6.39e-14 }  & {\bf 1.65e-13 }  \cr
NH$_{3}$	&    7.80e-11 &      6.46e-12    &      2.80e-13    &      3.08e-13    \cr
SO		&    1.34e-10 &      1.35e-15    &      1.91e-15    &      6.54e-15    \cr
\enddata  
\end{deluxetable} 

\clearpage

\begin{deluxetable}{ll}
\tablecaption{Summary of EC2 properties. \label{ec2sumprops}}
\tablewidth{0pt} 
\tablehead{ 
Property			&  Value
} 
\startdata  
Position A			&  $\alpha = 02^\mathrm{h}\,48^\mathrm{m}\,38.5^\mathrm{s}$, 
				   $\delta = 58\arcdeg\,28\arcmin\,28.1\arcsec$ (J2000)	\\
				&  $l = 137.76$, $b = -0.98$	\\
Radial velocity			&  $v_\mathrm{rad} = -103.70$ km\,s$^{-1}$	\\
Galactocentric distance		&  $R = 22$ to 28\,kpc	\\
Distance from Galatic plane	&  250 to 350\,pc	\\
Size				&  $\theta_\mathrm{EC2}$ = 30 to 40\,pc\\
Temperature			&  $T$ = 20\,K	\\
Density				&  $n(\mathrm{H}_{2}) \sim 10^{4}$\,cm$^{-3}$	\\
Cosmic ray ionisation rate	&  CRI = 10--20 $\times\ 1.3 \times 10^{-17}$\,s$^{-1}$	\\
UV photon field			&  UV = 10--20 $\times$ local ISM values	\\
Initial abundance		&  IA = 20 per cent of local ISM values	\\
Extinction			&  $A_\mathrm{V} < 4$\,mag	\\
Dust to gas ratio		&  $M_\mathrm{dust} / M_\mathrm{gas} = 0.001$--$0.015$	\\
Mass				&  $M_\mathrm{EC2} \approx 10^{4}\,M_{\sun}$	\\
\enddata  
\end{deluxetable}

\clearpage

\clearpage 

\clearpage

\clearpage

\clearpage


\end{document}